\documentclass[%
 reprint,
 superscriptaddress,
 amsmath,amssymb,
 aps,
 longbibliography
]{revtex4-2}

\usepackage[utf8]{inputenc}
\usepackage{lmodern}
\usepackage[ngerman,english]{babel}
\usepackage[table]{xcolor}
\usepackage{threeparttablex}

\usepackage{scalerel,stackengine,mathtools,amsmath,amsthm}
\usepackage{amssymb}
\usepackage{esint}
\usepackage{physics}
\usepackage{tensor}
\usepackage{array}
\usepackage{upgreek}
\usepackage{textcomp}
\usepackage{bm}

\usepackage{graphicx}
\usepackage{float}
\makeatletter
\let\newfloat\newfloat@ltx
\makeatother
\usepackage[margin=10pt, font=small, labelfont=bf, justification=justified]{caption}
\usepackage{subcaption}
\usepackage[pdfborderstyle={/S/U/W 1}]{hyperref}
\usepackage{booktabs}
\usepackage{tabularx}
\usepackage{siunitx}

\DeclareMathAlphabet{\mathdutchcal}{U}{dutchcal}{m}{n}
\SetMathAlphabet{\mathdutchcal}{bold}{U}{dutchcal}{b}{n}
\DeclareMathAlphabet{\mathdutchbcal}{U}{dutchcal}{b}{n}
\DeclareMathOperator\arctanh{arctanh}

\newcommand{\state}[4]{\text{${#1}\text{#2}_{#3\text{#4}}$}}
\newcommand{\sym}[3]{$\text{#1}_{#2\text{#3}}$}

\DeclareSIUnit\atomicunit{a.u.}

\newcommand{\im}{\mathrm i}

\newcommand{\udl}{\underline}

\newcommand{\imag}{\imaginary}
\newcommand{\mdc}{\mathdutchcal}

\usepackage{xspace}

\newcommand{\etal}{\textit{et al.}\xspace}
\newcommand{\apriori}{\textit{a priori}\xspace}
\newcommand{\ie}{i.e.,\xspace}
\newcommand{\eg}{e.g.,\xspace}

\newcolumntype{C}[1]{>{\centering\let\newline\\\arraybackslash\hspace{0pt}}m{#1}}
\newcolumntype{L}[1]{>{\raggedright\let\newline\\\arraybackslash\hspace{0pt}}m{#1}}

\makeatletter				
\def\input@path{{graphics/}{graphics/}}
\makeatother

\graphicspath{{graphics/}}

\usepackage{xtab}

\usepackage{orcidlink}

\begin{document}
\title{Quantum-inspired Large-Scale Calculations of Vibronic Spectra as Benchmarks for Gaussian Boson Sampling}
\author{Oliver Kreuz\,\orcidlink{0000-0002-6369-8528}}
\affiliation{Fachbereich Chemie, Philipps-Universität Marburg, Hans-Meerwein-Straße 4, 35032 Marburg, Germany}
\affiliation{Marburg Center for Quantum Materials and Sustainable Technologies, Hans-Meerwein-Straße 6, 35032 Marburg, Germany}
\author{Steffen M. Giesen\,\orcidlink{0000-0001-7853-8511}}
\affiliation{Fachbereich Chemie, Philipps-Universität Marburg, Hans-Meerwein-Straße 4, 35032 Marburg, Germany}
\affiliation{Marburg Center for Quantum Materials and Sustainable Technologies, Hans-Meerwein-Straße 6, 35032 Marburg, Germany}
\author{Joonsuk Huh\,\orcidlink{0000-0002-8792-5641}}
\email{joonsukhuh@yonsei.ac.kr}
\affiliation{Department of Quantum Information, Yonsei University, Incheon 21983, Republic of Korea}
\affiliation{Department of Chemistry, Yonsei University, Seoul 03722, Republic of Korea}
\affiliation{Department of Computational Science and Engineering, Yonsei University, Seoul 03722, Republic of Korea}
\author{Robert Berger\,\orcidlink{0000-0002-9107-2725}}
\email{robert.berger@uni-marburg.de}
\affiliation{Fachbereich Chemie, Philipps-Universität Marburg, Hans-Meerwein-Straße 4, 35032 Marburg, Germany}
\affiliation{Marburg Center for Quantum Materials and Sustainable Technologies, Hans-Meerwein-Straße 6, 35032 Marburg, Germany}
\date{\today}
\selectlanguage{english}
\begin{abstract}
Estimating vibronic spectra using Gaussian Boson Sampling (GBS) provides a
physically motivated benchmark for evaluating the fidelity and
programmability of continuous-variable photonic quantum processors.
Assessing device performance and potential quantum advantage, however,
requires obtaining on classical hardware complementary state-resolved
reference data, whose computation becomes challenging when higher-order
vibronic effects such as Herzberg--Teller (HT) contributions are included.
For this purpose, a unified quantum-inspired framework suitable for
classical hardware is developed herein for simulating
Bogoliubov-transformed bosonic systems with Duschinsky mode mixing,
finite-temperature environments, and quadratic HT contributions.
The framework combines an autocorrelation-based method in a
time-dependent (TD) formulation for computing spectral profiles with a
state-resolved method in a time-independent (TI) formulation. Within the
TI formulation, generalized prescreening conditions are
derived for products of complex Gaussian unitaries corresponding to GBS
architectures. These conditions provide \textit{a priori} bounds on Fock-space
truncation errors relative to total transition probabilities, enabling
the discarding of combinatorial sets of negligible transitions before
evaluating individual elements. Benchmark computations on coronene
(102 vibrational modes, expanding to 204 effective modes for thermal excitations)
demonstrate controlled accuracy, recovering over 98\% of
the zero-temperature spectral intensity (and at least 84\% at 100\,K)
as well as resolving spectral features unique to the quadratic HT
terms. The complementary TD formulation gives access to the full profile without
Fock-space truncation. The unified framework presented provides an efficient
diagnostic tool to assess the limits of Fock-space methods on classical
hardware and facilitates access to large-scale error-controlled reference data
for validating photonic quantum processors.

\end{abstract}
\maketitle
\section{Introduction}
\begin{figure*}
  \centering
  \includegraphics[width=0.95\linewidth]{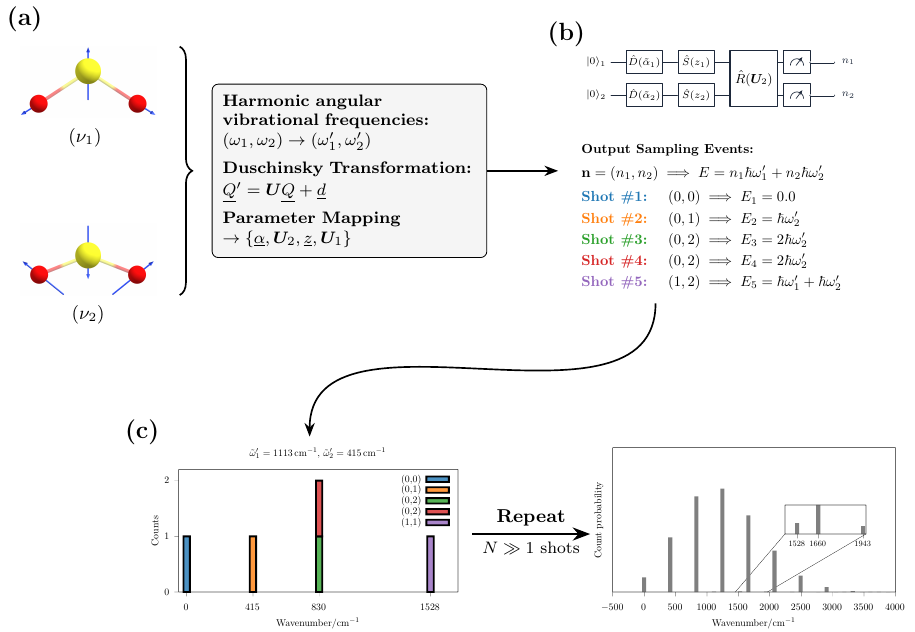}
  \caption{Illustration of Gaussian Boson Sampling (GBS) applied to the
  estimation of vibronic spectra within the Franck--Condon
  approximation.
  \textbf{(a)}~Vibrational parameters of modes $\nu_1$ and $\nu_2$
  (harmonic angular frequencies $\omega_i, \omega'_i$,
  Duschinsky rotation matrix $\mathbf{U}$, and normal-mode displacement $\udl{d}$)
  from electronic structure calculations are mapped onto Gaussian
  operations: single-mode displacements $\hat{D}(\udl{\tilde{\alpha}})$,
  single-mode squeezing $\hat{S}(\udl{z})$, and a final interferometer
  $\hat{R}(\bm{U}_2)$.
  \textbf{(b)}~The circuit is initialized in vacuum. Each shot is clocked
  by an active master trigger (e.g., optical pump sync or digital control
  cycle), actively defining each trial so that zero-photon outcomes $(0,0)$
  are recorded \cite{sparrow:2018,wang:2020}. Output configurations $\mathbf{n} = (n_1, n_2)$
  yield vibrational excess energies $E = \sum_i n_i \hbar \omega'_i$.
  \textbf{(c)}~Count histogram for an initial batch of shots, attributing
  discrete detection events to specific vibronic transitions (x-axis in $\mathrm{cm}^{-1}$, 
  where $\tilde{\omega}=\omega/(2\uppi c)$).
  \textbf{(d)}~Normalized distribution after $N \gg 1$ repetitions,
  converging to the Franck--Condon vibronic profile (inset: low-probability
  transitions). Vibrational parameters for the two-mode $\mathrm{SO}_2$ model
  are taken from Ref.~\cite{lee:2009b} at the B3LYP/6-311+G(3df) level.}
  \label{fig:gbs-illustration}
\end{figure*}

Photonic quantum processors operating with continuous variables have emerged
as leading platforms for specialized quantum simulations~\cite{zhong:2020,
madsen:2022,liu:2026,gong:2026}. Assessing the operational fidelity and progress of these
devices requires application-level benchmarks that go beyond synthetic
random-circuit sampling and encode structured, physically relevant
Hamiltonians~\cite{cao:2019,mcardle:2020}.
A key target in this domain is the estimation of vibronic spectra for which
the harmonic approximation is commonly invoked.
A time-independent (TI) framework~\cite{sharp:1964,doktorov:1975,doktorov:1977} requires evaluating
general matrix elements of the transition moment operator between vibronic wavefunctions;
the squared magnitudes of these transition matrix elements are proportional to transition probabilities.
In the Franck--Condon (FC) approximation the electronic part of the electric transition dipole
moment (TDM) is assumed to be constant with respect to displacements of the nuclei,
and the transition matrix elements reduce to evaluating
multidimensional overlap integrals between final and initial state vibrational wavefunctions.
These overlap integrals are called
Franck--Condon integrals (FCIs) and their squared magnitudes
are known as Franck--Condon factors (FCFs).
Because the FC approximation fails to capture vibronic intensity borrowing in 
many systems, the Herzberg--Teller (HT) formalism~\cite{herzberg:1933,teller:1937}
introduces higher-order terms in the TDM's Taylor expansion with respect to displacements
along normal coordinates. Incorporating these higher-order terms requires evaluating
linear combinations of FCIs.
When Duschinsky mixing of normal modes is taken into account~\cite{duschinsky:1937},
FCIs do not decompose into simple products of one-dimensional integrals.
An individual multidimensional FCI then takes the form of a loop hafnian of a matrix
defined by the Gaussian transformation \cite{quesada:2019}.
Consequently, the exact computation of an individual multidimensional FCI/FCF
is considered \#P-hard in the worst case~\cite{valiant:1979,kan:2008,huh:2011a,quesada:2019}.
Moreover, constructing a vibronic profile requires evaluating many such integrals,
whose number grows combinatorially with the number of modes and the chosen vibrational excitation cutoff.
Due to the computational complexity of calculating FCIs and their sheer
number, especially for polyatomic molecules, estimating vibronic spectra has
emerged as an application-level benchmark for photonic quantum
hardware~\cite{huh:2015,huh:2017a,shen:2018,zhu:2024,eickmann:2026}.
Rather than serving merely as synthetic sampling demonstrations, vibronic
spectra estimation challenges continuous-variable processors on structured,
parameterized Gaussian transformations across demanding multi-mode regimes.
To sample final vibrational states with probabilities given by the corresponding FCFs,
one can employ Gaussian Boson Sampling (GBS)~\cite{hamilton:2017,kruse:2019}.
For a fixed molecular transition, GBS implements
the transformation between the initial and final vibrational normal modes using
parametrized Gaussian gates acting on an initial vacuum state.
The resulting circuit defines a joint photon-number distribution over the output modes.
Sampling means measuring these modes in the photon-number basis and recording
the photon-number occupation pattern across all modes. In the vibronic mapping,
each such pattern corresponds to a final-state vibrational occupation pattern,
and the probability of observing it yields the corresponding FCF.
Thus, repeated samples from this distribution provide increasingly precise
additive-error estimates
of the FCFs of the observed vibronic transitions. Individual FCFs are thereby
estimated statistically rather than computed directly.
Figure~\ref{fig:gbs-illustration} illustrates the
mapping and workflow for estimating vibronic transition probabilities within
the FC approximation.

The efficient calculation of low-resolution spectral envelopes on classical hardware
via time-dependent correlation functions has long been established in molecular
physics~\cite{tannor:1982,huh:2011a,huh:2012}, including recent extensions
to quadratic Herzberg--Teller correlation functions~\cite{quinonero:2025}.
Whereas recent assessments of quantum advantage for vibronic GBS~\cite{oh:2024}
focused on energy-grouped transition probabilities rather than individual
outcomes, GBS devices equipped with photon-number-resolving detectors natively
produce samples from state-resolved photon-number distributions. A direct comparison
and validation of hardware outputs on equal footing therefore demands
efficient methods in the state-resolved picture. 

Early proposals and demonstrations primarily adopted the
Franck--Condon approximation~\cite{shen:2018}, in which a molecular transition
is represented by a single Gaussian circuit. Jnane \etal~\cite{jnane:2021}
subsequently developed a framework incorporating Herzberg--Teller effects
through second order by representing the non-unitary transition-dipole-moment
operator through auxiliary Gaussian unitary operations. The desired
non-Condon transition intensities are then recovered from a linear combination
of the output probabilities of several Gaussian circuits.

\begin{figure*}
  \centering
  \includegraphics[width=0.95\linewidth]{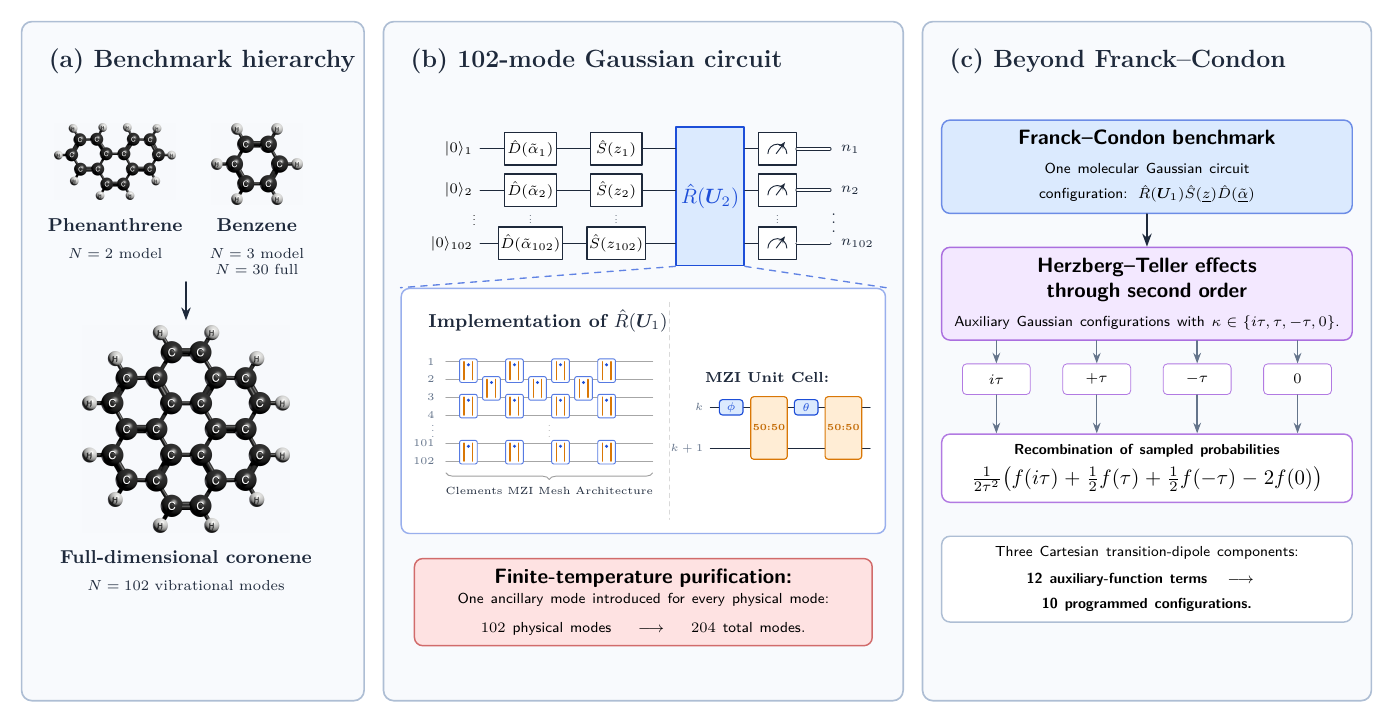}
  \caption{Hierarchy of application-level resources for Gaussian boson sampling
implementations of vibronic spectroscopy.
\textbf{(a)} Benchmark molecular systems, ranging from reduced-mode models
[phenanthrene ($N=2$) and benzene ($N=3$)] to full-dimensional benzene
($N=30$) and coronene ($N=102$ vibrational modes).
\textbf{(b)} 102-mode Gaussian circuit implementing the state preparation
$\ket{\psi} = \hat{R}(\bm{U}_2)\hat{S}(\udl{z})\hat{D}(\udl{\tilde{\alpha}})\ket{\udl{0}}$
for zero-temperature Franck--Condon vibronic profile estimation.
Input vacuum states $\ket{0}_k$ undergo mode-resolved
displacements $\hat{D}(\tilde{\alpha}_k)$, single-mode squeezing $\hat{S}(z_k)$, a final
interferometer $\hat{R}(\bm{U}_2)$, and photon-number-resolving (PNR) detection
measuring output photon numbers $n_k$.
Inset: Clements interferometer mesh implementing $\hat{R}(\bm{U}_2)$
using $N(N-1)/2$ Mach--Zehnder interferometers (MZIs). The two-mode MZI unit
cell (right) incorporates an input phase shifter $\phi$, two 50:50 directional
couplers, and an internal phase shifter $\theta$ acting on adjacent modes $k$
and $k+1$ \cite{clements:2016}.
Bottom: Finite-temperature ($T > 0$) state purification introduces one
ancillary mode per physical mode, scaling the coronene system to $2N = 204$ modes.
\textbf{(c)} Beyond-Franck--Condon pipeline accounting for Herzberg--Teller
effects up to second order. For each Cartesian transition-dipole component,
auxiliary Gaussian circuits parameterized by $\kappa \in \{\im\tau, \tau,
-\tau, 0\}$ are sampled and recombined to yield non-Condon transition
intensities. Across all three dipole components, the 12 required auxiliary
terms reduce to 10 distinct circuit configurations by evaluating the shared
$\kappa = 0$ reference configuration once.}
\label{fig:gbs-resource-hierarchy}
\end{figure*}

Progressing from Franck--Condon spectra to higher-order Herzberg--Teller
treatments and finite-temperature profiles substantially increases the
experimental requirements, as illustrated in
Fig.~\ref{fig:gbs-resource-hierarchy}. In parallel to vibronic spectra estimation,
the scale and capability of Gaussian photonic hardware have advanced
dramatically. Large-scale processors operating on temporal or spatial modes
have demonstrated multi-photon interference across hundreds of
modes~\cite{zhong:2020,madsen:2022,deng:2023,gong:2026}. However, bridging the
gap between synthetic sampling and vibronic spectra estimation requires high-fidelity
integration of single-mode squeezing, arbitrary linear-optical
transformations, displacements, and photon-number-resolving detection.

The auxiliary-function treatment of Herzberg--Teller effects replaces a
single Gaussian circuit by a linear combination of Gaussian circuits whose output
probabilities must be estimated and recombined on classical hardware, and
thermal purification further doubles the required optical modes. As physical
architectures scale toward hundreds of modes required to simulate polyatomic
molecules---such as coronene ($N = 102$ modes, or $204$ modes at finite
temperature, Fig.~\ref{fig:gbs-resource-hierarchy})---large-scale reference
calculations executed on classical hardware become vital to evaluate state-resolved
device performance and calibrate systematic hardware errors.

Although the auxiliary-function formalism~\cite{jnane:2021} was initially introduced for 
GBS, it can be naturally extended to computations on classical hardware.
The challenge of computing vibronic spectra on classical computers
has been recognized for decades~\cite{ruhoff:1994,berger:1997,berger:1998},
prompting the development of effective approximation techniques~\cite{santoro:2007,jankowiak:2007}
to address the combinatorial scaling of FCF calculations.
These methods typically involve reducing the number of FCFs to be computed
by restricting the maximum excitation level per
mode in Fock space and limiting the number of simultaneously excited modes~\cite{santoro:2007,jankowiak:2007}.

In particular, Ref.~\cite{jankowiak:2007} introduced rigorous prescreening conditions
that allow one to systematically restrict the Fock space by discarding entire batches of negligible FCFs.
These conditions also provide strict upper and lower bounds on the total FCF weight achievable
under a given truncation. In the complete, untruncated limit,
this weight is normalized to unity when summing over all final states for a fixed initial state,
or, at finite temperature, when averaging over normalized initial-state populations
and summing over all final states. The bounds therefore enable controlled approximations.
In this manner, a considerable speed-up in vibronic profile
calculations compared to no restriction can be achieved within the TI framework,
yet still capturing the essential spectral features.

The existing prescreening formulation, however, was derived for the
real-valued Gaussian transformation associated with the conventional
Doktorov operator. The auxiliary-function construction introduces complex
Gaussian parameters and therefore falls outside its scope. Extending the
prescreening conditions to products of complex Gaussian unitaries is thus
a viable path towards obtaining controlled, state-resolved benchmarks beyond
the Franck--Condon approximation.

Compared to previous prescreening approaches, the present method
extends the applicable Gaussian transformations from a single real-valued
Doktorov operator to products of complex fundamental Gaussian operators,
enabling efficient approximation of broader classes of state-resolved output distributions,
including those corresponding to quadratic HT effects and finite-temperature environments.
We also leverage connections between time-independent and time-dependent
frameworks~\cite{huh:2011a} to extend our formalism to the time-dependent domain,
encompassing both zero-temperature and thermally broadened vibronic spectra. 

We validate this framework on benchmark systems ranging from reduced-coordinate
models to full-dimensional coronene ($N = 102$ modes, scaling
to $204$ modes at finite temperature). These calculations provide
state-resolved reference distributions across the Franck--Condon and
Herzberg--Teller resource hierarchy with systematically tracked truncation
error. By establishing bounds on the retained probability weight,
our approach helps map the regimes accessible to controlled time-independent
methods executed on classical hardware and offers practical baselines for benchmarking future
quantum optical processors.

The paper is organized as follows. The theoretical foundations are presented
in Sec.~\ref{sec:theory}, and Sec.~\ref{sec:sumrules} shows the
working equations of the prescreening conditions generalized to
complex Gaussian operators. Sec.~\ref{sec:auxfunc} presents the formulae for applying
the prescreening conditions to the auxiliary function method as well as 
a time-dependent extension. In Sec.~\ref{sec:results} we present the main results
of the paper and discuss them. Sec.~\ref{sec:conclusion-outlook} concludes and gives an outlook.
\section{Theory}
\label{sec:theory}
\subsection{Gaussian operator decompositions}
\label{ssec:gaussops}
In this section we review the basic properties of Gaussian bosonic operations.
We define two general $N$-dimensional harmonic oscillators
\begin{equation}
  \begin{split}
    \hat{H}&=\sum_{j=1}^N\hbar\omega_j\left(\hat{a}_j^\dagger\hat{a}_j+1/2\right),\\
    \hat{H}^\prime&=\sum_{j=1}^N\hbar\omega_j'\left(\hat{a}_j^{\prime\dagger}\hat{a}_j^\prime+1/2\right),
\end{split}
\end{equation}
with $\omega_j$ and $\omega_j'$ as harmonic angular frequencies and where $\hat{a}_j^\dagger$, $\hat{a}_j$ and 
$\hat{a}_j^{\prime\dagger}$, $\hat{a}_j^{\prime}$
are the bosonic creation and annihilation operators of the unprimed and primed Hamiltonian, respectively.
The creation and annihilation operators satisfy the bosonic commutation relations
\begin{align}
  [\hat{a}_i,\hat{a}_j^\dagger]
  &= [\hat{a}_i^\prime,\hat{a}_j^{\prime\dagger}]
  = \delta_{i,j},\\
  [\hat{a}_i,\hat{a}_j]
  &= [\hat{a}_i^\dagger,\hat{a}_j^\dagger]
  = [\hat{a}_i^\prime,\hat{a}_j^\prime]
  = [\hat{a}_i^{\prime\dagger},\hat{a}_j^{\prime\dagger}]
  = 0,
\end{align}
where $[x,y]$ denotes the commutator of $x$ and $y$, and $\delta_{i,j}$
is the Kronecker delta.
We compactify the notation by
$\hat{\udl{a}}=(\hat{a}_1,\hat{a}_2,\dots,\hat{a}_N)^\intercal$, 
$\hat{\udl{a}}^\dagger=(\hat{a}_1^\dagger,\hat{a}_2^\dagger,\dots,\hat{a}_N^\dagger)^\intercal$,
$\hat{\udl{a}}'=(\hat{a}_1',\hat{a}_2',\dots,\hat{a}_N')^\intercal$ and
$\hat{\udl{a}}^{\prime\dagger}=(\hat{a}_1^{\prime\dagger},\hat{a}_2^{\prime\dagger},\dots,\hat{a}_N^{\prime\dagger})^\intercal$.
We consider both the initial $\hat{H}$ and final $\hat{H}'$ as belonging to the
dynamical symmetry group
$G=\mathrm{Sp}(2N,\mathbb{R})\wedge H(N)$, where $\wedge$ denotes the
semidirect product of the real symplectic group $\mathrm{Sp}(2N,\mathbb{R})$
and the Heisenberg--Weyl group $H(N)$~\cite{doktorov:1977}.
Consequently, the operators $\hat{\udl{a}}$, $\hat{\udl{a}}^{\dagger}$ are connected to
$\hat{\udl{a}}^{\prime}$ and $\hat{\udl{a}}^{\prime\dagger}$ by a
linear canonical transformation (also known as Bogoliubov transformation~\cite{bogoliubov:1958,braunstein:2005}), which 
takes the form~\cite{doktorov:1977}
\begin{equation}
  \begin{pmatrix}
    \hat{\udl{a}}^\prime\\
  \hat{\udl{a}}^{\prime\dagger}
  \end{pmatrix} = 
  \begin{pmatrix}
    \bm{E}&\bm{F}\\
    \bm{F}^*&\bm{E}^*
  \end{pmatrix}
  \begin{pmatrix}
    \hat{\udl{a}}\\\hat{\udl{a}}^{\dagger}
  \end{pmatrix} + 
  \begin{pmatrix}
    \udl{g}\\\udl{g}^*
  \end{pmatrix}.
  \label{eq:bogoliubov-trans}
\end{equation}
The linear canonical transformation matrices $\bm{E}$ and $\bm{F}$ satisfy the conditions~\cite{berezin:1966}
\begin{align}
  \bm{E}\bm{E}^\dagger - \bm{F}\bm{F}^\dagger &= \bm{I},\quad\bm{E}\bm{F}^\intercal - \bm{F}\bm{E}^\intercal = \bm{0},\\
  \bm{E}^\dagger\bm{E} - \bm{F}^\intercal\bm{F}^* &= \bm{I},\quad\bm{E}^\dagger\bm{F} - \bm{F}^\intercal\bm{E}^* = \bm{0}.
  \label{eq:canonical-commutation-relations}
\end{align}
In general, we are interested in the evaluation of the overlap integrals
\begin{equation}
  \braket{\udl{v}'}{\udl{v}}=\braket{\udl{v}',\text{f}}{\udl{v},\text{in}}=
\mel{\udl{v}',\text{f}}{\hat{U}_{\text{G}}}{\udl{v},\text{f}}=
  \mel{\udl{v}',\text{in}}{\hat{U}_{\text{G}}}{\udl{v},\text{in}}
  \label{eq:overlap-integrals}
\end{equation}
where the eigenbases of the initial and final Hamiltonians are denoted by in and f, respectively.
Furthermore, the initial and final states of the $N$-dimensional harmonic oscillator are defined by the
state vectors 
$\ket{\udl{v}}=\ket{v_1,v_2,\dots,v_N}$ and $\ket{\udl{v}^\prime}=\ket{v_1',v_2',\dots,v_N'}$,
respectively, where the indices denote the 
excitation levels of each mode. Transitions
to the final Fock state are evaluated using its corresponding
dual vector $\bra{\udl{v}^\prime}=\bra{v_1',v_2',\dots,v_N'}$.
The absolute squares of overlap integrals in Eq.~\eqref{eq:overlap-integrals}
correspond to the output probabilities of a Gaussian boson sampler (GBS).
In the Franck--Condon approximation they also determine physical observables
such as experimental one-photon vibronic absorption spectra.

The unitary Gaussian operator $\hat{U}_{\text{G}}\in G$ connects the 
creation and annihilation operators of the initial and final Hamiltonians~\cite[pp. 88--115]{berezin:1966}
\begin{equation}
  \hat{\udl{a}}^\prime =\hat{U}_{\text{G}}^\dagger\hat{\udl{a}}\hat{U}_{\text{G}},\quad
  \hat{\udl{a}}^{\prime\dagger} =\hat{U}_{\text{G}}^\dagger\hat{\udl{a}}^\dagger\hat{U}_{\text{G}}.
\end{equation}

The unitary Gaussian operator may be composed of products of the three fundamental
Gaussian unitary operators, the $N$-mode
displacement $\hat{D}(\udl{\alpha})$, squeezing $\hat{S}(\bm{\Xi})$
and rotation $\hat{R}(\bm{U})$ operators~\cite{ma:1990}, i.e.,
\begin{align}
  \hat{D}(\udl{\alpha})=&\exp(\udl{\alpha}^\intercal\hat{\udl{a}}^\dagger
  -\udl{\alpha}^\dagger\hat{\udl{a}})\label{eq:multimode-disp-op},\\
  \hat{S}(\bm{\Xi})=&\exp[1/2((\hat{\udl{a}}^\dagger)^\intercal\bm{\Xi}\hat{\udl{a}}^\dagger
  - \hat{\udl{a}}^\intercal\bm{\Xi}^\dagger\hat{\udl{a}})]\label{eq:multimode-squeeze-op},\\
  \hat{R}(\bm{U})=&\exp[(\hat{\udl{a}}^\dagger)^\intercal\ln(\bm{U})\hat{\udl{a}}],\label{eq:multimode-rot-op}
\end{align}
where $\udl{\alpha}\in\mathbb{C}^N$ and $\bm{\Xi}\in\mathbb{C}^{N\times N}$.
Moreover, $\bm{U}\in\mathbb{C}^{N\times N}$ is a unitary matrix. Furthermore,
the relations
$\hat{D}^\dagger(\udl{\alpha}) =
  \hat{D}^{-1}(\udl{\alpha})=\hat{D}(-\udl{\alpha})$, 
$\hat{S}^\dagger(\bm{\Xi}) =
\hat{S}^{-1}(\bm{\Xi})=\hat{S}(-\bm{\Xi})$,
$\hat{R}^\dagger(\bm{U}) =
\hat{R}^{-1}(\bm{U})=\hat{R}(\bm{U}^{-1})=\hat{R}(\bm{U}^\dagger)$ hold.
Ref.~\cite{ma:1990} gives the  action of these unitary operators on $\hat{\udl{a}}$ and $\hat{\udl{a}}^\dagger$ as
\begin{align}
  \hat{D}(\udl{\alpha})^\dagger\hat{\udl{a}}\hat{D}(\udl{\alpha})
  &= \hat{\udl{a}} + \udl{\alpha}\label{eq:ma-d-op-annihilation},\\
  \hat{D}(\udl{\alpha})^\dagger\hat{\udl{a}}^\dagger\hat{D}(\udl{\alpha})
  &= \hat{\udl{a}}^\dagger + \udl{\alpha}^*,\\
  \hat{R}(\bm{U})^\dagger\hat{\udl{a}}\hat{R}(\bm{U})
  &= \bm{U}\hat{\udl{a}},\\
  \hat{R}(\bm{U})^\dagger\hat{\udl{a}}^\dagger\hat{R}(\bm{U})
  &= \bm{U}^*\hat{\udl{a}}
  ^\dagger,\\
  \hat{S}(\bm{\Xi})^\dagger\hat{\udl{a}}\hat{S}(\bm{\Xi})
  &= \cosh(\bm{R})\hat{\udl{a}} + \sinh(\bm{R})\exp(\im\bm{T})\hat{\udl{a}}^\dagger,\\
  \hat{S}(\bm{\Xi})^\dagger\hat{\udl{a}}^\dagger\hat{S}(\bm{\Xi})
  &= \sinh(\bm{R}^\intercal)\exp(-\im(\bm{T})^\intercal)\hat{\udl{a}} +
  \cosh(\bm{R}^\intercal)\hat{\udl{a}}^\dagger\label{eq:ma-s-op-creation},
\end{align}
with the polar decomposition $\bm{\Xi}=\bm{R}\exp(\im\bm{T})$.
In this work we restrict ourselves to implementing one-mode and two-mode squeezing operations,
which imposes either a diagonal or a $2\times2$ block-diagonal structure
on $\bm{\Xi}$.
This structural simplification suffices for our current objectives,
because the applications within our immediate scope employ only up to two-mode
squeezing operations~\cite{huh:2017a,quesada:2019,wakefield:2024}.
Moreover, any unitary Gaussian operator can be decomposed into a linear
$N$-mode rotation, followed by a parallel set of $N$ single-mode squeezers,
an additional $N$-mode rotation, and, finally, an $N$-mode displacement operator.
This Gaussian operator decomposition is called Bloch--Messiah decomposition~\cite{bloch:1962,cariolaro:2016}.
In the one-mode squeezing case the matrix exponentials and inversions
are calculated trivially.
For an arbitrary number of combinations
of displacement, rotation and one-mode and two-mode squeezing operations, 
deriving the resulting $\bm{E}$, $\bm{F}$ matrices and $\udl{g}$ vector by hand is cumbersome
and error-prone. Instead, one may apply the relations
Eqs.~\eqref{eq:ma-d-op-annihilation}--\eqref{eq:ma-s-op-creation} such that $\bm{E}$, $\bm{F}$ and $\udl{g}$
are automatically constructed
given the operators and corresponding parameters. We implemented such a scheme
within the \texttt{hotFCHT}~\cite{berger:1998,jankowiak:2007,huh:2012} code.
\subsection{Integral calculation of Bogoliubov-transformed systems}
\label{ssec:rr}
To calculate the integral $\mel{\udl{v}',\text{in}}{\hat{U}_{\text{G}}}{\udl{v},\text{in}}$
one may calculate the $\mel{\udl{0}',\text{in}}{\hat{U}_{\text{G}}}{\udl{0},\text{in}}$ integral
and employ recurrence relations. The integral
$\braket{\udl{0}'}{\udl{0}}=\mel{\udl{0}',\text{in}}{\hat{U}_{\text{G}}}{\udl{0},\text{in}}$
utilizing the linear canonical transformation is given as~\cite{malkin:1973,dodonov:1994}:
\begin{equation}
  \braket{\udl{0}'}{\udl{0}} = \det(\bm{E})^{-1/2}
  \exp\left[-\frac{1}{2}\left(\udl{g}^\dagger\udl{g}-\udl{g}^\intercal\bm{F}^*\bm{E}^{-1}\udl{g}\right)\right].
  \label{eq:zero-zero-int}
\end{equation}
Because the square root of a complex number is not unique, $\det(\bm E)^{-1/2}$ requires a choice of branch. 
Although the time-dependent observables in the present method depend only on
the squared modulus of the integral in Eq.~\eqref{eq:zero-zero-int} (see Sec.~\ref{ssec:auxfunc-tdmethod}),
the integral itself should be evaluated using a time-continuous branch choice, \eg
by unwrapping the phase of $\det(\bm E)$, to avoid artificial phase jumps~\cite{littlejohn:1986,kay:1994,noid:2003}.
We discuss other phase factors appearing in the evaluation of complex FCIs in Appendix~\ref{sec:phase-factors}.

To utilize recurrence relations, we adapt the formalism established by
Fern\'andez and Tipping~\cite{fernandez:1989}, which is valid for any invertible
operator $\hat{U}$
that can be written as a product of exponentials of linear and quadratic terms
in the creation and annihilation operators. By specializing their 
expressions to the case of a unitary transformation, we obtain
the following coefficients for the recurrence relations:
\begin{equation}
  \begin{split}
    \tilde{\udl{g}}&=-\udl{g}+(\bm{E}^\dagger)^{-1}\bm{F}^\intercal
    \udl{g}^*,\\
    \tilde{\bm{E}}&=(\bm{E}^\dagger)^{-1},\\
    \tilde{\bm{F}} &= -(\bm{E}^\dagger)^{-1}\bm{F}^\intercal,\\
    \tilde{\udl{g}}^\prime &=-(\bm{E}^*)^{-1}\udl{g}^*,\\
    \tilde{\bm{E}}^\prime &= -\bm{F}^\dagger(\bm{E}^\dagger)^{-1},\\
    \tilde{\bm{F}}^\prime &=(\bm{E}^*)^{-1} = \tilde{\bm{E}}^\intercal.
  \end{split}
\label{eq:fernandez-coefficients-efg}
\end{equation}
Further details on this derivation are provided in Appendix~\ref{sec:fernandez-deriv}.
Moreover, for completeness, we also give the expression of the integral $\braket{\udl{0}'}{\udl{0}}$
using the coefficients for the recurrence relations:
\begin{equation}
  \braket{\udl{0}'}{\udl{0}} = \det(\tilde{\bm{E}}^\dagger)^{1/2}\exp\left(-\frac{1}{2}
  \udl{\tilde{g}}^\dagger(\tilde{\bm{E}}^\dagger)^{-1}(\udl{\tilde{g}}')^*\right).
\end{equation}

As a shorthand notation, we can group the state vectors of the initial and final states into a single collective index
vector $\udl{\tilde{v}}=(\udl{v},\udl{v}')^\intercal$ and assign
\begin{equation}
  \begin{split}
  \braket{\udl{v}'}{\udl{v}} &\rightarrow
  \expval{v_{1},\dots,v_{N},v_1',\dots,v_N'}\\
                             &=
  \expval{\tilde{v}_{1},\dots,
  \tilde{v}_{N},\tilde{v}_{N+1},\dots,\tilde{v}_{2N}}
  \end{split}
\end{equation}
Substituting the coefficients from Eq.~\eqref{eq:fernandez-coefficients-efg}
into the standard recurrence relations~\cite{willink:2005,huh:2011a},
we arrive at
\begin{equation}
  \begin{split}
  &\sqrt{\tilde{v}_k+1}\expval{\tilde{v}_1,\dots,\tilde{v}_k+1,\dots,\tilde{v}_{2N}}
  =\\
  &r_k\expval{\tilde{v}_1,\dots,\tilde{v}_{2N}}-\sum_{j=1}^{2N}W_{kj}\sqrt{\tilde{v}_j}
  \expval{\tilde{v}_1,\dots,\tilde{v}_{j}-1,\dots,\tilde{v}_{2N}}
  \end{split}
  \label{eq:rr}
\end{equation}
where
\begin{equation}
  \bm{W} =
  \begin{pmatrix}
    -\tilde{\bm{E}}'&-\tilde{\bm{F}}^\prime\\
    -\tilde{\bm{F}}^{\prime\intercal}&\tilde{\bm{F}}
  \end{pmatrix},\quad
  \udl{r}=
  \begin{pmatrix}
    \tilde{\udl{g}}'\\
    -\tilde{\udl{g}}
  \end{pmatrix}.
  \label{eq:w-r-fernandez}
\end{equation}
By employing Eq.~\eqref{eq:zero-zero-int} and the recurrence relations in Eq.~\eqref{eq:rr},
all matrix elements of products of Gaussian operators can be determined analytically.  
\section{Rigorous Prescreening Conditions}
\label{sec:sumrules}
In large molecular systems, the direct computation of
the full probability distribution of Franck--Condon factors (FCFs)
is computationally intractable because the number of relevant
Fock states grows combinatorially with the number of vibrational
modes and the chosen excitation cutoff.
In practical applications, one must restrict the active vibrational space \textit{a priori}.
In the context of Gaussian operators associated with the description
of vibronic transitions in molecules, a particularly effective Fock space restriction
relies on limiting the maximum excitation level in each normal mode (\ie $\tilde{v}_k^{\text{max}}$)
as well as the maximum number of simultaneously excited modes in the calculation
of the FCFs (\ie $M=\sum_{k=1}^{2N}(\tilde{v}_k>0)$).
Rigorous prescreening conditions provide a systematic means
for choosing the numerical values of these restrictions based on prescreening thresholds.
The prescreening relies on analytic sum rules~\cite{doktorov:1979,jankowiak:2007},
which yield the cumulative transition probability for specific subsets of vibrational excitations
and thus allow transparent error analysis when batches of integrals
are excluded from the calculation. In practice, this systematic Fock space restriction
results in a large number of integrals being excluded and therefore leads to speedups
of several orders of magnitude, especially for larger systems, without a large loss
in cumulative transition probability.
Such a restriction strategy has proven highly effective in approximating
state-resolved vibronic spectra~\cite{jankowiak:2007,huh:2011a}.

A powerful approach to deriving these sum rules, and indeed the
closed-form expressions for the Franck--Condon integrals (FCIs) themselves,
is the use of Generating Functions (GFs) formulated via coherent states.
Glauber~\cite{glauber:1963} popularized employing coherent states to treat oscillator modes.
Moreover, Malkin \etal~\cite{malkin:1973} explicitly formulated $N$-dimensional FCFs
in terms of coherent states. Doktorov \etal subsequently
exploited this formalism to obtain closed-form expressions for the FCIs and FCFs~\cite{doktorov:1975,doktorov:1977,doktorov:1977a,doktorov:1979}, yielding results equivalent to the recurrence relations
discussed in the previous section.

We employ a GF in the coherent state representation to extract information about the probability
distribution of the FCFs.
The coherent state overlap integral in the occupancy representation is given by~\cite{jankowiak:2007}
\begin{equation}
  \braket{\udl{\gamma}'}{\udl{\alpha}} = \exp(-\frac{1}{2}\udl{\alpha}^\dagger\udl{\alpha}
  -\frac{1}{2}\udl{\gamma}'^{\dagger}\udl{\gamma}')\sum_{\udl{v},\udl{v}'}
  \braket{\udl{v}'}{\udl{v}}\left[\prod_{k=1}^N
    \frac{\alpha_k^{v_k}(\gamma_k^{'*})^{v_k'}}{\sqrt{v_k!(v_k')!}}
  \right],
  \label{eq:coh-occupancy}
  \end{equation}

and in the spatial representation as 
\begin{equation}
  \braket{\udl{\gamma}'}{\udl{\alpha}} = \braket{\udl{0}'}{
  \udl{0}}\exp(-\frac{1}{2}\udl{\xi}^\dagger\udl{\xi})
  \exp(-\frac{1}{2}\udl{\xi}^\intercal\bm{W}\udl{\xi}+\udl{r}^\intercal
  \udl{\xi}),
  \label{eq:coh-ov-dok}
\end{equation}
with the $2N$-dimensional complex vector
\begin{equation}
  \udl{\xi}=
  \begin{pmatrix}
    \udl{\alpha}\\
    \udl{\gamma}^{\prime *}
  \end{pmatrix}.
  \label{eq:xi-jank}
\end{equation}

To extract the sum rules necessary for prescreening, we rely on the symmetric GF
for the FCFs introduced by Jankowiak \etal~\cite{jankowiak:2007}:
\begin{equation}
  \begin{split}
    G(\bm{Z}) = G(\bm{z}';\bm{z})=&\uppi^{-2N}\abs{\braket{\udl{0}'}{\udl{0}}}^{-2}
    \int\dd[2]{\udl{\alpha}}\dd[2]{\udl{\gamma}}'\\
    &\times\braket{\bm{z}'^*\udl{\gamma}'}{\bm{z}\udl{\alpha}}
      \braket{\bm{z}'\udl{\gamma}'}{\bm{z}^*\udl{\alpha}}^*,
  \label{eq:fcf-gf}
  \end{split}
\end{equation}
for $N$ vibrational modes. Here $\bm{z}$ and $\bm{z}'$ are diagonal matrices
consisting of the GF parameters,
\begin{align}
  \bm{z}&=\text{diag}(z_1,z_2,\cdots,z_N),\\
  \bm{z}'&=\text{diag}(z_1',z_2',\cdots,z_N'),
\end{align}
with $z_k$ and $z_k'$ of modulus unity, i.e.,
\begin{equation}
  z_k=\exp(\im\theta_k/2),\quad z_k'=\exp(\im\theta_k'/2),\quad \theta_k,\theta_k'\in\mathbb{R},
\end{equation}
for all $k=1,\cdots,N$, and $\bm{Z}=\text{bldiag}(\bm{z},\bm{z}')$,
where bldiag denotes a block-diagonal matrix.

Evaluating the GF at $z_k=1$ incorporates contributions from all values of
the unprimed quantum number $v_k$ for mode $k$,
whereas the limit $z_k\rightarrow 0$ isolates the zeroth-order term in $z_k$ and thus retains
only transitions with $v_k=0$. The primed quantum number $v_k'$ remains
unrestricted unless the analogous limit $z_k'\rightarrow 0$ is also taken.
Thus, only the simultaneous limits $z_k\rightarrow 0$ and
$z_k'\rightarrow 0$ restrict mode $k$ to the $0_k-0_k'$ contribution.
Furthermore, treating these generating parameters as formal variables
establishes a bridge to time-dependent (TD) formulations.
By restricting $z_k$ back to the unit circle and identifying the phase angle
with the product of the harmonic vibrational angular frequency and time,
such that $z_k(t)=\exp(\im\omega_k t/2)$ and $z_k'(t)=\exp(-\im\omega_k' t/2)$,
the GF $G(\bm{Z})$ becomes equivalent
to the TD wavepacket autocorrelation function \cite{huh:2011a,huh:2010thermal,huh:2012}.
Consequently, extracting specific transition probabilities via the power series
expansion of the GF in the energy domain is strictly analogous to obtaining
the FC profile via the Fourier transform of the correlation function in the time domain~\cite{huh:2011a,huh:2010thermal,huh:2012}.

To systematically exploit the projection property for prescreening,
Jankowiak \etal partitioned the GF Eq.~\eqref{eq:fcf-gf} into terms depending
on variables associated with orthogonal subspaces $X$ and $Y$.
The space $X$ contains modes for which quantum numbers $\udl{\tilde{v}}$
are kept fixed, whereas $Y$ is associated with modes that involve summation over all 
harmonic oscillator states.
The result is the FCF GF partitioned expression 

\begin{align}
  G(\bm{Z}_{XX},\bm{Z}_{YY}) = G_{X|Y}(\bm{Z}_{XX},\bm{Z}_{YY})G_Y(\bm{Z}_{YY}).
  \label{eq:fcf-gf-partitioned}
\end{align}

From this expression two explicit sum rules can be derived, the first for the case
where no excited modes are in the $X$ space, \ie all modes in the $X$ space remain in their ground state.
The second sum rule applies to the case where one excited mode is in the $X$ space.
Here, we present these sum rules generalized from
a single real Gaussian operator decomposition (the Doktorov operator will be defined below)
to any complex Gaussian operator decomposition.
The detailed derivation is provided in Appendix~\ref{sec:prescreening-derivation}.

In the special case where the summation runs over all final states $\ket{\udl{v}'}$
within the $Y$ subspace with the initial states remaining
in their ground state in the $X$ space, with notation defined in Sec.~\ref{ssec:rr}, we arrive at
\begin{equation}
  \begin{split}
    &G_Y(\bm{I}_{YY}) = \sum_{\udl{v}_Y'=0}^\infty q(\udl{v}_Y';\udl{0})
    = \\&\det((\bm{\mathcal{I}}+\bm{\mdc{\tilde{F}}})_{YY})^{-1/2}
    \exp(\udl{\mdc{\tilde{g}}}_Y^\intercal(\bm{\mathcal{I}}+\bm{\mdc{\tilde{F}}}_{YY})^{-1}
    \udl{\mdc{\tilde{g}}}_Y),
  \end{split}
  \label{eq:sum-rule-1}
\end{equation}
with
\begin{equation}
  \bm{\mdc{\tilde{F}}}_{YY} = 
  \begin{pmatrix}
    \real(\tilde{\bm{F}})_{YY} & -\imag(\tilde{\bm{F}})_{YY}\\
    -\imag(\tilde{\bm{F}})_{YY} &-\real(\tilde{\bm{F}})_{YY}
  \end{pmatrix},\quad
  \udl{\mdc{\tilde{g}}}_Y = 
  \begin{pmatrix}
    -\real(\tilde{\udl{g}})_Y\\
    \imag(\tilde{\udl{g}})_Y
  \end{pmatrix}.
  \label{eq:aug-f-tilde-aug-g-tilde}
\end{equation}
Here, the subscript $YY$ indicates that only values in Eq.~\eqref{eq:sum-rule-1}
are those associated with the modes in the $Y$ space. The sum rule in Eq.~\eqref{eq:sum-rule-1}
is identical to the sum rule from Jankowiak \etal in the limit where $\imag(\tilde{\bm{F}})_{YY} = \bm{0}$
and $\imag(\tilde{\udl{g}})_Y=\udl{0}$.

For the second sum rule, where a single mode $k$ occupies the $X$ space, we find
\begin{widetext}
\begin{equation}
  \begin{split}
    \sum_{\udl{v}_Y'|v_k'=n} q(\udl{v}';\udl{0}) =& \sum_{\udl{v}_Y'=0}^\infty q(\udl{v}_Y';\udl{0})\\
                                                  &\times
    \uppi^{-1}\sum_{l=0}^{n}\sum_{q=0}^{n-l}
    \sum_{\substack{m=0 \\ m+q\ \mathrm{even}}}^{2N_{lq}}
    \sum_{\substack{p=0 \\ p+q\ \mathrm{even}}}^{2N_{lq}-m}\\
                                                  &\times
      \frac{(-\tilde{W}_{kk}^+)^l(2\tilde{r}_k^+)^{m}
    (-\tilde{W}_{kk}^-)^{n-l-m/2-p/2-q}
      (2\tilde{r}_k^-)^{p}(-2\tilde{W}_{kk}^{\text{off}})^{q}}{l!m!(n-l-q-m/2-p/2)!p!q!}\\
                                                  &\times\Gamma(l+m/2+q/2+1/2)\Gamma(n-l-m/2-q/2+1/2),
  \end{split}
  \label{eq:sum-rule-2}
\end{equation}
\end{widetext}
where $N_{lq} = n-l-q$, and $\Gamma(x)$ refers to the Gamma function.
In the special case when the mode $k$ under consideration
corresponds to $v_k'$ and the space $Y$ is only taken from the final states
$\ket{\udl{v}'}$ , the parameters $\tilde{W}_{kk}^+$, $\tilde{W}_{kk}^-$,
$\tilde{W}_{kk}^{\text{off}}$, $\tilde{r}_k^+$ and $\tilde{r}_k^-$
that enter Eq.~\eqref{eq:sum-rule-2} are obtained
from the effective block resulting after summing over the modes in $Y$.
Specifically, these expressions are given by the $2\times 2$ matrix
\begin{equation}
  \begin{split}
    \tilde{\bm{W}}_{kk} &= 
  \begin{pmatrix}
    \tilde{W}_{kk}^+&\tilde{W}_{kk}^{\text{off}}\\
    \tilde{W}_{kk}^{\text{off}}&\tilde{W}_{kk}^-\\
  \end{pmatrix} \\
                        &= ( \bm{\mdc{\tilde{F}}}_{XX}- \bm{\mdc{\tilde{F}}}_{XY}
                        (\bm{\mathcal{I}}+\bm{\mdc{\tilde{F}}}_{YY})^{-1} \bm{\mdc{\tilde{F}}}_{YX})_{kk},
  \end{split}
\end{equation}
 and two-dimensional vector
  \begin{equation}
    \begin{split}
  \udl{\tilde{r}}_k = 
  \begin{pmatrix}
    \tilde{r}_k^+\\
    \tilde{r}_k^-
  \end{pmatrix} 
  &= ( \tilde{\mdc{\udl{g}}}_X- \bm{\mdc{\tilde{F}}}_{XY}
  (\bm{\mathcal{I}}+ \bm{\mdc{\tilde{F}}}_{YY})^{-1} \tilde{\mdc{\udl{g}}}_Y)_{k}.
    \end{split}
  \end{equation}
Here, $\bm{\mdc{\tilde{F}}}_{XX}\in\mathbb{R}^{2\times 2}$ contains
the $\pm$ real part of $\tilde{F}_{k,k}$ on the diagonal and its
imaginary part on the off-diagonal corresponding to
Eq.~\eqref{eq:aug-f-tilde-aug-g-tilde}. The real and imaginary parts
of $\tilde{\bm{F}}$ inserted into
$\bm{\mdc{\tilde{F}}}_{XY} = \bm{\mdc{\tilde{F}}}_{YX}^\intercal$ and $\bm{\mdc{\tilde{F}}}_{YY}$
are arranged identically to $\bm{\mdc{\tilde{F}}}_{XX}$. Moreover, $\bm{\mdc{\tilde{F}}}_{XY}\in\mathbb{R}^{2\times2(N-1)}$
contains all entries $\tilde{F}_{k,l\neq k}$, and $\bm{\mdc{\tilde{F}}}_{YY}$
contains all entries $\tilde{F}_{l\neq k,m\neq k}$. The vectors $\tilde{\mdc{\udl{g}}}_X\in\mathbb{R}^2$
and $\tilde{\mdc{\udl{g}}}_Y\in\mathbb{R}^{2(N-1)}$ contain the real and imaginary
parts of $\udl{\tilde{g}}$ for the mode $k$ and the remaining modes, respectively,
arranged according to Eq.~\eqref{eq:aug-f-tilde-aug-g-tilde}.
Similar to the first sum rule Eq.~\eqref{eq:sum-rule-1},
the second sum rule, Eq.~\eqref{eq:sum-rule-2}
becomes identical to the sum rule introduced by Jankowiak \etal~\cite{jankowiak:2007} in
the limit of $\imag(\tilde{\bm{F}})_{YY} = \bm{0}$ and $\imag(\tilde{\udl{g}})_Y=\udl{0}$.

The sum rules in Eqs.~\eqref{eq:sum-rule-1} and~\eqref{eq:sum-rule-2}
apply to the pure Gaussian unitary processes considered here,
including the Gaussian operator products arising in the vibronic, finite-temperature,
and auxiliary-function formulations.
Employing these sum rules leads to a Fock-space
restriction scheme that limits the maximum excitation level in each mode
as well as the maximum number of simultaneously excited modes.
This approach also provides \apriori rigorous error bounds on the minimum
and maximum total intensity before calculating
the required Franck--Condon integrals, because it allows one to estimate the
minimum and maximum fraction of the total intensity that
can be achieved when applying the corresponding restrictions
(see also Sec.~\ref{ssec:auxfunc-prescreening}).
\section{Auxiliary function method}
\label{sec:auxfunc}
We apply our framework to the computation of vibronic spectra,
modeling the initial and final states as multidimensional harmonic oscillators
linked by a linear canonical transformation. Under the assumption of Fermi's golden rule,
the profile is determined by the discrete probability distribution over final vibrational states,
defined as the zero-temperature non-Condon transition profile:
\begin{equation}
  P(\omega) = \frac{1}{\mathcal{N}}\sum_{\substack{\udl{v}'=\udl{0}\\r=x,y,z}}^\infty\abs{\mel{\udl{v}',\text{in}}
  {\hat{U}_{\text{Dok}}\hat{\mu}_r(\hat{\udl{Q}})}{\udl{0},\text{in}}}^2\delta(\omega-\udl{v}^{\prime\intercal}\udl{\omega}'),
    \label{eq:transition-profile}
\end{equation}
where $\mathcal{N} = \sum_{r=x,y,z}\mathcal{N}_r$ is a normalization constant
that depends on the electronic component of the 
electric transition dipole moment operator $\hat{\udl{\mu}}(\hat{\udl{Q}})
= (\hat{\mu}_x(\hat{\udl{Q}}),\hat{\mu}_y(\hat{\udl{Q}}),\hat{\mu}_z(\hat{\udl{Q}}))^\intercal$.
The operator $\hat{\mu}_r(\hat{\udl{Q}})$ is defined with respect to the mass-weighted normal coordinates of the initial state, 
$\hat{\udl{Q}}=(\hat{Q}_1,\cdots,\hat{Q}_N)^\intercal$, which are connected to the boson
annihilation and creation operators as follows:
\begin{equation}
  \hat{Q}_k = \sqrt{\frac{\hbar}{2\omega_k}}\left(\hat{a}_k+\hat{a}_k^\dagger\right).
\end{equation}
Moreover, $\omega$
is the transition angular frequency satisfying the resonance condition
$\omega=\udl{v}^{\prime\intercal}\udl{\omega}'$ and $\udl{\omega}'=
(\omega_1',\cdots,\omega_N')^\intercal$ is the corresponding vector of harmonic vibrational
angular frequencies of the final state. The term $\hat{U}_{\text{Dok}}$
denotes the Doktorov
operator~\cite{doktorov:1977}, which is defined below. 
We assume that the change
in normal coordinates from initial to final state can be represented by a linear transformation
and a shift~\cite{duschinsky:1937,doktorov:1977}
\begin{equation}
  \hat{\udl{Q}}'=\bm{S}\hat{\udl{Q}}+\udl{d},
\label{eq:duschinksy-transformation}
\end{equation}
where $\bm{S}$ denotes the Duschinsky transformation matrix
and $\udl{d}$ the displacement vector.
For $\bm{S}\neq\bm{I}$ the Duschinsky transformation
matrix introduces mode-mixing; therefore, the
multidimensional FCIs
$\mel{\udl{v}',\text{in}}{\hat{U}_{\text{Dok}}}{\udl{v},\text{in}}$ cannot generally be decomposed
into one-dimensional FCIs via a product ansatz.

To parametrize the Doktorov operator using two rotation operators, we
first form~\cite{doktorov:1977,jankowiak:2007}
\begin{align}
  \bm{J} &= \bm{\Omega}'\bm{S}\bm{\Omega}^{-1}\\
  \udl{\delta}&=\hbar^{-1/2}\bm{\Omega}'\udl{d}.
\end{align}
The diagonal vibrational angular frequency matrices are $\bm{\Omega}'
= \text{diag}\left((\omega_1')^{1/2},\cdots,(\omega_N')^{1/2}\right)$
and
$\bm{\Omega}= \text{diag}\left((\omega_1)^{1/2},\cdots,(\omega_N)^{1/2}\right)$,
respectively.
We subsequently employ the singular value decomposition (SVD)~\cite{huh:2015}:
\begin{equation}
  \bm{J}=\bm{U}_2\bm{L}\bm{U}_1,
\end{equation}
where $\bm{L}=\text{diag}(L_1,\cdots,L_N)\wedge L_k > 0$,
and the displacement parameters are calculated as
$\udl{\gamma}=\udl{\delta}/\sqrt{2}$~\cite{doktorov:1977}.
The resulting real-valued parameters represent the Doktorov operator as a
sequence of four fundamental Gaussian
unitary operators (see Eqs.~\eqref{eq:multimode-disp-op}--\eqref{eq:multimode-rot-op})~\cite{huh:2015}:

\begin{equation}
\hat{D}(\udl{\gamma})\hat{R}(\bm{U}_2)\hat{S}(\ln(\bm{L}))\hat{R}(\bm{U}_1).
\label{eq:doktorov-op}
\end{equation}
Other operator decompositions are also possible~\cite{doktorov:1975,doktorov:1977}.
For experimental realization, preparing the displaced state at the circuit
input allows the initial interferometer $\hat{R}(\bm{U}_1)$ and input
displacements $\hat{D}(\udl{\alpha})$ to be compressed into a single multi-mode
coherent displacement acting on vacuum,
$\hat{R}(\bm{U}_1)\hat{D}(\udl{\alpha})\ket{\udl{0}} =
\hat{D}(\udl{\tilde{\alpha}})\ket{\udl{0}}$, where $\udl{\tilde{\alpha}} =
\bm{U}_1\udl{\alpha}$ with $\udl{\alpha} = \bm{J}^{-1}\udl{\gamma}$
(aligned with Figs.~\ref{fig:gbs-illustration} and
\ref{fig:gbs-resource-hierarchy}). This eliminates the need for an initial
multi-mode interferometer mesh before the squeezing stage.

Because Eq.~\eqref{eq:transition-profile} 
decomposes the profile into three independent sums (one for each
polarization direction), we restrict the following discussion to a single component $r$ of the TDM operator.

The TDM operator component $\hat{\mu}_r$ is expanded as a function of normal mode coordinates $\hat{\udl{Q}}$
\begin{widetext}
\begin{equation}
  \begin{split}
    \hat{\mu}_r(\hat{\udl{Q}})&=\mu_r^{(0)}+\sum_{k=1}^N\mu_{r,k}^{(1)}\hat{Q}_k+\frac{1}{2}\sum_{k,l=1}^N
    \mu_{r,k,l}^{(2)}\hat{Q}_k\hat{Q}_l+\cdots\\
                              &= \mu_r^{(0)}+\udl{\chi}_r^\intercal\frac{(\hat{\udl{a}}+\hat{\udl{a}}^\dagger)}{\sqrt{2}}
    +\left(\frac{(\hat{\udl{a}}+\hat{\udl{a}}^\dagger)}{\sqrt{2}}\right)^\intercal
    \bm{X}_r\left(\frac{(\hat{\udl{a}}+\hat{\udl{a}}^\dagger)}{\sqrt{2}}\right)+\cdots,
  \end{split}
  \label{eq:tdm-expansion}
\end{equation}
where the expansion coefficients correspond to $\chi_{r,k}=\sqrt{\hbar/\omega_k}\mu_{r,k}^{(1)}$
and $X_{r,k,l}=\mu_{r,k,l}^{(2)}\left(\sqrt{\hbar/\omega_k}\sqrt{\hbar/\omega_l}\right)/2$,
utilizing the relation $\hat{Q}_k=\sqrt{\hbar/(2\omega_k)}(\hat{a}_k+\hat{a}_k^\dagger)$.
We consider the case of $\bm{X}_r$ being real and symmetric, 
as we expect it for real normal modes and transitions between non-degenerate (real) electronic states.
The normalization constant $\mathcal{N}_r$, if terms up to second order are included 
in the expansion of the TDM, is given as~\cite{jnane:2021}:
\begin{equation}
  \mathcal{N}_r = \left((\mu_r^{(0)})^2 + \frac{1}{2}\sum_{k=1}^N
    \chi_{r,k}^2+\mu_r^{(0)}\sum_{k=1}^NX_{r,k,k}+\frac{1}{4}
    \sum_{\substack{k,l=1\\k\neq l}}^NX_{r,k,k}X_{r,l,l}+\frac{1}{2}\sum_{\substack{k,l=1\\k\neq l}}^N
    X_{r,k,l}^2+\frac{3}{4}\sum_{k=1}^NX_{r,k,k}^2
  \right).
  \label{eq:norm-constant}
\end{equation}
\end{widetext}

Jnane~\etal~\cite{jnane:2021} developed an approximation scheme to evaluate transition
probabilities involving the non-unitary TDM operator using strictly unitary operations. 
This is achieved by defining an auxiliary function
$f_{\udl{v}'}(\kappa) = \abs{\mel{\udl{v}',\text{in}}{\hat{U}_{\text{Dok}}\exp(\kappa\hat{\mu})}{\udl{0},\text{in}}}^2$,
where $\hat{\mu}=\hat{\mu}^\dagger$ and $\kappa\in\mathbb{C}$.
By employing a specific linear combination of $f_{\udl{v}'}$ for distinct
values of $\kappa$ the target transition probability can be isolated:
\begin{widetext}
\begin{equation}
  \abs{\mel{
  \udl{v}',\text{in}}{\hat{U}_{\text{Dok}}\hat{\mu}}{\udl{0},\text{in}}}^2 =
  \frac{1}{2\tau^2}\left[f_{\udl{v}'}(\im\tau) + \frac{1}{2}f_{\udl{v}'}(\tau)
  + \frac{1}{2}f_{\udl{v}'}(-\tau)-2f_{\udl{v}'}(0)\right] + \mathcal{O}(\tau^2),
\label{eq:tdm-int-approx}
\end{equation}
where $\tau$ is a real positive parameter and herein referred to as the resolution of the auxiliary function.
The explicit Gaussian operator expression for $f_{\udl{v}'}$ is
\begin{equation}
  f_{\udl{v}'}(\kappa) = \abs{\mathcal{C}(\kappa)}^2
\Bigl|\vphantom{\hat{D}(\udl{\gamma})\hat{R}(\bm{U}_2)}\mel{\udl{v}',\text{in}}{
      \underbrace{\hat{D}(\udl{\gamma})\hat{R}(\bm{U}_2)\hat{S}(\ln(\bm{L}))\hat{R}(\bm{U}_1)}_{\hat{U}_{\text{Dok}}}
      \hat{R}(\bm{U})\hat{S}(\bm{\Xi}(\kappa))
    \hat{D}(\udl{\alpha}(\kappa))}{\udl{0},\text{in}}\Bigr|^2,
\label{eq:full-auxfunc}
\end{equation}
\end{widetext}
where $\mathcal{C}(\kappa)$ is a parameter-dependent scalar prefactor to be defined later.
Consequently, each auxiliary function is characterized by a unique Gaussian operator
$\hat{U}_{\text{aux}}(\kappa)$,
which comprises the Doktorov operator augmented by additional rotation, squeezing
and displacement operations.

Regarding the auxiliary Gaussian operators in Eq.~\eqref{eq:full-auxfunc},
we utilize the eigenvalue decomposition $\bm{X}=\bm{O}\bm{D}\bm{O}^\intercal$
with $\bm{D} = \text{diag}(d_1,\cdots,d_N)$ and the orthogonal
matrix $\bm{O}\in\mathbb{R}^{N\times N}$ (setting $\bm{U}=\bm{O}$).
For convenience, we define $\udl{b}=\bm{O}^\intercal\udl{\chi}$ and obtain for
the parameters associated with the auxiliary function method
\begin{widetext}
\begin{equation}
\bm{\Xi}(\kappa)=\bm{r}(\kappa)\exp(\im\bm{\theta}(\kappa)) = \text{diag}(\Xi_1(\kappa),\cdots,\Xi_N(\kappa))
=\text{diag}(r_1(\kappa)\exp(\im\theta_1(\kappa)),\cdots,r_N(\kappa)\exp(\im\theta_N(\kappa))),
\end{equation}
where the mode-specific functions are defined as~\cite{jnane:2021}:
\begin{align}
  t_k(\kappa)&=\kappa d_k/(1-\kappa d_k)\label{eq:auxfunc-t-k},\\
  \theta_k(\kappa)&=\arg(t_k(\kappa)),\\
  r_k(\kappa)&=\arctanh(\abs{t_k(\kappa)}),\\
  s_k(\kappa)&=\sech(\abs{\Xi_k(\kappa)}),\\
  \alpha_k(\kappa)&=\frac{\kappa b_k}{\sqrt{2}}(1+\Xi_k(\kappa)).
\end{align}
Finally, the mode-specific prefactors are given by
\begin{equation}
  C_k(\kappa)=\frac{1}{\sqrt{\abs{s_k(\kappa)}(1-\kappa d_k)}}\exp[\frac{(\kappa b_k)^2}{4}
    +\frac{(\kappa b_k)^2}{2}(1+\Xi_k^*(\kappa))\Xi_k(\kappa) + \frac{\abs{\kappa b_k(1+\Xi_k(\kappa))}^2}{4}],
  \label{eq:auxfunc-expansion-coefficients}
\end{equation}
which are combined into a single scalar prefactor
\begin{equation}
  \mathcal{C}(\kappa) = \exp(\kappa\mu^{(0)})\prod_{j=1}^NC_j(\kappa).
  \label{eq:auxfunc-expansion-coefficient}
\end{equation}
\end{widetext}
For the detailed derivation of $C_k(\kappa)$, we refer to the Supporting Information (SI) of Ref.~\cite{jnane:2021}.
Explicit expressions for the linear canonical transformation matrices $\bm{E}$ and $\bm{F}$,
and the transformation vector $\udl{g}$ for the auxiliary function method of Ref.~\cite{jnane:2021} are given in 
Appendix~\ref{sec:auxfunc-bogoliubov-explicit}.
By examining the denominators in Eqs.~\eqref{eq:auxfunc-t-k} and~\eqref{eq:auxfunc-expansion-coefficients}
we immediately recognize that the resolution $\tau$, which
directly affects $\kappa$, must be chosen carefully to avoid numerical instabilities.
We developed a procedure for determining the resolution $\tau$ that minimizes the 
approximation error of the auxiliary function,
and avoids potential instabilities, which is described in Appendix~\ref{sec:opt-tau}. 
Moreover, the evaluation of vibronic spectra at finite temperatures is often necessary to 
reproduce experimental results. Our approach for extending the auxiliary function method 
to finite temperatures is detailed in Appendix~\ref{ssec:finite-temp-extension}.
\subsection{Fock-space restriction for auxiliary function method}
\label{ssec:auxfunc-prescreening}
The auxiliary function method can be employed to calculate vibronic spectra
in the second-order HT-approximation. Within the TI framework, Fock-space restrictions
are necessary to treat large molecular systems. Here, we describe our approach to
deriving a systematic Fock-space restriction for vibronic spectra considering
the HT effect of up to second order via the auxiliary function method.

The strategy for prescreening employing auxiliary functions involves initializing twelve distinct 
prescreening procedures: three utilizing complex parameters and nine utilizing real parameters. 
These correspond to the four auxiliary function parametrizations (one complex and three real) 
required for each of the three Cartesian components of the electric transition dipole moment.
For the real parametrizations we employ the sum rules of Jankowiak \etal~\cite{jankowiak:2007},
whereas for the complex parametrizations we utilize the generalized sum rules Eqs.~\eqref{eq:sum-rule-1} and
\eqref{eq:sum-rule-2}.

We then use the linear combination of auxiliary functions Eq.~\eqref{eq:tdm-int-approx},
where instead of directly calculating the FCFs of a single transition 
via Eq.~\eqref{eq:full-auxfunc} we linearly combine the results
of the individual sum rules of the prescreenings
to approximate the true sum rules for the HT effect up to second order.

In this manner, we first determine the highest excitation $\tilde{v}_k^{\text{max}}$
required for each individual mode $k$
to satisfy the prescreening tolerance $t_{\text{m}}$.
The latter is the numerical threshold used in the prescreening procedure of Ref.~\cite{jankowiak:2007}
to truncate the one-mode excitation space. Excitations higher than $\tilde{v}_k^{\text{max}}$
are neglected once the cumulative contribution exceeds $1-t_\text{m}$.
Subsequently, we evaluate the importance of simultaneous excitations of up to $M$
out of the $N$ vibrational modes using the coupling prescreening tolerance $t_{\text{c}}$,
following Ref.~\cite{jankowiak:2007}.
The term coupling truncation error refers to the estimated error in the transition 
profile arising from neglecting multi-mode excitations beyond the coupling cutoff.

We only use the approximate values from the linear combination of the 12 separate prescreenings when comparing
to $t_{\text{m}}$ and $t_{\text{c}}$.
To prevent premature convergence we also subtract the estimated error of the auxiliary functions
$\epsilon_{\text{aux}}= f_{\text{err}}(\tau)$, whose calculation is detailed in Appendix~\ref{sec:opt-tau}.
We require $\epsilon_{\text{aux}}<t_{\text{m}}\wedge\epsilon_{\text{aux}}<t_{\text{c}}$ for the prescreening
to definitely converge. This condition is generally satisfied within relevant
tolerance regimes when employing the procedure for determining the resolution $\tau$ 
that minimizes the approximation error of the auxiliary function, described in Appendix~\ref{sec:opt-tau}.

We note that determining a suitable resolution for the prescreening procedure
is more complex than for the integral evaluation; a detailed discussion is provided
in Appendix~\ref{sec:auxfunc-prescreening-resolution}.
\subsection{Extension to Time-Dependent Method}
\label{ssec:auxfunc-tdmethod}
We begin by deriving the augmented thermal autocorrelation function.
Following Refs.~\cite{huh:2011a,huh:2010thermal,huh:2012}, the Franck--Condon factor weighted density
of states (FCW) is expressed as
\begin{equation}
  \begin{split}
    \varrho_{\text{FCW}}(\omega) = &\frac{1}{2\uppi\mathcal{N}}\abs{\braket{\udl{0}'}{\udl{0}}}^2
    \\&\times\int_{-\infty}^\infty\dd{t}G^K(\bm{Z}(t))\exp(\im(\omega-\omega_0)t),
  \end{split}
    \label{eq:fcw}
\end{equation}
where $\omega_0$ is the $0'-0$ electronic transition angular frequency and $G^K$
is the generating function with thermal kernel $K$ (which reduces to $G$ at \SI{0}{\kelvin}).
In Eq.~\eqref{eq:fcw}, we have also introduced the time-dependent variables
\begin{align}
  \bm{Z}(t)&=\text{bldiag}(\bm{z}(t),\bm{z}'(t)),\\
  \bm{z}(t)&=\text{diag}(z_1(t),\cdots,z_N(t)),\\
  \bm{z}'(t)&=\text{diag}(z_1'(t),\cdots,z_N'(t)),\\
  z_i(t)&=\exp(\im\epsilon_i t/(2\hbar)),\\
  z_i'(t)&=\exp(-\im\epsilon_i't/(2\hbar)).
\end{align}
A convolution of $\varrho_{\text{FCW}}(\omega)$ with a Lorentzian line shape function 
of full width at half maximum (FWHM) $\Gamma$ can be realized by introducing
$L(t)=\exp(-\Gamma\abs{t}/2)$ in Eq.~\eqref{eq:fcw}.
The working equation for the time-dependent (TD) method for complex
Gaussian unitary operators is
\begin{widetext}
\begin{equation}
  \begin{split}
    \varrho_{\text{FCW},L}(\omega) &= \frac{1}{2\uppi\mathcal{N}}\abs{\braket{\udl{0}'}{\udl{0}}}^2
    \int_{-\infty}^\infty\dd{t}\det(\bm{\mathcal{I}}+\bm{\mdc{Z}}(t)\bm{\mdc{W}}
    \bm{\mdc{Z}}(t))^{-1/2}\\
    &\times
    \exp(\udl{\mdc{r}}^\intercal\bm{\mdc{Z}}(t)
    (\bm{\mathcal{I}}+\bm{\mdc{Z}}(t)\bm{\mdc{W}}\bm{\mdc{Z}}(t))^{-1}
    \bm{\mdc{Z}}(t)\udl{\mdc{r}})L(t)\exp(\im(\omega-\omega_0)t),
\end{split}
\label{eq:aug-tdmethod}
\end{equation}
\end{widetext}
where we have defined the $4N$-dimensional quantities
\begin{equation}
  \begin{split}
    \bm{\mathcal{I}} &= 
  \begin{pmatrix}
    \bm{I}&0\\
    0&\bm{I}
  \end{pmatrix},
  \bm{\mdc{W}} = 
  \begin{pmatrix}
    \real(\bm{W})&-\imag(\bm{W})\\
    -\imag(\bm{W}) &-\real(\bm{W})
  \end{pmatrix},\\
    \udl{\mdc{r}} &= 
  \begin{pmatrix}
    \real(\udl{r})\\
    -\imag(\udl{r})
  \end{pmatrix},
  \bm{\mdc{Z}}(t) = 
  \begin{pmatrix}
    \bm{Z}(t)&\bm{0}\\
    \bm{0}&\bm{Z}(t)
  \end{pmatrix}.
\end{split}
\end{equation}
To extend the TD method to the auxiliary function ansatz, we employ a strategy
analogous to the auxiliary function prescreening, exploiting the established link
between the TI and TD frameworks~\cite{huh:2012}.
We approximate the exact autocorrelation function via a linear combination
of the autocorrelation functions $G^K(\bm{Z}(t),\kappa)$,
corresponding to the twelve individual parametrizations of the auxiliary function method,
inserted into Eq.~\eqref{eq:aug-tdmethod}.
Subsequently, the transition profile is obtained from the approximate
autocorrelation function via the Fast Fourier transform (FFT).
The extension to the finite-temperature case is given in Appendix~\ref{ssec:finite-temp-extension}.
\section{Select application results}
\label{sec:results}

In this section, we demonstrate the efficacy of the proposed framework by 
applying it to molecular systems of increasing complexity. We demonstrate the 
accuracy of our method through benchmark applications to phenanthrene and 
benzene---including the full 30-mode benzene system---which are detailed in 
Appendix~\ref{sec:examples}. With the validity of the approach illustrated 
against these standards, we focus in this section only on the 102-mode coronene system. 

Computational details for all calculations are provided in 
Appendix~\ref{ssec:comp-details}, and the full set of parameters required to 
reproduce these results is available in the Supplementary Material.
Additionally, Appendix~\ref{ssec:phases-determination} outlines the 
procedure employed to resolve the phase ambiguity inherent in evaluating 
transition properties on a numerical grid.
The coronene system challenges the capabilities of state-of-the-art photonic hardware for vibronic
spectroscopy~\cite{madsen:2022,yu:2023,deng:2023,zhu:2024,yu:2025}.
The limiting factor is not only the number of modes; it is the required circuit
programmability. A device would have to implement mode-resolved squeezing,
an arbitrary interferometer encoding the 102-mode Duschinsky rotation,
and the additional Gaussian operations of the auxiliary-function method.
Moreover, at finite temperature, the interferometer acts on 204 modes. 
Additionally, stringent requirements on photon loss and photon-number-resolving detection apply.
Therefore, coronene serves as an ideal molecular testbed to demonstrate the
scalability of our framework and establish state-resolved reference data at the
100-mode scale.

Symmetry selection rules determine which transitions can gain intensity only at second order in the HT expansion, irrespective of the level of theory
used for the electronic-structure treatment. The intensities of these features,
however, depend on the quality of the computed HT parameters. As with any effective
Hamiltonian, quantitative accuracy is contingent on the underlying
parameter set and on the approximations inherent to the model.
Although higher-level electronic structure calculations would likely enhance agreement
with experimental data, they entail a substantial increase in computational cost,
thereby limiting accessible system sizes. Consequently, our primary
objective is to demonstrate that the extension of the time-dependent and
time-independent methods to the second-order Herzberg--Teller treatment---via the auxiliary function method---
can elucidate vibrational transitions that were neglected in previous approaches,
rather than solely maximizing agreement with experiment.

Orlandi and Zerbetto~\cite{orlandi:1988} previously analyzed the absorption
and emission spectra of the S$_1$ ($^1$B$_{2\mathrm{u}}$)--S$_0$ ($^1$A$_{1\mathrm{g}}$)
transition of coronene \cite{babkov:1973,ohno:1973,ohno:1975,bermudez:1986},
demonstrating that for this transition,
which is Franck--Condon forbidden in the electric dipole approximation,
a linear HT ansatz alone fails to account for specific 
spectral features~\cite{orlandi:1988}. Our present work enables investigating
the effect of second order HT contributions. By employing the auxiliary function method with prescreening, we can evaluate for this purpose the full 102-mode system.

\begin{figure*}\centering
  \begin{minipage}{.45\linewidth}
    \includegraphics[width=\linewidth]{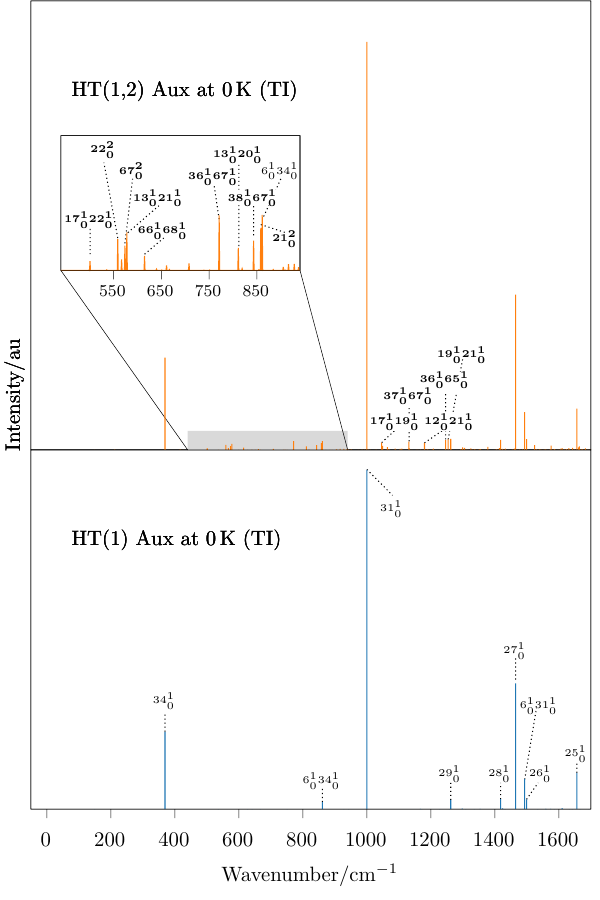}
  \end{minipage}
  \begin{minipage}{.45\linewidth}
    \includegraphics[width=\linewidth]{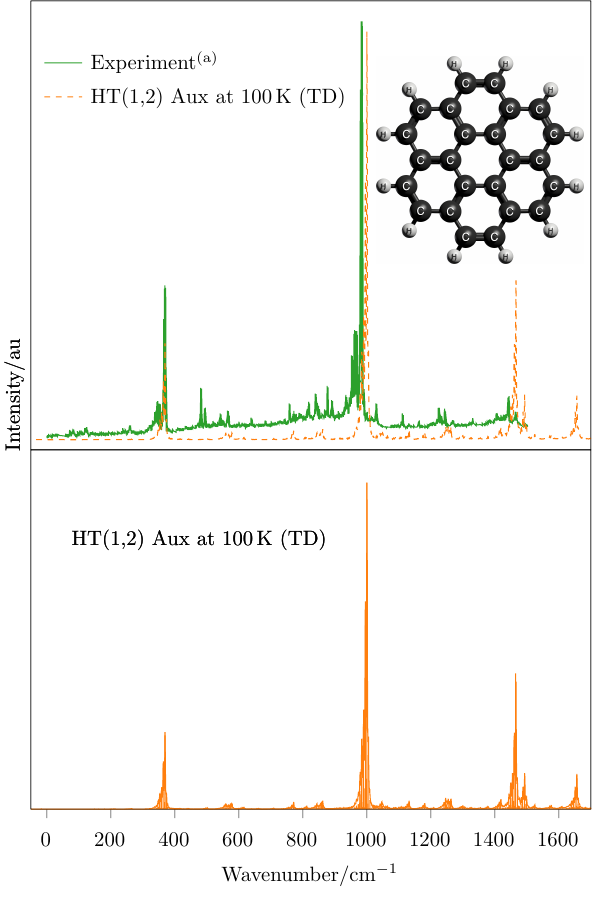}
  \end{minipage}
    \caption{Computed absorption profiles of the $\mathrm{S}_1\leftarrow \mathrm{S}_0$ transition of coronene at various orders of Herzberg--Teller (HT)
  coupling employing the auxiliary function method within the time-independent (TI) framework
  at \SI{0}{\kelvin} (left) as well as TI and time-dependent (TD) frameworks at \SI{100}{\kelvin} (right),
  compared to the digitized, mirrored experimental fluorescence excitation spectrum $(\mathrm{a})$ of Kunishige \etal
  from Ref.~\cite{kunishige:2017} with a small y-axis offset.
  Additionally, the TD \SI{100}{\kelvin} absorption profile is superimposed onto the experimental
  spectrum with a different small y-axis offset in the top right.
  As scale on the x-axis the wavenumber in excess of the $0'-0$ transition is chosen.
  The experimental spectrum was normalized with respect to the most intense peak.
  The fluorescence-excitation spectrum is used here as an experimental reference
for the absorption profile, assuming that the fluorescence quantum yield varies
only weakly over the displayed spectral range.
The experimental spectrum was truncated in Ref.~\cite{kunishige:2017} at approximately \SI{1450}{\per\centi\meter}
because the short-wavelength limit of the dye laser scan (\SI{395}{\nano\meter})
restricted the accessible vibrational excess energy above the electronic origin (\SI{420}{\nano\meter}).
For the TI framework a wavenumber graining of \SI{1}{\per\centi\meter} was employed. 
  For the TD framework a Lorentzian line shape with a FWHM of $\SI{10}{\per\centi\meter}$,
  a Fast Fourier Transform (FFT) grid size of $2^{14}$, a time increment
  of $\Delta t\sim\SI{2}{\femto\second}$, and a time interval of $(-\SI{16.677187}{\pico\second},\SI{16.677187}{\pico\second})$ were employed.
Peak assignments correspond to the notation of Tab.~\ref{tab:coronene-energies-comparison}.
Treatment of the Herzberg--Teller (HT) effect of first order is denoted HT(1), of first and
second order HT(1,2).
Select transitions due to second-order HT effects are denoted in bold letters.
The shaded region highlights a magnified view of smaller spectral features.}
  \label{fig:coronene}
\end{figure*}

We compare our results against the experimental fluorescence excitation spectrum
measured by Kunishige \etal~\cite{kunishige:2017}, which represents 
a comparatively well-resolved gas-phase spectrum of jet-cooled
coronene.
The corresponding vibronic spectral profiles are displayed in Fig.~\ref{fig:coronene}.
Before discussing the agreement and assignment of individual transitions,
we analyze in general terms the role of individual normal modes and their excitations in inducing vibronic intensity.

As usual, the vibrational normal modes are grouped according to their irreducible 
representations and arranged in descending order of their
vibrational energy. Degenerate modes of vibrations are not counted doubly.
In Tab.~\ref{tab:coronene-energies-comparison} we compare harmonic vibrational wavenumbers
computed on the density functional theory (DFT) level for the electronic ground state of coronene
and on the time-dependent DFT level for the lowest electronically excited singlet state
of this molecule to available experimental vibrational transition energies.
\begin{table}\centering
  \caption{Selected harmonic vibrational wavenumbers $\tilde{\omega}$ of coronene for the ground
    ($\mathrm{S}_0$, \sym{A}{1}{g} symmetry) and excited state
    ($\mathrm{S}_1$, \sym{B}{2}{u} symmetry) arranged in descending order within each irreducible representation $\Gamma$.
    Experimental values of corresponding fundamental transition wavenumbers $\tilde{\nu}$ from Ref.~\cite{kunishige:2017} are provided in parentheses.
    Stated experimental uncertainties are $\pm\SI{12}{\per\centi\meter}$ and
  $\pm\SI{1}{\per\centi\meter}$ for $\mathrm{S}_0$ and $\mathrm{S}_1$ respectively.
The full list of computed harmonic vibrational wavenumbers is given in the Supplementary Material.
}
  \label{tab:coronene-energies-comparison}
  \begin{tabular}{cc
      S[table-format=4.2]
      c
      S[table-format=4.2]
      c
    }\toprule
    &&\multicolumn{2}{c}{S$_0$}&\multicolumn{2}{c}{S$_1$}\\
  Mode&$\Gamma$&\multicolumn{2}{c}{{$\tilde{\omega}$/cm$^{-1}$} }&
  \multicolumn{2}{c}{{$\tilde{\omega}'$/cm$^{-1}$} }\\\midrule
  $\nu_{3}$&a$_{1\mathrm{g}}$&1376.0&(1355)&1385.56&(1379)\\
  $\nu_{6}$&a$_{1\mathrm{g}}$& 494.6&(488)&492.76&(482)\\
  $\nu_{12}$&b$_{1\mathrm{g}}$&774.07&&751.34&\\
  $\nu_{13}$&b$_{1\mathrm{g}}$&165.65&&149.03&\\
  $\nu_{17}$&b$_{2\mathrm{g}}$&227.44&&221.62&\\
  $\nu_{19}$&e$_{1\mathrm{g}}$&858.24&&824.92&\\
  $\nu_{20}$&e$_{1\mathrm{g}}$&686.44&&662.84&\\
  $\nu_{21}$&e$_{1\mathrm{g}}$&459.47&&429.43&\\
  $\nu_{22}$&e$_{1\mathrm{g}}$&297.50&&279.53&\\
  $\nu_{25}$&e$_{2\mathrm{g}}$&1654.8&&1656.30&(1637)\\
  $\nu_{26}$&e$_{2\mathrm{g}}$&1487.74& (1458)&1499.09&(1441)\\
  $\nu_{27}$&e$_{2\mathrm{g}}$&1470.22&&1465.81&\\
  $\nu_{28}$&e$_{2\mathrm{g}}$&1428.3&(1421)&1418.89&(1404)\\
  $\nu_{29}$&e$_{2\mathrm{g}}$&1254.8&&1262.93&(1221)\\
  $\nu_{30}$&e$_{2\mathrm{g}}$&1185.4&&1172.08&(1159)\\
  $\nu_{31}$&e$_{2\mathrm{g}}$&1018.6& (996)&1000.35&(981)\\
  $\nu_{32}$&e$_{2\mathrm{g}}$& 695.5&&693.66&(682)\\
  $\nu_{33}$&e$_{2\mathrm{g}}$& 502.31& &499.74 &(483)\\
  $\nu_{34}$&e$_{2\mathrm{g}}$& 373.08& (371)&369.14&(367)\\
  $\nu_{36}$&a$_{1\mathrm{u}}$& 532.61& &483.83&\\
  $\nu_{37}$&a$_{2\mathrm{u}}$& 874.98& &844.63&\\
  $\nu_{38}$&a$_{2\mathrm{u}}$& 571.69&&556.47&\\
  $\nu_{49}$&b$_{2\mathrm{u}}$& 1219.29&&1265.70&\\
  $\nu_{65}$&e$_{2\mathrm{u}}$&  789.31&& 762.36&\\
  $\nu_{66}$&e$_{2\mathrm{u}}$&  561.37&& 532.01&\\
  $\nu_{67}$&e$_{2\mathrm{u}}$&  303.22&& 287.24&\\
  $\nu_{68}$&e$_{2\mathrm{u}}$&  88.79&& 83.61&(87) \\
  \bottomrule
  \end{tabular}
\end{table}
The electronic component of the electric transition dipole moment operator $\udl{\hat{\mu}}(\udl{\hat{Q}})$ is expanded in
the mass-weighted normal coordinates $\udl{\hat{Q}}$ of the ground (initial)
electronic state. With this choice of coordinate system, the terms entering the
total transition intensity $\mathcal{N}$ in Eq.~\eqref{eq:norm-constant} can be
assigned to individual ground-state normal modes, or to pairs of modes in the
case of quadratic Herzberg--Teller (HT) terms.

\begin{figure*}\centering
  \begin{minipage}{.45\linewidth}
    \includegraphics[width=\linewidth]{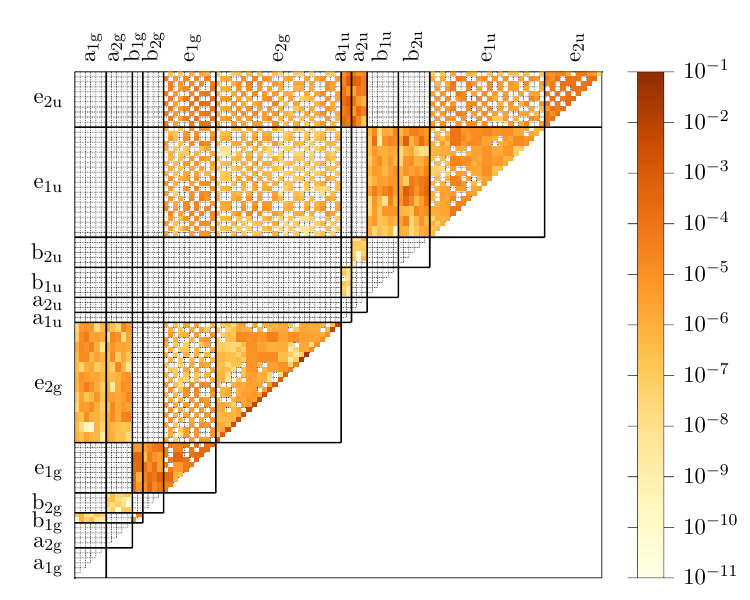}
  \end{minipage}
  \begin{minipage}{.45\linewidth}
    \includegraphics[width=\linewidth]{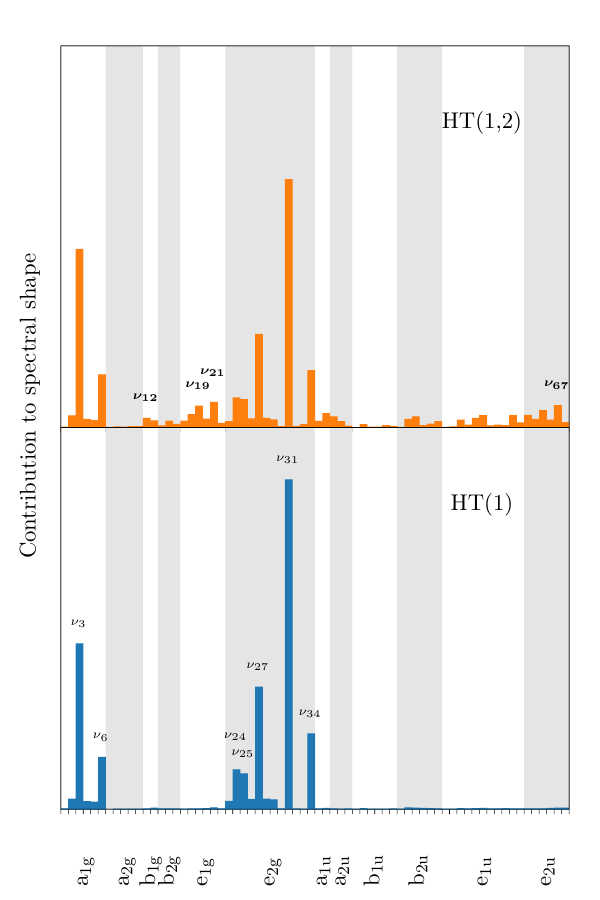}
  \end{minipage}
  \caption{
        Left: Ground state mode-resolved contributions to the total intensity derived 
  from Eq.~\eqref{eq:norm-constant}, with $\mu_r^{(0)}=0$ due to the Franck--Condon
  symmetry-forbidden nature of the ($\mathrm{S}_1\leftarrow \mathrm{S}_0$) transition in
  coronene. Linear terms are added to the diagonal elements. The vibrational 
  modes are arranged in descending order of their harmonic vibrational wavenumbers within each 
  irreducible representation.
  Solid lines mark the boundaries between different
  irreducible representations; dotted lines indicate individual vibrational 
  modes. Right: Mode contributions calculated by evaluating the auxiliary 
  function linear combination of Eq.~\eqref{eq:sum-rule-2} with $n=0$ for 
  coronene at \SI{0}{\kelvin}. Contributions from degenerate modes with identical 
  harmonic vibrational wavenumbers are combined.
  Alternating background shading indicates the different irreducible representations;
  axis ticks represent individual vibrational modes.
  Treatment of the linear Herzberg--Teller (HT) 
  effect is denoted HT(1); inclusion of both linear and quadratic effects is denoted HT(1,2).
  Select mode contributions with considerable increase due to second-order HT effects are denoted in bold.
}
  \label{fig:coronene-mode-contributions}
\end{figure*}

The corresponding mode-resolved contributions are shown in the left panel of
Fig.~\ref{fig:coronene-mode-contributions}. In this representation, modes
belonging to the same irreducible representation are ordered by decreasing
vibrational energy. Degenerate modes are plotted separately rather than combined,
so that quadratic HT couplings within the degenerate subspace remain visible.
As expected, the dominant contribution to the total intensity arises from
ground-state modes of \sym{e}{2}{g} symmetry, in particular from mode $\nu_{31}$
(\sym{e}{2}{g}). However, additional symmetry-allowed contributions that appear
only when HT terms up to second order are included also make a non-negligible
contribution to the total summed intensity.
The total intensity at \SI{0}{\kelvin} considering first-order HT
effects is computed to be $\SI{1.018E-01}{\mathrm{D}^2}$, and considering up to second-order HT
effects is computed to be $\SI{1.340E-01}{\mathrm{D}^2}$. Quadratic and bilinear HT effects
therefore account for about 24\% of the total intensity in an HT expansion of up to second-order.

This ground-state decomposition of $\mathcal{N}$ should be distinguished from
the analysis of the vibronic absorption profile. The latter also depends on 
the normal modes of the excited (final) electronic state. Therefore,
the modes that induce intensity in the ground-state expansion of the transition
dipole moment are not necessarily identical to the excited-state modes that shape
the observable profile.

To identify which excited-state modes are responsible for the spectral profile,
we employed the second sum rule, Eq.~\eqref{eq:sum-rule-2}. For each excited-state
normal mode, we evaluated how the calculated profile changes when progressions
in that mode are excluded, both for HT expansions truncated after first order
and after second order. This analysis measures the importance of each final-state
mode for the distribution of intensity over the vibronic transitions. Because a
single vibronic transition can contain simultaneous excitations in several
normal modes, these mode-specific contributions are not mutually exclusive.
Consequently, the individual ``mode-excluded'' contributions are not expected to
sum to unity. The resulting excited-state mode contributions are shown in the
right panel of Fig.~\ref{fig:coronene-mode-contributions}.
The largest contributions to the spectral shape 
originate from modes $3$ (\sym{a}{1}{g}) and $31$ (\sym{e}{2}{g})
in the final state.
We anticipate that progressions of the totally symmetric \sym{a}{1}{g} modes of
the excited state, built upon HT-induced so-called false origins, will be dominant. 
Notably, we observe a substantial increase in mode contributions when including second-order 
HT coupling, as shown in the right panel of
Fig.~\ref{fig:coronene-mode-contributions}. This behavior can be rationalized from
the structure of the HT(2) contribution. In the auxiliary-function formulation,
the second-order terms are decomposed into squeezing and rotation operations
[Eq.~\eqref{eq:full-auxfunc}], in addition to the rotations already present in
the Doktorov transformation [Eq.~\eqref{eq:doktorov-op}]. These additional
operations do not necessarily enhance the Duschinsky-induced mode mixing in all
cases. Depending on the transition and on the signs and magnitudes of the
couplings, they may either increase or partly compensate existing mode mixing.
For the present transition in coronene, however, the HT(2) terms lead to a net
increase in mode mixing. Considering Duschinsky mixing,
when the molecular point group is conserved, the Duschinsky matrix only couples
normal modes belonging to the same irreducible representation. By contrast,
quadratic HT terms contain products of normal-mode operators and can therefore
couple modes of different symmetry, provided that the overall product satisfies
the symmetry selection rules of the electric transition dipole moment. Thus, HT(2)
introduces additional intensity pathways that are absent at the FC/HT(1) level.
This conclusion does not depend on the auxiliary-function representation. The same
mode-coupling terms are already present in the ladder-operator form of the
second-order electric transition dipole expansion [Eq.~\eqref{eq:tdm-expansion}].
Based on this decomposition and the results shown in Fig.~\ref{fig:coronene-mode-contributions}
we can expect noticeable differences between absorption spectra computed with first-
versus second-order HT coupling and anticipate that the vibronic absorption profile of coronene exhibits observable
spectral features attributable exclusively to second-order HT coupling.

With reference to Fig.~\ref{fig:coronene}, we note that establishing
the precise vibrational temperature of the molecules within the 
supersonic jet expansion~\cite{engelking:1991} presents a non-trivial challenge; 
standard empirical approximations~\cite{depaul:1993} did not improve visual agreement 
with experimental observations in the present study. To address this model insufficiency,
we exploited the computational efficiency of the TD method, employing a narrow FWHM,
to perform a parameter sweep over the 
temperature range $0\text{--}\SI{300}{\kelvin}$ in steps of \SI{25}{\kelvin} (not shown). 
We subsequently selected the vibrational temperature that minimized the deviation 
from the experimental spectral profile by visual inspection (\SI{100}{\kelvin}).

We note that this approach allows for individual mode temperatures $T_k$ \cite{huh:2010thermal,huh:2011a,huh:2012}, which may prove
useful especially in jet-cooling experiments, where modes are not necessarily
cooled on similar timescales~\cite{engelking:1991,depaul:1993}.
We reserve a detailed analysis for future work.

Finally, the state-resolved TI formalism was applied to resolve the specific vibronic
transitions at this chosen temperature. 
In order to compute a state-resolved vibronic absorption profile for coronene, we resort to
our prescreening conditions. For the tolerances employed here $t_\mathrm{m}=0.0001$, $t_\mathrm{c}=0.1$
(see Appendix~\ref{ssec:int-prescreening-bounds} for more information regarding these tolerances)
our rigorous bounds guarantee that the truncated Fock space captures a fraction of the total 
intensity in [0.955,\,0.957] for the HT(1) and [0.984,\,0.986]
for the HT(1,2) \SI{0}{\kelvin} calculations, with only two and three simultaneously excited modes, respectively.
We evaluated 3,200 and 40,660 integrals up to a wavenumber of \SI{2000}{\per\centi\meter}
in excess of the $0'-0$ transition, for the HT(1) and HT(1,2) \SI{0}{\kelvin} calculations respectively,
out of \num{2.4e6} energetically accessible final vibrational states
(determined via the energy-grained Beyer--Swinehart algorithm \cite{beyer:1973,stein:1973}).
Employing the prescreening conditions led to a reduction of overall integral
calculations by a factor of 750 and 59, respectively. 
Our prescreening conditions indicated that the finite-temperature profile of coronene is congested, 
capturing a fraction of the total intensity in [0.847,\,0.850] for the HT(1,2) calculations with five simultaneously excited modes.
We evaluated 167,771,086 integrals up to a wavenumber of \SI{2000}{\per\centi\meter}
in excess of the $0'-0$ transition for the HT(1,2) \SI{100}{\kelvin} calculation,
out of about \num{9.9e10} thermally accessible transitions (for an initial-state Boltzmann cutoff of 
$\ge 10^{-5}$ \cite{berger:1997}).
Employing the prescreening conditions led to a reduction of overall integral
calculations by a factor of 590. Larger computational savings may be realized
with different tolerance settings \cite{jankowiak:2007}.

At higher temperatures, the intensity is spread out over more vibronic transitions,
making peak assignment less sought after. Moreover, because these ``hot'' band transitions
are often energetically close to each other, false mappings of GBS devices are harder to detect, rendering
these vibronic spectra less suitable for benchmarking. 
Therefore, although the \SI{100}{\kelvin} profile was also evaluated within the
TI framework for comparison, peak assignments were carried out on the
\SI{0}{\kelvin} TI profile, whereas the complementary TD formalism provides the
untruncated finite-temperature spectral envelope. 
The resulting absorption spectra are shown in Fig.~\ref{fig:coronene}, compared to the experimental spectrum.
The band shapes are dominated by the first-order HT effect. The second-order HT effect contributes less overall.
However, we observe that some vibronic transitions in specific spectral regions are almost entirely
caused by second-order coupling features,
notably around \SIrange{600}{800}{\per\centi\meter} and \SIrange{1000}{1200}{\per\centi\meter}. The state-resolved framework facilitates access to information regarding individual transitions,
thereby enabling the assignment of specific peaks. The assignments attributed 
to either HT(1) or HT(2) effects are annotated
in Fig.~\ref{fig:coronene}.
In these spectral regions, the HT(1)-weighted density of states is low.
Consequently, a second-order treatment is indicated to elucidate the experimental data.

The assignments above also serve the purpose of providing benchmark data for GBS. Each
assigned transition corresponds to a photon-number pattern that a GBS device
would have to sample with resolvable frequency. The features induced solely by
second-order HT coupling define state-resolved targets beyond the common 
FC level circuits realized so far. The spectra of Fig.~\ref{fig:coronene} thus
constitute reference data against which future photonic implementations can be measured.

Regarding the HT(1) features, when comparing our assignments with those of Kunishige \etal~\cite{kunishige:2017},
we see that the $6_0^131_0^1$, $28_0^1$, $29_0^1$, $31_0^1$ and $34_0^1$ assignments match,
whereas we assigned the peak at $\sim\SI{1440}{\per\centi\meter}$ to $26_0^1$ instead of $27_0^1$.
Moreover, we acknowledge that the series of peaks in the \SIrange{450}{700}{\per\centi\meter}
range, particularly the feature at \SI{500}{\per\centi\meter}, is effectively absent
in our computed spectra. Kunishige \etal assigned this feature to the $33_0^1$ transition \footnote{We
use the conventional notation for vibronic transitions, where 
the principal numbers refer to the vibrations excited and the superscripts and subscripts
refer to the number of quanta involved in the final and initial states, respectively~\cite{brand:1983}.},
which aligns wavenumber-wise with our computations, as there is an HT-active $\text{e}_{2\text{g}}$ mode with
harmonic vibrational wavenumber \SI{499.74}{\per\centi\meter} in this energy region; however, the
computed linear and quadratic HT parameters are too small for the resulting
peak to become discernible. It is well-established~\cite{coriani:2010}
that the computed HT parameters are highly sensitive to the choice of density functional,
and, more generally, to the quality of the underlying electronic structure calculations.
We also note that the present framework employs the Born--Oppenheimer (BO) approximation \cite{born:1927}, and there is discussion in the literature \cite{sharf:1971,orlandi:1972} that the neglected non-Born--Oppenheimer effects could be of similar magnitude to the leading HT contributions. 
The primary objective of this study, however, is not to maximize agreement with experiment,
but rather to validate the extension of the TI and TD methods to include second-order
HT effects. Furthermore, for systems where non-BO coupling is negligible,
this framework may enable more rigorous benchmarking of density functionals, as it can provide access 
to second-order HT spectral features previously unavailable to comparison with experiment.

\section{Conclusion and outlook}
\label{sec:conclusion-outlook}
In this work, we presented a comprehensive framework for handling 
products of fundamental Gaussian operators efficiently on classical hardware
and derived rigorous prescreening conditions for the associated transition probabilities.
We applied our developments to the computation of vibronic spectral profiles,
a common benchmark for advancements in Gaussian Boson Sampling (GBS) devices.
Employing the auxiliary function method introduced in Ref.~\cite{jnane:2021},
our quantum-inspired formalism extended the prescreening criteria
to account for the Herzberg--Teller (HT) effect up to second order.
Furthermore, we generalized this approach to treat finite-temperature environments.

We validated the approach on reduced-dimensionality model systems of phenanthrene and benzene,
and demonstrated its scalability on larger systems, including full 30-mode benzene and 102-mode coronene.
The results indicate that including second-order HT effects is essential for the 
accurate description of spectral features in these highly symmetric molecules
in regions that are essentially uncongested by transitions induced by first-order HT terms.
For high-temperature regimes or systems where peak assignment is less critical,
we utilized the connection described in Ref.~\cite{huh:2011a} between time-independent
and time-dependent formulations to extend the auxiliary function method into the
time-dependent domain, thereby facilitating the analysis of even larger molecular systems.

From the GBS benchmarking perspective, the present results show that
state-resolved vibronic distributions with realistic Herzberg--Teller and
finite-temperature effects can be approximated on classical hardware with 
intensity bounds for systems that challenge current photonic implementations of vibronic spectroscopy.
The computational cost of state-resolved vibronic spectra calculation is therefore
dictated not merely by mode count, but by the displacement and squeezing
parameters, thermal occupation, and the convergence rate of the prescreening
bounds.

The framework presented herein is adaptable to other vibronic processes,
including optical techniques such as resonance Raman spectroscopy~\cite{santoro:2011}
as well as non-radiative decay via intersystem crossing~\cite{etinski:2011,khoroshkin:2026}
and internal conversion~\cite{peng:2007,roiz:2026}.

Beyond optical spectroscopy and non-radiative decay, the generalized prescreening conditions are formally applicable
to a broad range of problems expressible as Gaussian operator decompositions,
such as classes of non-unitary processes~\cite{wakefield:2024},
vibronic excitations~\cite{jahangiri:2020} and vibronic dynamics~\cite{sparrow:2018}.
Finally, because prescreening is substantially less demanding than evaluating full distributions,
the presented framework establishes a versatile and efficient diagnostic tool to assess the limits
of tractability for controlled Fock-space methods on classical hardware.
For parameter regimes where truncations remain compact, this approach yields error-controlled
benchmarks on realistic molecular models with up to 102 modes
(204 effective modes at finite temperature) to validate subsequent GBS hardware implementations.
Conversely, parameter regimes where compact representations break down identify
concrete computational targets for evaluating photonic devices against advanced classical
heuristics~\cite{phillips:2019,deshpande:2022}.

\begin{acknowledgments}
  J.H. acknowledges the following funding resources: (1) Basic Science Research
  Program through the National Research Foundation of Korea (NRF), funded by the
  Ministry of Science and ICT (RS-2023-NR068116, RS-2025-03532992);
  (2) Institute for Information \& Communications Technology Promotion (IITP)
  grant funded by the Korea government (MSIP) (No. 2019-0-00003), focusing on
  research and development of core technologies for programming, running,
  implementing, and validating fault-tolerant quantum computing systems;
  (3) Yonsei University Research Fund (Project No. 2025-22-0140).
  Support by the Deutsche Forschungsgemeinschaft (DFG, German Research Foundation) through
  project number 223848855 - CRC 1083 ``structure and dynamics of internal interfaces'' 
  within subproject B8 is gratefully acknowledged.
  For the computational work, we acknowledge computing time provided at the NHR
  Center NHR@SW at Goethe-University Frankfurt. This was funded by the
  Federal Ministry of Education and Research and the state
  governments participating on the basis of the resolutions
  of the GWK for national high-performance computing at universities
  (http://www.nhr-verein.de/unsere-partner).
  We are indebted to Dieter Schuch, Sabrina Marquardt, Jason Stuber and Hans-Christian
  Jankowiak for discussions.
\end{acknowledgments}
\appendix
\section{Derivation of the recurrence relation coefficients}
\label{sec:fernandez-deriv}
The following relations hold when an invertible operator $\hat{U}$ can be expressed as a product
of exponentials of operators containing only linear and quadratic terms in the
creation and annihilation operators. 
Fern\'andez and Tipping defined a transformation relating the final state ladder
operators to the initial state operators
\begin{equation}
  \hat{a}_j = \hat{U}^{-1}\hat{a}_j^\prime\hat{U}\quad (\hat{a}_j^\dagger)^{\sim} = 
  \hat{U}^{-1}\hat{a}_j^{\prime\dagger}\hat{U},
  \label{eq:sim-trans}
\end{equation}
which can be written in matrix form as~\cite{fernandez:1989}:
\begin{equation}
  \begin{pmatrix}
    \hat{\udl{a}}\\
    (\hat{\udl{a}}^\dagger)^{\sim}
  \end{pmatrix} = 
  \begin{pmatrix}
    \bm{U}_-&\bm{U}_+\\
    \bm{V}_-&\bm{V}_+
  \end{pmatrix}
  \begin{pmatrix}
    \hat{\udl{a}}^\prime\\
    \hat{\udl{a}}^{\prime\dagger}
  \end{pmatrix} + 
  \begin{pmatrix}
    \udl{U}_0\\
    \udl{V}_0
  \end{pmatrix}.
  \label{eq:inverse-bogoliubov-trans}
\end{equation}
We can relate these coefficients to our linear canonical transformation
coefficients that describe a transformation from the initial to the final state
via 
\begin{align}
  \tilde{\bm{E}}&=\bm{U}_-^{-1},\quad&&\tilde{\bm{E}}^\prime = \bm{V}_-\bm{U}_-^{-1}\\
  \tilde{\bm{F}}&= \bm{U}_-^{-1}\bm{U}_+,\quad&&\tilde{\bm{F}}^\prime=(\bm{U}_-^\intercal)^{-1} = \tilde{\bm{E}}^\intercal\\
  \tilde{\udl{g}}&=\bm{U}_-^{-1}\udl{U}_0,\quad&&\tilde{\udl{g}}^\prime=\udl{V}_0-\bm{V}_-\bm{U}_-^{-1}\udl{U}_0.
\label{eq:fernandez-coefficients}
\end{align}
As indicated in Eq.~\eqref{eq:sim-trans}, the Fern\'andez and Tipping relations are valid
for general similarity transforms. However, because
our operator $\hat{U}_{\text{G}}$ is unitary, the condition 
$\hat{U}^{-1}=\hat{U}^\dagger$ holds. Moreover, 
$(\udl{\hat{a}}^\dagger)^{\sim}=\udl{\hat{a}}^\dagger$ and the Fern\'andez and Tipping
coefficients satisfy the following relations:
\begin{equation}
  \udl{V}_0=\udl{U}_0^*,\quad\bm{V}_-=\bm{U}_+^*,\quad\bm{V}_+=\bm{U}_-^*.
  \label{eq:fernandez-unitary-operator-relations}
\end{equation}
To derive these coefficients, we recall Eq.~\eqref{eq:bogoliubov-trans} and 
we can subsequently express the inverse Bogoliubov transformation from Eq.~\eqref{eq:inverse-bogoliubov-trans}
using the quantities defined in  Eq.~\eqref{eq:bogoliubov-trans}:
\begin{equation}
  \begin{split}
  \begin{pmatrix}
    \hat{\udl{a}}\\\hat{\udl{a}}^{\dagger}
  \end{pmatrix} &= 
  \begin{pmatrix}
    \bm{E}&\bm{F}\\
    \bm{F}^*&\bm{E}^*
  \end{pmatrix}^{-1}\left(
  \begin{pmatrix}
    \hat{\udl{a}}^\prime\\\hat{\udl{a}}^{\prime\dagger}
  \end{pmatrix} - 
  \begin{pmatrix}
    \udl{g}\\\udl{g}^*
\end{pmatrix}\right) \\&= 
  \begin{pmatrix}
    \bm{E}^\dagger &-\bm{F}^\intercal \\
    -\bm{F}^\dagger &\bm{E}^\intercal 
  \end{pmatrix}
  \begin{pmatrix}
    \hat{\udl{a}}^\prime\\\hat{\udl{a}}^{\prime\dagger}
  \end{pmatrix} - 
  \begin{pmatrix}
    \bm{E}^\dagger &-\bm{F}^\intercal \\
    -\bm{F}^\dagger &\bm{E}^\intercal 
  \end{pmatrix}
  \begin{pmatrix}
      \udl{g}\\\udl{g}^*
  \end{pmatrix}
\end{split}
\end{equation}
Comparing the result with Eq.~\eqref{eq:inverse-bogoliubov-trans}, we identify
the relations
\begin{align}
  \udl{U}_0 = -\bm{E}^\dagger\udl{g}
  +\bm{F}^\intercal \udl{g}^*,\quad
  \bm{U}_-=\bm{E}^\dagger ,\quad
  \bm{U}_+=-\bm{F}^\intercal.
\end{align}
Substituting these expressions into Eqs.~\eqref{eq:fernandez-unitary-operator-relations} and
\eqref{eq:fernandez-coefficients} yields Eq.~\eqref{eq:fernandez-coefficients-efg}.
\section{A note on phase factors}
\label{sec:phase-factors}
For complex Franck--Condon integrals, it is important to note that the standard 
relations in Eqs.~\eqref{eq:ma-d-op-annihilation}--\eqref{eq:ma-s-op-creation} do 
not cover the contraction of products of Gaussian operators of the same fundamental type.
All contraction rules for these operators are provided, for instance, in Ref.~\cite{ma:1990}:
\begin{align}
  \hat{D}(\udl{\alpha})\hat{D}(\udl{\beta}) &= \exp[1/2(\udl{\beta}^\dagger\udl{\alpha}
  -\udl{\alpha}^\dagger\udl{\beta})]
  \hat{D}(\udl{\alpha}+\udl{\beta}),\label{eq:d-op-prod}\\
  \hat{R}(\bm{U}_1)\hat{R}(\bm{U}_2) &= \hat{R}(\bm{U}_1\bm{U}_2),\label{eq:r-op-prod}\\
  \hat{S}(\bm{\Xi}_1)\hat{S}(\bm{\Xi}_2) &=\exp[\im/2\trace(\bm{\Phi}_3)]\hat{S}(\bm{\Xi}_3)\hat{R}(\exp(\im\bm{\Phi}_3)),\label{eq:s-op-prod}
\end{align}
where $\bm{\Xi}_3$ and $\bm{\Phi}_3$ are defined via
\begin{align}
  \bm{T}_3 = \bm{S}_1^{-1}(\bm{T}_1+\bm{T}_2)(\bm{I}+\bm{T}_1^\dagger\bm{T}_2)^{-1}\bm{S}_1^\intercal,\\
  \exp(\im\bm{\Phi}_3) = \bm{U}_3 = \bm{S}_3^{-1}\bm{S}_1(\bm{I}+\bm{T}_2\bm{T}_1^\dagger)^{-1}\bm{S}_2,
\end{align}
with $\bm{\Xi}_k\equiv\bm{r}_k\exp(\im\bm{\theta}_k)$, $\bm{T}_k\equiv\tanh(\bm{r}_k)\exp(\im\bm{\theta}_k)$,
$\bm{S}_k\equiv\sech(\bm{r}_k)$ and $k=1,2,3$.

Because we treat arbitrary combinations of fundamental unitary Gaussian operators
with complex parameters, we must address 
how the phase factors in Eqs.~\eqref{eq:d-op-prod} and~\eqref{eq:s-op-prod} influence 
the final result when composing multiple operations.

To resolve apparent phase discrepancies during such operator contractions,
it is crucial to distinguish between the classical symplectic group and its
quantum mechanical double cover, the metaplectic group~\cite{littlejohn:1986,arvind:1995}.
Classical linear canonical transformation matrices (see Eq.~\eqref{eq:bogoliubov-trans}) accurately track
the transformation of phase space coordinates but are entirely blind to the global quantum phase.
Conversely, the unitary squeeze and rotation operators---such as the operators
defined by Ma and Rhodes~\cite{ma:1990}---form a representation of the metaplectic group. 

The composition of the unitary operators along the group manifold generates an
accumulated topological phase, precisely manifesting as the
explicit $\exp[\im/2\trace(\bm{\Phi})]$ term when multiplying two
multimode squeeze operators~\cite{ma:1990}.
Therefore, all contractions must be evaluated explicitly at the operator
level---carefully commuting intermediate rotation operators through
subsequent squeeze operators~\cite{ma:1990,fan:2003}
(\ie resolving the chain $\hat{S}_1\hat{S}_2\hat{S}_3
\rightarrow\hat{S}_4\hat{R}_1\hat{S}_3\rightarrow\hat{S}_4\hat{S}_3\hat{R}_2
\rightarrow\hat{S}_5\hat{R}_3\hat{R}_2$)---to maintain phase consistency.
The switching rules required are given in Ref.~\cite{ma:1990}:
\begin{align}
    \hat{D}(\udl{\alpha})\hat{S}(\bm{\Xi}) &= \hat{S}(\bm{\Xi})\hat{D}
    \label{eq:switching-rule-1}
    (\cosh(\bm{r})\udl{\alpha}-\sinh(\bm{r})\exp(\im\bm{\theta})\udl{\alpha}^*),\\
    \label{eq:switching-rule-2}
    \hat{D}(\udl{\alpha})\hat{R}(\bm{U}) &= \hat{R}(\bm{U})\hat{D}(\bm{U}^\dagger\udl{\alpha}),\\
    \hat{S}(\bm{\Xi})\hat{R}(\bm{U}) &= \hat{R}(\bm{U})\hat{S}(\bm{U}^\dagger\bm{\Xi}\bm{U}^*)
    \label{eq:switching-rule-3}
\end{align}

A proposed algorithm for maintaining phase consistency for a Gaussian operator
decomposition $\hat{U}_{\text{G}}$ is as follows:
First, employ the switching rules detailed in Eqs.~\eqref{eq:switching-rule-1}
and~\eqref{eq:switching-rule-2} to commute
all displacement operators to the right and combine them via Eq.~\eqref{eq:d-op-prod}.
Second,  commute all rotation operators
to the left via Eq.~\eqref{eq:switching-rule-3} and combine them using Eq.~\eqref{eq:r-op-prod}.
Third, contract all squeezing operators via Eq.~\eqref{eq:s-op-prod} from 
right to left such that $\hat{S}_1,\cdots,\hat{S}_N\rightarrow \hat{S}_1'\hat{R}_2,\cdots\hat{R}_N$,
and combine the resulting rotation operators using
Eq.~\eqref{eq:r-op-prod}. Fourth, factorize the $N$-mode squeezing
operation into two rotations and a single-mode squeezing operation
\begin{equation}
\hat{S}_1'(\bm{\Xi})=\hat{R}(\bm{U}_1)\hat{S}(\bm{\xi})\hat{R}(\bm{U}_1^\intercal),
\end{equation}
by Takagi--Autonne factorization~\cite{takagi:1924,autonne:1915}
\begin{equation}
  \bm{\Xi} = \bm{U}_1\bm{\xi}\bm{U}_1^\intercal,
\end{equation}
where $\bm{\xi} = \text{diag}(\xi_1,\cdots,\xi_N)$. Finally, combine all accumulated phases
by the contraction into a single global phase.
The resulting operator
\begin{equation}
  \hat{U}_{\text{G}}'=\hat{R}(\bm{U}_1)\hat{S}(\bm{\xi})\hat{R}(\bm{U}_2)\hat{D}(\udl{\gamma})
\end{equation}
can then be mapped to the linear canonical transformation matrices
$\bm{E}$, $\bm{F}$ and $\udl{g}$ by employing
Eqs.~\eqref{eq:ma-d-op-annihilation}--\eqref{eq:ma-s-op-creation}.

To demonstrate that the discussed phase ambiguity does not affect our specific application, 
we revisit the symmetric Franck--Condon factor (FCF) generating function (GF) 
ansatz Eq.~\eqref{eq:fcf-gf}
from Ref.~\cite{jankowiak:2007}, which underpins both the TI and TD approaches. 
Writing the Gaussian operators explicitly:
\begin{equation}
  \begin{split}
    &G(\bm{Z}) = \uppi^{-2N}(\mel{\udl{0}',\text{in}}{\hat{U}_{\text{G}}}{\udl{0},\text{in}}^*
      \mel{\udl{0}',\text{in}}{\hat{U}_{\text{G}}}{\udl{0},\text{in}})^{-1}\\
       &\times\int\dd{\udl{\alpha}\dd{\udl{\gamma}}'}\mel{\bm{z}'^*\udl{\gamma}',\text{in}}{\hat{U}_{\text{G}}}{\bm{z}\udl{\alpha},\text{in}}
      \mel{\bm{z}'\udl{\gamma}',\text{in}}{\hat{U}_{\text{G}}}{\bm{z}^*\udl{\alpha},\text{in}}^*.
  \end{split}
  \label{eq:gf-detail}
\end{equation}
In Eq.~\eqref{eq:gf-detail}, the same Gaussian operator $\hat{U}_{\text{G}}$
is inserted in both matrix elements and acts on the ket labeled by the
unprimed state in each case. The complex conjugation in the second factor is
applied to the scalar matrix element as a whole. Consequently, the phase
factors generated by the contraction order cancel exactly. For Gaussian Boson
Samplers this cancellation also holds, because they sample from the same
probability distribution encoded by the GF in Eq.~\eqref{eq:gf-detail}.
Similarly, the cancellation holds for the augmented 
GFs of Huh~\cite{huh:2011a}, which generalize to non-Condon cases. Even though the specific 
integrals may differ (\eg in the decomposition of the non-Condon GF into FC/FC, 
FC/HT, and HT/HT terms), the definition of the Gaussian operator remains consistent 
across the bra and ket vectors. Therefore, application of the relations in
Eqs.~\eqref{eq:ma-d-op-annihilation}--\eqref{eq:ma-s-op-creation} suffices.
\section{Rigorous Prescreening Conditions Derivation}
\label{sec:prescreening-derivation}
In the following, we derive the rigorous prescreening conditions of
Jankowiak \etal~\cite{jankowiak:2007} extended to the case
$\bm{W}\in\mathbb{C}^{2N\times 2N},\udl{r}\in\mathbb{C}^{2N}$.
Inserting the spatial representation of the coherent state overlap integral
Eq.~\eqref{eq:coh-ov-dok} into the FCF GF Eq.~\eqref{eq:fcf-gf} yields
\begin{equation}
  \begin{split}
    G(\bm{Z}) = &\uppi^{-2N}\int\dd[2]{\udl{\alpha}}\dd[2]{\udl{\gamma}'}\exp(-\udl{\xi}^\dagger
    \udl{\xi})\\
                &\times\exp(-\frac{1}{2}\udl{\xi}^\intercal\bm{Z}\bm{W}\bm{Z}\udl{\xi}
  + \udl{r}^\intercal\bm{Z}\udl{\xi})\\
                &\times\exp(-\frac{1}{2}\udl{\xi}^\dagger\bm{Z}\bm{W}^*\bm{Z}\udl{\xi}^*
    + \udl{r}^\dagger\bm{Z}\udl{\xi}^*).
    \label{eq:fcf-gf-spatial}
  \end{split}
\end{equation}
We decompose $\udl{\xi}$ into its real and imaginary components,
\begin{equation}
  \udl{\xi} = \real(\udl{\xi}) + \im\imag(\udl{\xi})
  ,\quad\udl{\xi}^* = \real(\udl{\xi}) - \im\imag(\udl{\xi})
\end{equation}
and obtain
\begin{equation}
  -\udl{\xi}^\dagger\udl{\xi} = -\real(\udl{\xi})^\intercal\bm{I}\real(\udl{\xi})
  - \imag(\udl{\xi})^\intercal\bm{I}\imag(\udl{\xi}).
  \label{eq:xi-part}
\end{equation}
Upon expanding the exponents in Eq.~\eqref{eq:fcf-gf-spatial}, we observe that the 
real and imaginary parts of $\udl{\xi}$ are coupled.
The FCF GF~\eqref{eq:fcf-gf-spatial} can be reformulated as
\begin{equation}
  \begin{split}
    G(\bm{Z}) =&\uppi^{-2N}\int\dd[2]{\udl{\alpha}}\dd[2]{\udl{\gamma'}}\\
               &\times\exp(-\real(\udl{\xi})^\intercal
    (\bm{I}+\bm{Z}\real(\bm{W})\bm{Z})\real(\udl{\xi})
    +2\real(\udl{r})^\intercal\bm{Z}\real(\udl{\xi}))\\
               &\times\exp(-\imag(\udl{\xi})^\intercal
               (\bm{I}-\bm{Z}\real(\bm{W})\bm{Z})\imag(\udl{\xi})
      -2\imag(\udl{r})^\intercal\bm{Z}\imag(\udl{\xi}))\\
               &\times\exp(\real(\udl{\xi})^\intercal\bm{Z}\imag(\bm{W})\bm{Z}\imag(\udl{\xi}))\\
               &\times\exp(\imag(\udl{\xi})^\intercal\bm{Z}\imag(\bm{W})\bm{Z}\real(\udl{\xi})).
  \end{split}
\end{equation}
This expression is compactly written in block-matrix form as
\begin{equation}
  G(\bm{Z}) =\uppi^{-2N}\int\dd[2]{\udl{\alpha}}\dd[2]{\udl{\gamma'}}
  \exp(-\udl{\zeta}^\intercal(\bm{\mathcal{I}} + \bm{\mdc{Z}}\bm{\mdc{W}}
  \bm{\mdc{Z}})\udl{\zeta}
  + 2\udl{\mdc{r}}^\intercal\bm{\mdc{Z}}\udl{\zeta}),
 \label{eq:gf-int-reform}
\end{equation}
where we have defined the $4N$-dimensional quantities
\begin{equation}
  \begin{split}
    \bm{\mathcal{I}} &= 
  \begin{pmatrix}
    \bm{I}&0\\
    0&\bm{I}
  \end{pmatrix},
  \bm{\mdc{W}} = 
  \begin{pmatrix}
    \real(\bm{W})&-\imag(\bm{W})\\
    -\imag(\bm{W}) &-\real(\bm{W})
  \end{pmatrix},
  \udl{\zeta} =
  \begin{pmatrix}
    \real(\udl{\xi})\\
    \imag(\udl{\xi})
  \end{pmatrix},\\
    \udl{\mdc{r}} &= 
  \begin{pmatrix}
    \real(\udl{r})\\
    -\imag(\udl{r})
  \end{pmatrix},
  \bm{\mdc{Z}} = 
  \begin{pmatrix}
    \bm{Z}&\bm{0}\\
    \bm{0}&\bm{Z}
  \end{pmatrix}.
\end{split}
\label{eq:four-dim-quantities}
\end{equation}

The remaining derivation of the sum rules follows the approach by Jankowiak \etal
taking into account the coupled real and imaginary parts of the quantities
defined in Eq.~\eqref{eq:four-dim-quantities}. 

To evaluate the FCF GF, we utilize the general solution for an $M$-dimensional
Gaussian integral
\begin{equation}
  \begin{split}
  \mathcal{I}_M[\bm{A},\udl{b}]\equiv&\uppi^{-M/2}\int\dd{\udl{x}}\exp(-\udl{x}^\intercal
  \bm{A}\udl{x} + 2\udl{b}^\intercal\udl{x})\\ =& \det(\bm{A})^{-1/2}\exp(\udl{b}^\intercal
  \bm{A}^{-1}\udl{b}),
  \end{split}
  \label{eq:general-gauss-int}
\end{equation}
which converges, provided that the eigenvalues of the symmetric matrix
$\bm{A}$ possess positive real parts. Applying Eq.~\eqref{eq:general-gauss-int}
to Eq.~\eqref{eq:gf-int-reform} in terms of $4N$-dimensional quantities yields
\begin{equation}
  G(\bm{Z}) = \det(\bm{\mathcal{I}}+\bm{\mdc{Z}}\bm{\mdc{W}}\bm{\mdc{Z}})^{-1/2}
  \exp(\udl{\mdc{r}}^\intercal\bm{\mdc{Z}}
  (\bm{\mathcal{I}}+\bm{\mdc{Z}}\bm{\mdc{W}}\bm{\mdc{Z}})^{-1}
  \bm{\mdc{Z}}\udl{\mdc{r}}).
  \label{eq:gf-int}
\end{equation}
This result directly reduces to Eq.~(46) of~\cite{jankowiak:2007} in the limit
where $\bm{W}$ and $\udl{r}$ are strictly real-valued.
\subsection{Partitioning Scheme}
We proceed to derive the partitioning scheme generalized for complex-valued $\bm{W}$ and $\udl{r}$. 
The generating function in Eq.~\eqref{eq:gf-int} is partitioned into orthogonal subspaces $X$ 
and $Y$ with dimensions $\dim(X)=N_X$ and $\dim(Y) = N_Y = 2N-N_X$.
We aim to express the partitioned FCF generating function as
\begin{equation}
  G(\bm{Z}_{XX},\bm{Z}_{YY}) = G_{X|Y}(\bm{Z}_{XX},\bm{Z}_{YY})G_Y(\bm{Z}_{YY}).
\end{equation}
Utilizing the decomposition approach from Eq.~(53) of~\cite{jankowiak:2007}, we
factorize Eq.~\eqref{eq:gf-int} into contributions from the two subspaces:
\begin{equation}
  \begin{split}
    &G(\bm{Z}_{XX},\bm{Z}_{YY}) = 
  \det((\bm{\mathcal{I}}+\bm{\mdc{Z}}\bm{\mdc{W}}\bm{\mdc{Z}})_{YY})^{-1/2}\\
    &\times\exp(\udl{\mdc{r}}_Y^\intercal\bm{\mdc{Z}}_{YY}(\bm{\mathcal{I}}+\bm{\mdc{Z}}
    \bm{\mdc{W}}\bm{\mdc{Z}})_{YY}^{-1}
      \bm{\mdc{Z}}_{YY}\udl{\mdc{r}}_Y)\\
    &\times\det(\bm{\mathcal{I}}_{XX}+\bm{\mdc{Z}}_{XX}\tilde{\bm{\mdc{W}}}_{XX}\bm{\mdc{Z}}_{XX})^{-1/2}\\
    &\times\exp(\udl{\tilde{\mdc{r}}}_{X}^\intercal\bm{\mdc{Z}}_{XX}
    (\bm{\mathcal{I}}_{XX}+\bm{\mdc{Z}}_{XX}\tilde{\bm{\mdc{W}}}_{XX}\bm{\mdc{Z}}_{XX})^{-1}
    \bm{\mdc{Z}}_{XX}\udl{\tilde{\mdc{r}}}_{X}).
  \end{split}
  \label{eq:gf-int-part}
\end{equation}
The quantities associated with the $X$ subspace are defined as
\begin{equation}
  \begin{split}
    \tilde{\bm{\mdc{W}}}_{XX} &=
    \begin{pmatrix}
    \tilde{\bm{W}}_{XX}^+&\tilde{\bm{W}}_{XX}^{\text{off}}\\
    \tilde{\bm{W}}_{XX}^{\text{off}}&\tilde{\bm{W}}_{XX}^-
  \end{pmatrix} \\&= \bm{\mdc{W}}_{XX}-\bm{\mdc{W}}_{XY}\bm{\mdc{Z}}_{YY}
  (\bm{\mathcal{I}}+\bm{\mdc{Z}}
      \bm{\mdc{W}}\bm{\mdc{Z}})_{YY}^{-1}\bm{\mdc{Z}}_{YY}
     \bm{\mdc{W}}_{YX}\\
    \udl{\tilde{\mdc{r}}}_X &=
    \begin{pmatrix}
  \udl{\tilde{r}}_X^+\\
    \udl{\tilde{r}}_X^-
  \end{pmatrix}\\
                      &= \udl{\mdc{r}}_X-\bm{\mdc{W}}_{XY}\bm{\mdc{Z}}_{YY}
  ((\bm{\mathcal{I}}+\bm{\mdc{Z}}\bm{\mdc{W}}\bm{\mdc{Z}})_{YY})^{-1}\bm{\mdc{Z}}_{YY}\udl{\mdc{r}}_Y.
  \end{split}
\end{equation}
because $\bm{I}_{XY} = 0$. Consequently, the GF decomposes into the marginal
distribution for $Y$:
\begin{align}
  &G_Y(\bm{Z}_{YY}) =\det((\bm{\mathcal{I}}+\bm{\mdc{Z}}\bm{\mdc{W}}\bm{\mdc{Z}})_{YY})^{-1/2}\nonumber\\
                    &\times\exp(\udl{\mdc{r}}_Y^\intercal\bm{\mdc{Z}}_{YY}(\bm{\mathcal{I}}+
                    \bm{\mdc{Z}}\bm{\mdc{W}}\bm{\mdc{Z}})_{YY}^{-1}
    \bm{\mdc{Z}}_{YY}\udl{\mdc{r}}_Y),
  \label{eq:gf-y-int}
\end{align}
and the conditional distribution for $X$:
\begin{align}
    &G_{X|Y}(\bm{Z}_{XX},\bm{Z}_{YY}) = \det(\bm{\mathcal{I}}_{XX}+\bm{\mdc{Z}}_{XX}\tilde{\bm{\mdc{W}}}_{XX}\bm{\mdc{Z}}_{XX})^{-1/2}\nonumber\\
                                &\times\exp(\udl{\tilde{\mdc{r}}}_{X}^\intercal\bm{\mdc{Z}}_{XX}
                                (\bm{\mathcal{I}}_{XX}+\bm{\mdc{Z}}_{XX}\tilde{\bm{\mdc{W}}}_{XX}\bm{\mdc{Z}}_{XX})^{-1}
                                \bm{\mdc{Z}}_{XX}\udl{\tilde{\mdc{r}}}_{X}).
  \label{eq:gf-xy-int}
\end{align}
\subsection{Sum Rules}
\label{ssec:sum-rules}
\subsubsection{No Excited Modes in the X Space}
Setting $\bm{Z}_{YY}=\bm{I}_{YY}$, Eq.~\eqref{eq:gf-y-int} reduces to
\begin{equation}
  \begin{split}
  &G_Y(\bm{I}_{YY}) = \sum_{\udl{v}_Y',\udl{v}_Y=0}^\infty q(\udl{v}_Y';\udl{v}_Y)\\
  &= \det((\bm{\mathcal{I}}+\bm{\mdc{W}})_{YY})^{-1/2}
  \times\exp(\udl{\mdc{r}}_Y^\intercal((\bm{\mathcal{I}}+
  \bm{\mdc{W}})_{YY})^{-1}
    \udl{\mdc{r}}_Y).
  \label{eq:gyiyy}
\end{split}
\end{equation}
In the special case where the summation runs over all final Fock states within
the $Y$ subspace, and utilizing the notation in Sec.~\ref{ssec:rr}, Eq.~\eqref{eq:gyiyy} becomes
\begin{equation}
  \begin{split}
    &G_Y(\bm{I}_{YY}) = \sum_{\udl{v}_Y'=0}^\infty q(\udl{v}_Y';\udl{0})\\
    &= \det((\bm{\mathcal{I}}+\bm{\mdc{\tilde{F}}})_{YY})^{-1/2}
    \exp(\udl{\mdc{\tilde{g}}}_Y^\intercal((\bm{\mathcal{I}}+\bm{\mdc{\tilde{F}}}_{YY})^{-1}
                       \udl{\mdc{\tilde{g}}}_Y)),
  \label{eq:sum-rule-1-si}
  \end{split}
\end{equation}
where the relevant matrices and vectors are defined as
\begin{equation}
  \bm{\mdc{\tilde{F}}}_{YY} = 
  \begin{pmatrix}
    \real(\tilde{\bm{F}})_{YY} & -\imag(\tilde{\bm{F}})_{YY}\\
    -\imag(\tilde{\bm{F}})_{YY} &-\real(\tilde{\bm{F}})_{YY}
  \end{pmatrix},\quad
  \udl{\mdc{\tilde{g}}}_Y = 
  \begin{pmatrix}
    -\real(\tilde{\udl{g}})_Y\\
    \imag(\tilde{\udl{g}})_Y
  \end{pmatrix}.
\end{equation}
We observe that from the components of $\bm{W}$ and $\udl{r}$ in Eq.~\eqref{eq:w-r-fernandez}, 
only $\tilde{\bm{F}}$ and $-\tilde{\udl{g}}$ yield non-zero contributions. 
Eq.~\eqref{eq:sum-rule-1-si} is identical to Eq.~\eqref{eq:sum-rule-1} in the main text; 
it corresponds to Eq.~(60) in Ref.~\cite{jankowiak:2007} and reduces to it in the limit 
of real-valued $\bm{W}$ and $\udl{r}$.

For finite-temperature evaluations, one must employ the thermally weighted versions 
of $\bm{W}$ and $\udl{r}$ provided in Ref.~\cite{huh:2012} in conjunction
with Eq.~\eqref{eq:gf-int-reform}.
\subsubsection{One Excited Mode in the X space}
Proceeding with the assumption that $\bm{Z}_{YY}=\bm{I}_{YY}$, the effective quantities 
for the $X$ subspace reduce to
\begin{equation}
  \begin{split}
  \tilde{\bm{\mdc{W}}}_{XX} &= \begin{pmatrix}
    \tilde{\bm{W}}_{XX}^+&\tilde{\bm{W}}_{XX}^{\text{off}}\\
    \tilde{\bm{W}}_{XX}^{\text{off}}&\tilde{\bm{W}}_{XX}^-
  \end{pmatrix} \\&= \bm{\mdc{W}}_{XX}-\bm{\mdc{W}}_{XY}
  (\bm{\mathcal{I}}+\bm{\mdc{W}}_{YY})^{-1}\bm{\mdc{W}}_{YX},\\
  \udl{\tilde{\mdc{r}}}_X &= \begin{pmatrix}
  \udl{\tilde{r}}_X^+\\
    \udl{\tilde{r}}_X^-
  \end{pmatrix} \\
  &= \udl{\mdc{r}}_X-\bm{\mdc{W}}_{XY}
  (\bm{\mathcal{I}}+\bm{\mdc{W}}_{YY})^{-1}\udl{\mdc{r}}_Y.
  \end{split}
\end{equation}
Consider the case in which only mode $k$ in the $X$ subspace is excited, with
final quantum number $v_k'$, whereas all other modes in $X$ are fixed to zero
occupation and hence are not summed over. The $Y$ subspace contains the
remaining final-state modes $\ket{\udl{v}'}$.
The relevant scalar components are given by
\begin{equation}
  \begin{split}
  \begin{pmatrix}
    \tilde{W}_{kk}^+&\tilde{W}_{kk}^{\text{off}}\\
    \tilde{W}_{kk}^{\text{off}}&\tilde{W}_{kk}^-\\
  \end{pmatrix} &= ( \bm{\mdc{\tilde{F}}}_{XX}- \bm{\mdc{\tilde{F}}}_{XY}
  (\bm{\mathcal{I}}+\bm{\mdc{\tilde{F}}}_{YY})^{-1} \bm{\mdc{\tilde{F}}}_{YX})_{kk}\\
  \begin{pmatrix}
    \tilde{r}_k^+\\
    \tilde{r}_k^-
  \end{pmatrix} 
  &= ( \tilde{\mdc{\udl{g}}}_X- \bm{\mdc{\tilde{F}}}_{XY}
  (\bm{\mathcal{I}}+ \bm{\mdc{\tilde{F}}}_{YY})^{-1} \tilde{\mdc{\udl{g}}}_Y)_{k}.
  \end{split}
\end{equation}
Expressing the integral form of Eq.~\eqref{eq:gf-xy-int} (using the definition 
in Eq.~\eqref{eq:general-gauss-int}) for this single-mode $k$ case yields
\begin{widetext}
\begin{equation}
  \begin{split}
    G_{X|Y}(\bm{Z}_{XX},\bm{I}_{YY}) = &\uppi^{-N_X}\int\dd{\udl{x}}\dd{\udl{y}}\exp(-\udl{x}^\intercal\bm{I}\udl{x})
                                       \exp(-\real(x_k)Z_{kk}\tilde{W}_{kk}^+Z_{kk}\real(x_k))
    \exp(2Z_{kk}\tilde{r}_k^+\real(x_k))\\
                                       &\times\exp(-\udl{y}^\intercal\bm{I}\udl{y})
                                       \exp(-\imag(x_k)Z_{kk}\tilde{W}_{kk}^-Z_{kk}\imag(x_k))
    \exp(2Z_{kk}\tilde{r}_k^-\imag(x_k))\\
                                       &\times\exp(-\real(x_k)Z_{kk}\tilde{W}_{kk}^\text{off}Z_{kk}\imag(x_k))
                                       \exp(-\imag(x_k)Z_{kk}\tilde{W}_{kk}^\text{off}Z_{kk}\real(x_k))\\
                                       &=\uppi^{-N_X}\int\dd{\udl{x}}\dd{\udl{y}}\exp(-\udl{x}^\intercal\bm{I}\udl{x})
                                       \exp(-\real(x_k)Z_{kk}\tilde{W}_{kk}^+Z_{kk}\real(x_k))
    \exp(2Z_{kk}\tilde{r}_k^+\real(x_k))\\
                                       &\times\exp(-\udl{y}^\intercal\bm{I}\udl{y})
                                       \exp(-\imag(x_k)Z_{kk}\tilde{W}_{kk}^-Z_{kk}\imag(x_k))
    \exp(2Z_{kk}\tilde{r}_k^-\imag(x_k))\\
                                       &\times\exp(-2\real(x_k)Z_{kk}\tilde{W}_{kk}^\text{off}Z_{kk}\imag(x_k))
  \end{split}
\end{equation}
Rather than evaluating this integral directly---which would recover Eq.~\eqref{eq:gf-xy-int} 
with $\bm{Z}_{YY}=\bm{I}_{YY}$---we employ the series expansion 
$\exp(A) = \sum_{n=0}^\infty A^n/n!$ for the interaction terms:
\begin{equation}
  \begin{split}
    G_{X|Y}(\bm{Z}_{XX},\bm{I}_{YY}) = &\uppi^{-N_X}\int\dd{\real(\udl{x})}
    \dd{\imag(\udl{x})}
  \exp(-{\real(\udl{x})}^\intercal\bm{I}\real(\udl{x}))\exp(-\imag(\udl{x})^\intercal\bm{I}
  \imag(\udl{x}))\\
                                       &\times\sum_{l=0}^\infty\frac{\real(x_k)^{2l}z_k^{2l}(-\tilde{W}_{kk}^+)^l}{l!}
                                     \sum_{m=0}^\infty\frac{\real(x_k)^{m}z_k^{m}(2\tilde{r}_k^+)^{m}}{(m)!}
        \sum_{n=0}^\infty\frac{\imag(x_k)^{2n}z_k^{2n}(-\tilde{W}_{kk}^-)^n}{n!}\\
                                       &\times
        \sum_{p=0}^\infty\frac{\imag(x_k)^{p}z_k^{p}(2\tilde{r}_k^-)^{p}}{p!}
        \sum_{q=0}^\infty\frac{\real(x_k)^q\imag(x_k)^qz_k^{2q}(-2\tilde{W}_{kk}^\text{off})^q}{q!}.
  \end{split}
  \label{eq:gf-xy-exp}
\end{equation}
We observe that the integration over the complex domain vanishes if the total power 
of the integration variables is odd. Specifically, the integral yields a non-zero 
contribution only if the conditions ($m+q$ is even) and ($p+q$ is even) are 
simultaneously satisfied.
Utilizing the multidimensional generalization of the Gamma function integral~\cite{gradshtein:2014,ruffa:2021},
\begin{equation}
  \begin{split}
    \int_0^\infty&\dd{u}\int_0^\infty\dd{v}\,u^{2l+m+q}v^{2n+p+q}\exp(-(u^{a}+v^b)^c)\\
    =&\Gamma((2l+m+q+1)/a)\Gamma((2n+p+q+1)/b)\\
     &\times\Gamma((1/c)((2l+m+q+1)/a
    + (2n+p+q+1)/b))\\
     &\times(abc\Gamma((2l+m+q+1)/a
      + (2n+p+q+1)/b))^{-1}.
  \end{split}
\end{equation}
With $a=b=2$ and $c=1$, symmetry over the four quadrants contributes $4P(m,q)P(p,q)$, canceling $(abc)^{-1}=1/4$ and yielding
\begin{equation}
  \begin{split}
    G_{X|Y}(\bm{Z}_{XX},\bm{I}_{YY})= &\uppi^{-1}\sum_{l,m,n,p,q=0}^\infty\frac{(-\tilde{W}_{kk}^+)^l
      (2\tilde{r}_k^+)^{m}(-\tilde{W}_{kk}^-)^n
    (2\tilde{r}_k^-)^{p}(-2\tilde{W}_{kk}^{\text{off}})^{q}}{l!m!n!p!q!}\\
                                      &\times P(m,q)P(p,q)\\
                                      &\times\Gamma(l+m/2+q/2+1/2)\Gamma(n+p/2+q/2+1/2)\\
                                      &\times z_k^{(2l+m+2n+p+2q)}.\\
    \end{split}
\end{equation}
where $P(a,b) = [1+(-1)^{a+b}]/2$ serves as a parity selector.
The variable $z_k$ acts as a bookkeeping parameter for the excitation level. 
We restrict our analysis to terms of order $z_k^{2n}$, which corresponds to the 
$n$-th excitation. This is enforced by substituting $n \to n-l-m/2-p/2-q$ and 
imposing the constraint $l+m/2+p/2+q = n$ on the summations. With $N_{lq}=n-l-q$,
the final expression is thus
\begin{equation}
  \begin{split}
    \sum_{\udl{v}_Y'|v_k'=n} q(\udl{v}';\udl{0}) =& \sum_{\udl{v}_Y'=0}^\infty q(\udl{v}_Y';\udl{0})\\
                                                  &\times
    \uppi^{-1}\sum_{l=0}^{n}\sum_{q=0}^{n-l}
    \sum_{\substack{m=0 \\ m+q\ \mathrm{even}}}^{2N_{lq}}
    \sum_{\substack{p=0 \\ p+q\ \mathrm{even}}}^{2N_{lq}-m}\\
                                                  &\times
      \frac{(-\tilde{W}_{kk}^+)^l(2\tilde{r}_k^+)^{m}
    (-\tilde{W}_{kk}^-)^{n-l-m/2-p/2-q}
      (2\tilde{r}_k^-)^{p}(-2\tilde{W}_{kk}^{\text{off}})^{q}}{l!m!(n-l-q-m/2-p/2)!p!q!}\\
                                                  &\times\Gamma(l+m/2+q/2+1/2)\Gamma(n-l-m/2-q/2+1/2),
  \end{split}
  \label{eq:sum-rule-2-si}
\end{equation}
This result is identical to Eq.~\eqref{eq:sum-rule-2} in the main text; it corresponds 
to Eq.~(66) of Ref.~\cite{jankowiak:2007} and reduces to it in the limit of real-valued 
$\bm{W}$ and $\udl{r}$.
\end{widetext}
\subsection{Integral Prescreening and intensity bounds}
\label{ssec:int-prescreening-bounds}
In this section, for completeness, we present the integral prescreening procedure
of Jankowiak \etal~\cite{jankowiak:2007} that employs the sum rules derived in Sec.~\ref{ssec:sum-rules}. 

The objective of the numerical prescreening procedure~\cite{jankowiak:2007}
is to define a truncated vibrational subspace, restricted by quantum excitation limits
as well as number of simultaneously excited modes, such that the 
accumulated sum of FCFs satisfies a specified convergence 
criterion prior to the explicit evaluation of individual integrals.

For a specified error tolerance $\epsilon_{\text{max}}^{\text{tol}}$ in the FC total
intensity (the integrated FC profile), the vibrational occupation number vector (ONV)
subspace $\mathcal{S}$ is systematically expanded until the calculated FC intensity
error $\epsilon_{\text{tot}}$ is smaller than the desired intensity error $\epsilon_{\text{max}}^{\text{tol}}$
such that 
\begin{equation}
  \epsilon_{\text{tot}} = 1-F_{\text{FC}}^{\text{tot}}\leq\epsilon_{\text{max}}^{\text{tol}},
\end{equation}
where the total accumulated intensity within the ONV subspace is given by
\begin{equation}
  F_{\text{FC}}^{\text{tot}}=\sum_{\udl{v},\udl{v}'\in\mathcal{S}}\abs{\braket{\udl{v}'}{\udl{v}}}^2.
\end{equation}
Rather than utilizing $\epsilon_{\text{max}}^{\text{tol}}$ directly, it is convenient
to employ auxiliary parameters~\cite{jankowiak:2007} $t_{\mathrm{m}}$ and $t_{\mathrm{c}}$ satisfying
\begin{equation}
  \epsilon_{\text{max}}^{\text{tol}}=1-\prod_{\gamma}^{N_{\text{irrep}}}[1-(t_{\mathrm{c}}+N_\gamma t_{\mathrm{m}})],\quad 0\leq t_{\mathrm{c}}+N_\gamma t_{\mathrm{m}}\leq1,
\end{equation}
where $t_{\mathrm{m}}$ and $t_{\mathrm{c}}$ are associated with the mode and coupling tolerances, respectively,
$N_\gamma$ denotes the number of vibrational modes in each irreducible representation
(irrep) where $N_{\text{irrep}}$ is the number of irreducible representations.
\subsubsection{Vibrational mode excitation error}
We define a mode contribution $F_{\text{FC};\mathrm{m}}^{(k)}(n)$ associated with a fixed vibrational
quantum number $v_k'=n\neq0$ for mode $k$, marginalized over all other modes:
\begin{equation}
  F_{\text{FC};\mathrm{m}}^{(k)}(n)=\sum_{\udl{v},\udl{v}'|v_k'=n}\abs{\braket{\udl{v}'}{\udl{v}}}^2.
\end{equation}
This contribution is evaluated via a finite series summation in Eq.~\eqref{eq:sum-rule-2-si}.
To establish the truncation limit for a specific mode, we identify the minimum
quantum number $v_k^{\prime\text{max}}$ that satisfies the condition
\begin{equation}
\epsilon_{\mathrm{m}}^{(k)}=\left(1-F_{\text{FC};v_k^{\prime\text{max}}}\right)<t_{\mathrm{m}},
\end{equation}
where $\epsilon_{\mathrm{m}}^{(k)}$ denotes the corresponding vibrational mode excitation error and
\begin{equation}
  F_{\text{FC};v_k^{\prime\text{max}}}=\sum_{n=0}^{v_k^{\prime\text{max}}}F_{\text{FC};\mathrm{m}}^{(k)}(n).
\end{equation}
\subsubsection{Vibrational mode coupling error}
By iteratively applying Eq.~\eqref{eq:sum-rule-1-si}, we determine the minimum 
simultaneous excitation level $M_{\text{max}}$ required to meet the coupling tolerance $t_{\mathrm{c}}$. 
Integrals involving more than $M_{\text{max}}$ simultaneously excited modes are discarded, 
yielding a substantial reduction in computational cost~\cite{jankowiak:2007}.
The maximum number of simultaneously excited modes is incremented from $M=0$ 
until
\begin{equation}
  \epsilon_{\mathrm{c}}=1-\sum_{m=0}^{M}F_{\text{FC};\mathrm{c}}^{(m)}=1-\tilde{F}_{\text{FC};\mathrm{c}}^{(M)}<t_{\mathrm{c}},
\end{equation}
where $F_{\text{FC};\mathrm{c}}^{(m)}$ represents the incremental contribution of $m$ simultaneously
excited modes to the total intensity, and $\tilde{F}_{\text{FC};\mathrm{c}}^{(M)}=\sum_{m=0}^{M}F_{\text{FC};\mathrm{c}}^{(m)}$
is the cumulative sum.
Each increment is determined from the intermediate quantity
\begin{equation}
  B_{\text{FC}}^{(M)}=\abs{\mel{\udl{0}',\text{in}}{\hat{U}_{\text{G}}}{\udl{0},\text{in}}}^2
\sum_{Y\in C_{M}^{N}}G_Y^{K}(\bm{I}_{YY}),
\end{equation}
where the summation runs over $C_{M}^N$, the set of all unique combinations of
$M$ modes chosen from the $N$-mode system.
To correct for overcounting, we apply the recursive relation
\begin{equation}
  F_{\text{FC};\mathrm{c}}^{(M)}=B_{\text{FC}}^{(M)}-\sum_{i=1}^M\binom{N-M+i}{i}F_{\text{FC};\mathrm{c}}^{(M-i)},
\end{equation}
with the initial condition $F_{\text{FC};\mathrm{c}}^{(0)}=\abs{\braket{\udl{0}'}{\udl{0}}}^2$.
\subsubsection{Error bound condition}
Jankowiak \etal derived conservative lower and upper bounds
for the total truncation error introduced by the Fock-space restriction:
Adopting this error analysis framework we define the quantities of
$\epsilon_{\mathrm{m}}$ and $\epsilon_{\mathrm{c}}$
as $\epsilon_{\mathrm{m}}^{(k,\gamma)}$ and $\epsilon_{\mathrm{c}}^{(\gamma)}$
respectively, corresponding to the irreducible representations.
From these quantities an upper bound for the error made by a single mode is
\begin{equation}
  \epsilon_{\mathrm{m}}^{(\gamma)} = \max_k\epsilon_{\mathrm{m}}^{(k,\gamma)},
\end{equation}
within a given irreducible representation. A bound on the minimum error made for a
given irreducible representation is
\begin{equation}
  \epsilon_{\text{min}}^{(\gamma)}=\max(\epsilon_{\mathrm{m}}^{(\gamma)},\epsilon_{\mathrm{c}}^{(\gamma)}),
\end{equation}
whereas the upper bound error is
\begin{equation}
  \epsilon_{\text{max}}^{(\gamma)}=\epsilon_{\mathrm{c}}^{(\gamma)} + \sum_{k\in\gamma}\epsilon_{\mathrm{m}}^{(k,\gamma)}
\end{equation}
In the event that $\epsilon_{\text{max}}^{(\gamma)}\geq 1$, we set $\epsilon_{\text{max}}^{(\gamma)}=1$.
The total error bounds, aggregated over all representations, are subsequently
\begin{align}
  \epsilon_{\text{min}} &= 1-\prod_{\gamma}(1-\epsilon_{\text{min}}^{(\gamma)}),\\
  \epsilon_{\text{max}} &= 1-\prod_{\gamma}(1-\epsilon_{\text{max}}^{(\gamma)}),
\end{align}
These bounds have an error range of
\begin{equation}
  \delta\epsilon=\epsilon_{\text{max}}-\epsilon_{\text{min}},
\end{equation}
and the total error is strictly bounded by
\begin{equation}
\epsilon_{\text{min}}\leq\epsilon_{\text{tot}}\leq\epsilon_{\text{max}}\leq\epsilon_{\text{max}}^{\text{tol}}.
\end{equation}
A rigorous proof of the validity of these bounds is provided in the Appendix of Ref.~\cite{jankowiak:2007}.
\section{Explicit expressions for the linear canonical transformation of the auxiliary function method}
\label{sec:auxfunc-bogoliubov-explicit}
To evaluate the Franck--Condon integrals $\mel{\udl{v}',\text{in}}{\hat{U}_{\text{aux}}(\kappa)}{\udl{v},\text{in}}$ 
associated with the auxiliary function in Eq.~\eqref{eq:full-auxfunc}, we recall the operator sequence
from the auxiliary function method in the original parametrization~\cite{jnane:2021}
\begin{equation}
  \begin{split}
    \hat{U}_{\text{aux}}(\kappa) =& \hat{R}(\bm{U}_2)\hat{S}(\ln(\bm{L}))\hat{R}(\bm{U}_1)
    \hat{D}((\bm{U}_2\bm{L}\bm{U}_1)^{-1}\udl{\gamma})\\
                                  &\times\hat{R}(\bm{U})\hat{S}(\bm{\Xi}(\kappa))
    \hat{D}(\udl{\alpha}(\kappa)),
  \end{split}
  \label{eq:op-kappa}
\end{equation}
We apply the standard relations from Eqs.~\eqref{eq:ma-d-op-annihilation}--\eqref{eq:ma-s-op-creation}, 
which simplify considerably when the arguments of the Gaussian operators are real. 
In the present formalism, the parameters $\udl{\gamma}$, $\bm{U}_2$, $\ln(\bm{L})$, $\bm{U}_1$, 
and $\bm{U}$ are strictly real-valued, whereas $\bm{\Xi}(\kappa)$ and $\udl{\alpha}(\kappa)$ are complex-valued. 
Because $\bm{\Xi}(\kappa)$ is defined as a diagonal matrix, its constituent magnitude and phase 
components, $\bm{r}(\kappa)$ and $\bm{\theta}(\kappa)$, are likewise diagonal.
The resulting transformation matrices and displacement vector are given by:
\begin{align}
  \bm{E}^*(\kappa) &= \bm{U}_2\cosh(\ln(\bm{L}))\bm{U}_1\bm{U}\cosh(\bm{r}(\kappa))\nonumber\\
                   &+ \bm{U}_2\sinh(\ln(\bm{L}))\bm{U}_1\bm{U}\sinh(\bm{r}(\kappa))\exp(-\im\bm{\theta}(\kappa))
  \label{eq:ykappa},\\
  \bm{F}^*(\kappa) &= \bm{U}_2\sinh(\ln(\bm{L}))\bm{U}_1\bm{U}\cosh(\bm{r}(\kappa))\nonumber\\
                   &+ \bm{U}_2\cosh(\ln(\bm{L}))\bm{U}_1\bm{U}\sinh(\bm{r}(\kappa))\exp(\im\bm{\theta}(\kappa))
  \label{eq:xkappa},\\
  \udl{g}^*(\kappa) &= \bm{F}^*(\kappa)\udl{\alpha}(\kappa) + \bm{E}^*(\kappa)\udl{\alpha}^*(\kappa)
  + \udl{\gamma}\label{eq:zkappa}.
\end{align}
We explicitly present the complex conjugates of $\bm{E}$, $\bm{F}$, and $\udl{g}$ to 
facilitate direct comparison with the results in the Supporting Information of Jnane~\etal~\cite{jnane:2021}. 
It is important to note that the expressions provided in Ref.~\cite{jnane:2021} are valid 
only for $\kappa \in \mathbb{R}$; their expressions for imaginary $\kappa$ 
violate the canonical commutation relations established in Eq.~\eqref{eq:canonical-commutation-relations}.
\section{Extension to finite temperatures}
\label{ssec:finite-temp-extension}
Jnane~\etal~\cite{jnane:2021} suggested the possibility of incorporating 
finite-temperature effects within the auxiliary function framework. Here, we provide 
the explicit formalism for this extension. The finite temperature transition profile of Eq.~\eqref{eq:transition-profile} can be described as
\begin{widetext}
\begin{equation}
  \begin{split}
    P(\omega) =& \frac{1}{\mathcal{N}(\udl{T})}\sum_{\substack{\udl{v}'=0\\r=x,y,z}}^\infty
  \abs{\mel{\udl{v}',\text{in}}
  {\hat{U}_{\mathrm{Dok}}\hat{\mu}_r(\hat{\udl{Q}})}{\udl{v},\text{in}}}^2P_{\text{in}}(\udl{v})\delta(\omega-\udl{v}^{\prime\intercal}\udl{\omega}'+\udl{v}^\intercal\udl{\omega}),
  \end{split}
  \label{eq:transition-profile-temp-naive}
\end{equation}
\end{widetext}
where $P_{\text{in}}$ is the initial distribution of the modes based on the thermal
state $\hat{\rho}_{\text{th}}=\exp(-\hat{H})/\text{Tr}\left(\exp(-\hat{H})\right)$ with
$\hat{H}=\sum_{k=1}^N\beta_k\hbar\omega_k\hat{a}_k^\dagger\hat{a}_k$ and $\beta_k=1/(k_{\text{B}}T_k)$
with the Boltzmann constant $k_{\text{B}}$ and mode-specific temperature $T_k$.
Moreover, the total summed intensity of the transition profile, expressed as
the normalization constant $\mathcal{N}(\udl{T})$, is temperature-dependent.

Huh and Yung~\cite{huh:2017a} have demonstrated that one can purify every Boltzmann distributed thermal state
with ancillary modes, \ie $\ket{\udl{0}(\udl{\beta})}=\sum_{\udl{v}=\udl{0}}^{\udl{\infty}}
\sqrt{\mel{\udl{v}}{\hat{\rho}_{\text{th}}}{\udl{v}}}\ket{\udl{v}}\otimes\ket{\udl{v}}_\mathrm{C}$, where
the ancillary Hilbert space 'C' has been introduced. Further, 
$\ket{\udl{0}(\udl{\beta})}=\hat{T}(\udl{\beta})\ket{\udl{0}}\otimes\ket{\udl{0}}_\mathrm{C}$,
where Huh and Yung introduced a product of two-mode squeezing operators,
whose squeezing parameters $t_k$ are 
temperature-dependent and given by~\cite{huh:2017a}
\begin{equation}
  t_k = \arctanh(\exp(-1/2\hbar\beta_k\omega_k)).
  \label{eq:auxfunc-twomode-squeezing}
\end{equation}
The transformation of the ladder operators by a
two-mode squeezing operator $\hat{T}(t)$ is expressed as:
\begin{equation}
  \begin{split}
    \hat{T}(t)^\dagger\hat{a}_1\hat{T}(t)&=\cosh(r)\hat{a}_1+\sinh(r)\exp(\im\theta)\hat{a}_2^\dagger,\\
    \hat{T}(t)^\dagger\hat{a}_2\hat{T}(t)&=\cosh(r)\hat{a}_2+\sinh(r)\exp(\im\theta)\hat{a}_1^\dagger,\\
    \hat{T}(t)^\dagger\hat{a}_1^\dagger\hat{T}(t)&=
    \sinh(r)\exp(-\im\theta)\hat{a}_2+\cosh(r)\hat{a}_1^\dagger,\\
    \hat{T}(t)^\dagger\hat{a}_2^\dagger\hat{T}(t)&=
    \sinh(r)\exp(-\im\theta)\hat{a}_1+\cosh(r)\hat{a}_2^\dagger,
\label{eq:ma-two-mode-s-op-creation}
  \end{split}
\end{equation}
with the polar decomposition $t=r\exp(\im\theta)$.

Because the auxiliary function method assumes a pure vacuum state as an initial state, 
we map the problem onto an effective zero-temperature system, whose transition
profile is 
\begin{widetext}
\begin{equation}
  P(\omega) =\frac{1}{\mathcal{N}(\udl{T})}\sum_{\substack{\udl{v}',\udl{v}_{\mathrm{C}}=\udl{0}\\r=x,y,z}}^\infty
    \abs{\mel{\udl{v}',\udl{v}_{\mathrm{C}},\text{in}}
    {\hat{\mdc{U}}_{\mathrm{Dok}}\hat{\mu}_{\text{th},r}
(\hat{\bm{\mdc{Q}}})\mdc{\hat{T}}(\udl{T})}{\udl{0},\udl{0}_{\mathrm{C}},\text{in}}}^2\\
\delta(\omega-\udl{v}^{\prime\intercal}\udl{\omega}'+\udl{v}_{\mathrm{C}}^\intercal\udl{\omega}),
  \label{eq:transition-profile-temp-twomode-squeeze}
\end{equation}
\end{widetext}
where we augmented the Doktorov operator $\hat{U}_{\mathrm{Dok}}$ acting on 
the physical modes to also act on the ancillary modes via the identity operation,
denoted $\hat{\mdc{U}}_{\mathrm{Dok}}=\hat{U}_{\mathrm{Dok}}\otimes\hat{I}_{N}$.
Although the extended collective normal coordinates $\bm{\mdc{Q}}$ span both the physical
and ancillary domains, the electric transition dipole moment operator remains functionally dependent
only on the physical modes.
Moreover, $\mdc{\hat{T}}(\udl{T}) = \prod_{k=1}^N\hat{T}(t_k) = \hat{T}(t_1)\cdots\hat{T}(t_N)$
denotes a composite two-mode squeezing operator that couples 
each physical mode $k$ to a fictitious ancillary mode $k+N$ within the Hilbert space 
'C' via the thermal squeezing parameter $t_k$. The additional modes are denoted
with the subscript C. Note that the initial states $\udl{v}$ have been mapped 
to the larger final state in the integral. 

Within the extended $2N$-dimensional mode space utilized for finite-temperature calculation,
we must distinguish between the $N$ physical modes and the $N$ ancillary modes.
The purification of the thermal state introduces these ancillary degrees of freedom 
purely as a mathematical construct to trace out the correct Boltzmann statistics.
Because the external driving field couples exclusively to physical charge
distributions---specified by the spatial coordinates of the actual electrons and nuclei---the
electric transition dipole moment operator, $\hat{\mu}$, possesses no physical interaction with,
nor functional dependence on, the ancillary bath.

Consequently, the total dipole operator acts strictly as the identity on the ancillary subspace,
such that $\hat{\udl{\mu}}_{\text{th}} = \hat{\udl{\mu}} \otimes \udl{\hat{I}}_{N}$.
It follows directly that in any non-Condon treatment, such as a HT Taylor expansion
of $\hat{\udl{\mu}}_{\text{th}}$ with respect to the extended collective normal coordinates
$\udl{\mdc{Q}}$, all partial derivatives evaluated with respect to the
ancillary coordinates are identically zero.
The spatial dependence of the electric transition dipole moment is therefore exclusively
confined to the physical molecular subspace.

In order to apply the auxiliary function ansatz to finite temperatures, we transform the 
electric transition dipole operator $\hat{\udl{\mu}}_{\text{th}}(\hat{\bm{\mdc{Q}}})$, truncated to second order, 
by inserting the identity operator $\hat{1}=\mdc{\hat{T}}(\udl{T})\mdc{\hat{T}}^\dagger(\udl{T})$
into the transition profile:
\begin{widetext}
\begin{align}
  P(\omega) &= \frac{1}{\mathcal{N}(\udl{T})}\sum_{\substack{\udl{v}',\udl{v}_{\mathrm{C}}=\udl{0}\\r=x,y,z}}^\infty\abs{\mel{\udl{v}',\udl{v}_{\mathrm{C}},\text{in}}
  {\hat{\mdc{U}}_{\mathrm{Dok}}\mdc{\hat{T}}(\udl{T})\mdc{\hat{T}}^{\dagger}(\udl{T})\hat{\mu}_{r,\text{th},\text{quad}}(\hat{\udl{\mdc{Q}}})\mdc{\hat{T}}(\udl{T})}{\udl{0},\udl{0}_{\mathrm{C}},\text{in}}}^2\delta(\omega-\udl{v}^{\prime\intercal}
  \udl{\omega}'+\udl{v}_{\mathrm{C}}^\intercal\udl{\omega}),\\
            &= \frac{1}{\mathcal{N}(\udl{T})}\sum_{\substack{\udl{v}',\udl{v}_{\mathrm{C}}=\udl{0}\\r=x,y,z}}^\infty\abs{\mel{\udl{v}',\udl{v}_{\mathrm{C}},\text{in}}
            {\hat{\mdc{U}}_{\mathrm{Dok}}\mdc{\hat{T}}(\udl{T})\hat{\mu}_{r,\text{th},\text{quad}}'(\hat{\udl{\mdc{Q}}})}{\udl{0},\udl{0}_{\mathrm{C}},\text{in}}}^2\delta(\omega-\udl{v}^{\prime\intercal}\udl{\omega}'+\udl{v}_{\mathrm{C}}^\intercal\udl{\omega}),
  \label{eq:transition-profile-temp-trans}
\end{align}
where $\hat{\mu}_{r,\text{th},\text{quad}}'(\hat{\udl{\mdc{Q}}})$ contains the transformed parameters
truncated to quadratic terms which now operate in the expanded $2N$-dimensional space
\begin{equation}
  \hat{\mu}_{r,\text{th},\text{quad}}(\hat{\udl{\mdc{Q}}}) = \mu^{(0)}+\udl{\mdc{\chi}}_{\text{th}}^{\intercal}\frac{(\hat{\udl{\mdc{c}}}+\hat{\udl{\mdc{c}}}^\dagger)}{\sqrt{2}}+\left(\frac{(\hat{\udl{\mdc{c}}}+\hat{\udl{\mdc{c}}}^\dagger)}{\sqrt{2}}\right)^\intercal \bm{\mdc{X}}_{\text{th}}\left(\frac{(\hat{\udl{\mdc{c}}}+\hat{\udl{\mdc{c}}}^\dagger)}{\sqrt{2}}\right),
\label{eq:tdm-purified}
\end{equation}
where we defined the collective $2N$-dimensional ladder operator vectors
$\udl{\hat{c}} = \udl{\hat{a}}\oplus\udl{\hat{a}}_{\mathrm{C}}\equiv(\udl{\hat{a}}^\intercal,\udl{\hat{a}}_{\mathrm{C}}^\intercal)^\intercal$
and 
$\udl{\hat{c}}^\dagger = \udl{\hat{a}}^\dagger\oplus\udl{\hat{a}}_{\mathrm{C}}^\dagger\equiv
((\udl{\hat{a}}^\dagger)^\intercal,(\udl{\hat{a}}_{\mathrm{C}}^\dagger)^\intercal)^\intercal$,
which comprise both physical and auxiliary modes.
Because the external field strictly couples to the physical subspace,
the $2N$-dimensional parameter vector $\udl{\mdc{\chi}}_{\text{th}}$
and matrix $\bm{\mdc{X}}_{\text{th}}$ for the untransformed electric transition dipole
operator are zero-padded in the ancillary dimensions; therefore,
$\udl{\mdc{\chi}}_{\text{th}} = \udl{\chi}\oplus\udl{0}\equiv (\chi_1,\cdots,\chi_N,0_1,\cdots,0_N)^\intercal$
and $\bm{\mdc{X}}_{\text{th}} = \bm{X}\oplus\bm{0}\equiv\text{bldiag}(\bm{X},\bm{0})$.

The transformation is performed by evaluating the action of the
thermal squeezing operator $\hat{\mdc{T}}(\udl{T})$ on the ladder operators via the linear canonical relation
\begin{equation}
  \mdc{\hat{T}}^\dagger(\udl{T})(\hat{\udl{\mdc{c}}}+\hat{\udl{\mdc{c}}}^\dagger)\mdc{\hat{T}}(\udl{T}) = (\bm{\mdc{E}}(\udl{T})+\bm{\mdc{F}}(\udl{T}))(\hat{\udl{\mdc{c}}}+\hat{\udl{\mdc{c}}}^\dagger),
\end{equation}
utilizing Eq.~\eqref{eq:ma-two-mode-s-op-creation}.
Substituting this relation into the dipole expansion yields the transformed parameters:
\begin{align}
  \udl{\mdc{\chi}}_{\text{th}}^{\prime\intercal}(\hat{\udl{\mdc{c}}}+\hat{\udl{\mdc{c}}}^\dagger)&=\udl{\mdc{\chi}}_{\text{th}}^\intercal
  (\bm{\mdc{E}}(\udl{T})+\bm{\mdc{F}}(\udl{T}))(\hat{\udl{\mdc{c}}}+\hat{\udl{\mdc{c}}}^\dagger),\\
  (\hat{\udl{\mdc{c}}}+\hat{\udl{\mdc{c}}}^\dagger)^\intercal\bm{\mdc{X}}_{\text{th}}'(\hat{\udl{\mdc{c}}}+\hat{\udl{\mdc{c}}}^\dagger)
                                                                    &=
((\hat{\udl{\mdc{c}}}+\hat{\udl{\mdc{c}}}^\dagger))^\intercal((\bm{\mdc{E}}(\udl{T})+
\bm{\mdc{F}}(\udl{T})))^\intercal\bm{\mdc{X}}_{\text{th}}(\bm{\mdc{E}}(\udl{T})+\bm{\mdc{F}}(\udl{T}))(\hat{\udl{\mdc{c}}}+\hat{\udl{\mdc{c}}}^\dagger).
\end{align}

Consequently, the transformed quadratic dipole operator acting on the input vacuum state in Eq.~\eqref{eq:transition-profile-temp-trans} is given by:
\begin{equation}
\hat{\mu}_{r,\text{th},\text{quad}}'(\hat{\udl{\mdc{Q}}}) = \mu^{(0)}+\udl{\mdc{\chi}}_{\text{th}}^{\prime\intercal}\frac{(\hat{\udl{\mdc{c}}}+\hat{\udl{\mdc{c}}}^\dagger)}{\sqrt{2}}+\left(\frac{(\hat{\udl{\mdc{c}}}+\hat{\udl{\mdc{c}}}^\dagger)}{\sqrt{2}}\right)^\intercal \bm{\mdc{X}}_{\text{th}}^\prime \left(\frac{(\hat{\udl{\mdc{c}}}+\hat{\udl{\mdc{c}}}^\dagger)}{\sqrt{2}}\right).
\label{eq:tdm-transform}
\end{equation}
By mapping the thermal problem onto an effective 
zero-temperature profile, the auxiliary function expressions derived by Jnane \etal 
become directly applicable. The modified auxiliary function $f_{\udl{v}'}(\kappa)$ is expressed as
\begin{equation}
  f_{\udl{v}'}(\kappa) = \abs{\mathcal{C}(\kappa)}^2
  \left|\mel{\udl{v}',\udl{v}_{\mathrm{C}},\text{in}}{\hat{D}(\udl{\mdc{\gamma}}')\hat{R}(\bm{\mdc{U}}_2)\hat{S}(\ln(\bm{\mdc{L}}))\hat{R}(\bm{\mdc{U}}_1)
  \mdc{\hat{T}}(\udl{T})\hat{R}(\bm{\mdc{U}'})\hat{S}(\bm{\Xi}'(\kappa))
\hat{D}(\udl{\mdc{\alpha}}'(\kappa))}{\udl{0},\udl{0}_{\mathrm{C}},\text{in}}\right|^2,
\label{eq:full-auxfunc-temp}
\end{equation}
\end{widetext}
This Gaussian unitary operator product is subsequently resolved into the transformation matrices 
$\bm{\mdc{E}}, \bm{\mdc{F}} \in \mathbb{C}^{2N\times 2N}$ and vector $\udl{\mdc{g}} \in \mathbb{C}^{2N}$ 
by applying the relations in Eqs.~\eqref{eq:ma-d-op-annihilation}--\eqref{eq:ma-s-op-creation}
and~\eqref{eq:ma-two-mode-s-op-creation}. 
The quantities associated with the Doktorov operator ($\udl{\mdc{\gamma}}'$, $\bm{\mdc{U}}_2$, 
$\bm{\mdc{L}}$, $\bm{\mdc{U}}_1$) act trivially on the auxiliary subspace. Conversely, 
the parameters arising from the approximate decomposition of $\udl{\mdc{\chi}}_{\text{th}}'$ and $\bm{\mdc{X}}_{\text{th}}'$ 
($\bm{\mdc{U}}'$, $\bm{\Xi}'$, $\udl{\mdc{\alpha}}'$) explicitly couple the physical and auxiliary modes.
The evaluation of the normalization constant $\mathcal{N}(\udl{T})$ proceeds analogously 
to the zero-temperature derivation detailed in the Supporting Information of Ref.~\cite{jnane:2021},
and which is given in the main text as Eq.~\eqref{eq:norm-constant},
utilizing the transformed parameters ($\udl{\chi}_{\text{th}}'$, $\bm{X}_{\text{th}}'$) over the expanded $2N$-dimensional space.

After the transformation outlined in this section, the sum rules and TD method
in the main text can be employed for the finite-temperature case.
\section{Minimizing resolution errors}
\label{sec:opt-tau}
Although the approximation error formally decays quadratically as $\mathcal{O}(\tau^2)$, 
previous error analysis~\cite{jnane:2021} demonstrated a deviation from this expected 
trend for a three-mode benzene model. Here, we investigate this deviation to mitigate 
the errors inherent to implementing the auxiliary function method in finite-precision 
arithmetic.

Because the Herzberg--Teller (HT) effect is approximated by a finite-difference 
scheme, its computational implementation is sensitive to floating-point round-off errors. 
We note that this issue persists even when employing analog Gaussian Boson 
Samplers~\cite{madsen:2022,yu:2023}, as their outputs must ultimately be linearly 
combined using classical hardware. If the resolution parameter $\tau$ is chosen too small, 
these numerical inaccuracies accumulate rapidly, resulting in catastrophic cancellation. 
Furthermore, the magnitudes of the linear and quadratic HT parameters directly influence 
the auxiliary function approximation error; smaller magnitudes yield reduced errors.
This behavior is evident from the Taylor series expansion of the auxiliary function
up to fourth order [Eq.~\eqref{eq:auxfunc-expansion}]~\footnote{Eq.~\eqref{eq:auxfunc-expansion} corrects a typographical error (factor of 2) in the $\real(\kappa^2\kappa^*)$ term of Eq.~(12) in Ref.~\cite{jnane:2021}.}:
\begin{widetext}
\begin{equation}
  \begin{split}
    f_{\udl{v}'}(\kappa) =&
                         \abs{\mel{\udl{v}',\text{in}}{\hat{U}_{\mathrm{Dok}}}{\udl{0},\text{in}}}^2
                         +2\real(\kappa)\mel{\udl{v}',\text{in}}{\hat{U}_{\mathrm{Dok}}\hat{\mu}}{\udl{0},\text{in}}
                         \mel{\udl{v}',\text{in}}{\hat{U}_{\mathrm{Dok}}}{\udl{0},\text{in}}^*\\
                          &+\real(\kappa^2)\mel{\udl{v}',\text{in}}{\hat{U}_{\mathrm{Dok}}\hat{\mu}^2}{\udl{0},\text{in}}
    \mel{\udl{v}',\text{in}}{\hat{U}_{\mathrm{Dok}}}{\udl{0},\text{in}}^*+
                         \abs{\kappa}^2\mel{\udl{v}',\text{in}}{\hat{U}_{\mathrm{Dok}}\hat{\mu}}{\udl{0},\text{in}}
                         \mel{\udl{v}',\text{in}}{\hat{U}_{\mathrm{Dok}}\hat{\mu}^\dagger}{\udl{0},\text{in}}^*\\
                          &+\frac{1}{3}\real(\kappa^3)\mel{\udl{v}',\text{in}}{\hat{U}_{\mathrm{Dok}}\hat{\mu}^3}{\udl{0},\text{in}}
                         \mel{\udl{v}',\text{in}}{\hat{U}_{\mathrm{Dok}}}{\udl{0},\text{in}}^*
    +\real(\kappa^2\kappa^*)\mel{\udl{v}',\text{in}}{\hat{U}_{\mathrm{Dok}}\hat{\mu}^2}{\udl{0},\text{in}}
    \mel{\udl{v}',\text{in}}{\hat{U}_{\mathrm{Dok}}\hat{\mu}^\dagger}{\udl{0},\text{in}}^*\\
                          &+ \frac{1}{12}
    \real(\kappa^4)\mel{\udl{v}',\text{in}}{\hat{U}_{\mathrm{Dok}}\hat{\mu}^4}{\udl{0},\text{in}}
    \mel{\udl{v}',\text{in}}{\hat{U}_{\mathrm{Dok}}}{\udl{0},\text{in}}^*
    +\frac{1}{3}\real(\kappa^3\kappa^*)\mel{\udl{v}',\text{in}}{\hat{U}_{\mathrm{Dok}}\hat{\mu}^3}{\udl{0},\text{in}}
    \mel{\udl{v}',\text{in}}{\hat{U}_{\mathrm{Dok}}\hat{\mu}^\dagger}{\udl{0},\text{in}}^*\\
                          &+ \frac{1}{4}\abs{\kappa}^4\mel{\udl{v}',\text{in}}{\hat{U}_{\mathrm{Dok}}\hat{\mu}^2}{\udl{0},\text{in}}
    \mel{\udl{v}',\text{in}}{\hat{U}_{\mathrm{Dok}}\left(\hat{\mu}^\dagger\right)^2}{\udl{0},\text{in}}^* + \mathcal{O}(\kappa^5).
  \end{split}
  \label{eq:auxfunc-expansion}
\end{equation}
Upon applying the linear combination of auxiliary functions for real $\tau$, the odd terms vanish,
leaving only the $\abs{\kappa}^2$, $\abs{\kappa}^4$, and $\real(\kappa^4)$ terms.
\begin{equation}
  \begin{split}
  \abs{\mel{
  \udl{v}',\text{in}}{\hat{U}_{\mathrm{Dok}}\hat{\mu}}{\udl{0},\text{in}}}^2 =&
  \frac{1}{2\tau^2}\left[f_{\udl{v}'}(\im\tau) + \frac{1}{2}f_{\udl{v}'}(\tau)
  + \frac{1}{2}f_{\udl{v}'}(-\tau)-2f_{\udl{v}'}(0)\right]\\
                                                                              &- \frac{1}{4}\tau^2\mel{\udl{v}',\text{in}}{\hat{U}_{\mathrm{Dok}}\hat{\mu}^2}{\udl{0},\text{in}}\mel{\udl{v}',\text{in}}{\hat{U}_{\mathrm{Dok}}\left(\hat{\mu}^\dagger\right)^2}{\udl{0},\text{in}}^*\\
                                                                              &- \frac{1}{12}\tau^2\mel{\udl{v}',\text{in}}{\hat{U}_{\mathrm{Dok}}\hat{\mu}^4}{\udl{0},\text{in}}\mel{\udl{v}',\text{in}}{\hat{U}_{\mathrm{Dok}}}{\udl{0},\text{in}}^* + \mathcal{O}(\tau^4).
  \end{split}
\label{eq:tdm-int-approx2}
\end{equation}
\end{widetext}
Summing Eq.~\eqref{eq:tdm-int-approx2} over all final states $\udl{v}'$ and using
completeness together with the unitarity of $\hat{U}_{\mathrm{Dok}}$ and Hermiticity
of $\hat{\mu}$, the two $\mathcal{O}(\tau^2)$ error terms combine into
$\frac{1}{3}\tau^2\mel{\udl{0},\text{in}}{\hat{\mu}^4}{\udl{0},\text{in}} \geq 0$.
Consequently, the auxiliary-function approximation overestimates the total
normalization constant $\mathcal{N}$ to leading order in $\tau^2$.
To quantify this total error, 
we leverage the analytic expression for the normalization constant $\mathcal{N}$, valid 
for HT effects up to second order (see Appendix A of Ref.~\cite{jnane:2021}) and 
extended to finite temperatures using the formalism in Appendix~\ref{ssec:finite-temp-extension}.

\begin{figure}
  \includegraphics[width=0.8\linewidth]{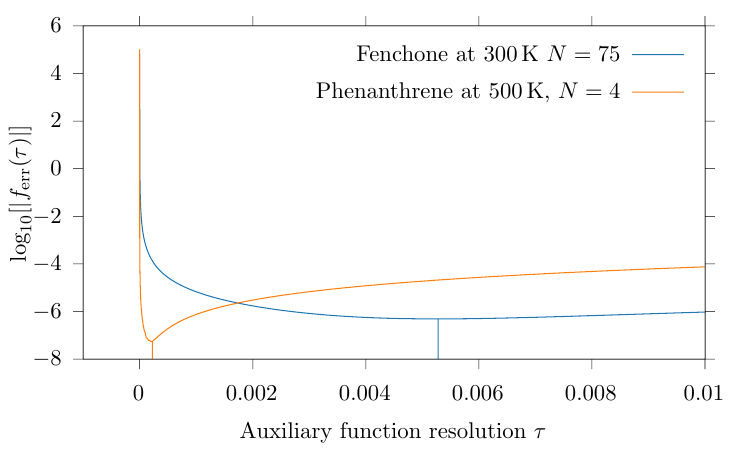}
  \caption{Auxiliary function error function $f_{\text{err}}(\tau)$ for the
    $(\mathrm{S}_1\leftarrow \mathrm{S}_0)$ transition of (1R,4S)-(-)-fenchone
at \SI{300}{\kelvin} from Ref.~\cite{kerber:2025}
and for the ($\mathrm{S}_1\leftarrow \mathrm{S}_0$) transition 2-band (a,b) phenanthrene system from
Ref.~\cite{small:1971} at $T=\SI{500}{\kelvin}$. The optimal auxiliary function resolutions $\tau_{\text{opt}}^{\text{Fenchone}}=\num{5.3E-03}$
and $\tau_{\text{opt}}^{\text{Phenanthrene}}=\num{2.2E-04}$
were determined by minimizing $f_{\text{err}}(\tau)$ via Brent's method and are indicated by vertical lines.}
  \label{fig:auxfuncerr}
\end{figure}

For a given $\tau$, we evaluate the approximate normalization constant and compute
the relative error to the exact analytical result. This comparison requires an estimate of the numerical 
precision limit of the underlying linear algebra subroutines, denoted as $\epsilon_{\text{eff}}$. 
A reasonable estimate for $\epsilon_{\text{eff}}$ can be obtained by evaluating Eq.~\eqref{eq:gyiyy} 
with all final-state modes assigned to the $Y$ subspace; in exact arithmetic, this evaluates to unity. 
The deviation from unity reflects the truncation error of the inversion routine, 
which may exceed round-off error due to internal tolerance settings. 
Incorporating a suitable $\epsilon_{\text{eff}}$, we obtain the expression
\begin{equation}
  \begin{split}
    \tilde{\mathcal{N}}(\tau) =& \frac{1}{2\tau^2}\Bigl[\abs{\mathcal{C}(\im\tau)}^2(1+\epsilon_{\text{eff}})
                               + \frac{1}{2}\Bigl(\abs{\mathcal{C}(\tau)}^2(1+\epsilon_{\text{eff}})\\
                               &+\abs{\mathcal{C}(-\tau)}^2(1+\epsilon_{\text{eff}})\Bigr)
                               -2(1-\epsilon_{\text{eff}})\Bigr],
  \end{split}
  \label{eq:approx-normalization}
\end{equation}
which allows us to formulate the relative error function:
\begin{equation}
f_{\text{err}}(\tau) = \tilde{\mathcal{N}}(\tau)/\mathcal{N}-1.
\label{eq:auxfunc-error-function}
\end{equation}
Figure~\ref{fig:auxfuncerr} displays the base-10 logarithm of the error function
$\log_{10}f_{\text{err}}(\tau)$, for two distinct systems: 
the $(\mathrm{S}_1\leftarrow \mathrm{S}_0)$ transition of (1R,4S)-(-)-fenchone at \SI{300}{\kelvin}~\cite{kerber:2025}, 
and the ($^1\text{A}-^1\text{A}$) transition of the 2-mode (a,b) phenanthrene model 
at $T=\SI{500}{\kelvin}$~\cite{small:1971}.
Plotted on a semi-log scale ($\text{log}_{10}$ vertical axis), these results are
consistent with the expected competition between analytical
truncation and numerical round-off errors. For sufficiently large $\tau$, the
analytical truncation error dominates. For small $\tau$, the error increases steeply due to finite-precision
arithmetic.
Notably, the precision limit of the inversion routine scales with increasing system size.
For the 75-mode fenchone system, 
$\epsilon_{\text{eff}}^{\text{Fenchone}} \approx \num{8.5e-15}$, whereas for the 2-mode
phenanthrene model (4 modes at finite temperatures), it is only $\epsilon_{\text{eff}}^{\text{Phenanthrene}} \approx \num{6.7e-16}$.

Assuming the error function is smooth, determining the optimal auxiliary 
function resolution $\tau_{\text{opt}}$ reduces to finding the global minimum of
Eq.~\eqref{eq:auxfunc-error-function}.  This univariate, derivative-free minimization
is carried out efficiently using Brent's method~\cite{brent:1971}.
We have implemented this automated  $\tau$-optimization routine in the \texttt{hotFCHT} software package.
For numerical reasons we confine the optimization to the interval $[\num{E-8},\min(0.01,\tau_{\text{max}})]$. 
The upper bound $\tau_{\text{max}}$ is dictated by the domain of the $\operatorname{arctanh}$ 
function utilized to evaluate $r_j(\kappa)$ for $\Xi_j(\kappa)$ in
Eq.~\eqref{eq:auxfunc-expansion-coefficients}. 
Specifically, the argument $t_j(\kappa) = \kappa d_j/(1-\kappa d_j)$ must satisfy 
$-1 < t_j(\kappa) < 1$. Solving this inequality for the maximum permissible $\kappa$ yields 
$\kappa_{\text{max}} = 1/(2\abs{d_{j}}_{\text{max}})$, where $\abs{d_{j}}_{\text{max}}$ is 
the spectral radius (maximum absolute eigenvalue) of the matrix $\bm{D}$. This directly defines 
$\tau_{\text{max}}$, thereby establishing an upper limit for the minimization interval,
which avoids the poles at $\kappa d_k=1$ of the mode-specific function $t_k(\kappa)$ in Eq.~\eqref{eq:auxfunc-t-k} as well as the scalar prefactors given by Eq.~\eqref{eq:auxfunc-expansion-coefficients}.
\section{Determining a suitable resolution for the auxiliary function prescreening procedure}
\label{sec:auxfunc-prescreening-resolution}
The prescreening procedure consists of two independent steps: first, determining 
the maximum excitation required for each mode to achieve a specific mode tolerance; 
second, evaluating the cumulative importance of simultaneous excitations across 
the entire transition profile. Crucially, these two steps possess distinct optimal 
resolutions $\tau$.

In the first step, a low truncation error of the auxiliary function is desired. We 
accept a higher sensitivity to the effective precision error $\epsilon_{\text{eff},1}$, 
as these errors remain on the order of machine precision. 
Conversely, in the second step, the effective precision error $\epsilon_{\text{eff},2}$ 
can be many orders of magnitude larger for extended systems. This catastrophic 
amplification arises because each step involves a summation over $\binom{N}{M}$ terms.
Although exact summation algorithms~\cite{neal:2015} 
can completely eliminate the floating-point round-off error introduced during the 
addition process itself, the individual summands possess intrinsic numerical errors 
dictated by the preceding inversion routine. Consequently, even an exact sum over
$\tilde{F}_{\mathrm{c}}^{(M)}$ over all $0\leq M\leq N$ terms will not evaluate exactly
to unity.

Thus, for the second step, a coarser resolution (\ie a relatively larger $\tau$) 
is preferable, provided its associated auxiliary function approximation error $f_{\text{err}}(\tau)$ 
remains strictly smaller than the coupling tolerance $t_{\mathrm{c}}$. Because $\epsilon_{\text{eff}}$ 
for the second step cannot be computed \apriori, we introduce 
a safety margin, constraining $f_{\text{err},2}(\tau)$ to be two orders of magnitude 
smaller than the target $t_{\mathrm{c}}$. Consequently, we solve for $\tau_{\text{opt},2}$
via Brent's method. This heuristic is reasonable for large systems: 
the combinatorial scaling of evaluating $\tilde{F}_{\mathrm{c}}^{(M)}$ for large $M$ imposes 
a computational bottleneck long before numerical precision limits are reached. 
For instance, a coarse resolution of $\tau=0.01$ yields an approximation error 
proportional to $\tau^2$. Assuming an order-unity prefactor, this is approximately $10^{-4}$, 
which permits the use of $t_{\mathrm{c}}\approx 0.01$. 
In practice, $t_{\mathrm{c}}$ will be substantially larger for large systems.

To justify our choice of constraining $f_{\text{err,2}}(\tau)$
to be two orders of magnitude smaller than the target $t_{\mathrm{c}}$, we quantify
the sensitivity of $f_{\text{err}}(\tau,\epsilon_{\text{eff}})$ 
to truncation error by differentiating Eq.~\eqref{eq:auxfunc-error-function} with 
respect to $\epsilon_{\text{eff}}$, yielding:
\begin{equation}
  \begin{split}
    \pdv{f_{\text{err}}}{\epsilon_{\text{eff}}} &=\frac{1}{2\mathcal{N}\tau^2}\Bigl[\abs{\mathcal{C}(\im\tau)}^2
                                                 + \frac{1}{2}\Bigl(\abs{\mathcal{C}(\tau)}^2
                                                 +\abs{\mathcal{C}(-\tau)}^2\Bigr)+2\Bigr].
  \end{split}
  \label{eq:sensitivity}
\end{equation}
We note that Eq.~\eqref{eq:sensitivity} is valid only in the regime where the 
error function is not dominated by numerical precision errors (\ie $\tau \geq \tau_{\text{opt}}$).

Utilizing the sensitivity derived in Eq.~\eqref{eq:sensitivity}, and 
employing a first-order Taylor expansion, we can explicitly estimate an
upper bound ($\tilde{F}_{\mathrm{c}}^{(M)}\leq 1$)
of the maximum permissible truncation error $\epsilon_{\text{eff, max}} > 0$ 
required to preserve a desired maximum error $\Delta\tilde{F}_{\mathrm{c},\text{max}}^{(M)}$
for the target quantity:
\begin{equation}
\Delta \tilde{F}_{\mathrm{c},\text{max}}^{(M)} = \pdv{f_{\text{err}}}{\epsilon_{\text{eff}}}\epsilon_{\text{eff, max}}
  \label{eq:max_tolerance}
\end{equation}
During the prescreening phase, we compute the total accumulated
truncation error, assuming that all round-off errors also accumulate, resulting in the
bound $\mathcal{E}_{\text{bound}}=\sum_{i=0}^M\binom{N}{i}\epsilon_{\text{eff}}$.
This error bound combined with the maximum permissible truncation error serves 
as a diagnostic for identifying numerical instabilities within the procedure.
Employing our safety margin, we have not observed these numerical instabilities.
\section{Computational details}
\label{ssec:comp-details}
The equilibrium structures and harmonic vibrational force fields of the 
full-dimensional benzene and coronene systems were determined using
the Turbomole software package~\cite{turbomole7.7} (Version 7.7).
Molecular structures were energy-minimized to a gradient norm below the convergence threshold of \SI{E-4}{\hartree\per\bohr} using density functional theory (DFT) for the electronic ground state and time-dependent DFT (TD-DFT) for electronically excited states at the B3LYP/def2-TZVP level of theory~\cite{becke:1988,lee:1988,weigend:2005}
with the resolution of the identity method for the Coulomb part (RI-J)~\cite{eichkorn:1995} enabled on 
the numerical integration grid 4 with D3 dispersion correction~\cite{grimme:2010} with damping (BJ)~\cite{grimme:2011}.
The self-consistent field convergence criterion was chosen to be \SI{E-9}{\hartree}.
More details and results of the quantum chemical calculations are given in the supplementary material.
The vibrational frequencies and normal modes were obtained in the harmonic approximation
of the respective potential energy surfaces at their respective equilibrium structures.
All vibronic structure results presented in this paper were obtained with a modified version of the \texttt{hotFCHT} code~\cite{berger:1998,jankowiak:2007,huh:2012}.
\subsection{Procedure for determining phases of transition matrix moments}
\label{ssec:phases-determination}
To obtain the derivative parameters $\udl{\chi}$ 
and $\bm{X}$, we resort to numerical differentiation. Determining 
$\udl{\chi}$ requires at least $N n_{\text{gp}}$ individual electronic structure 
calculations, where $n_{\text{gp}}>1$ denotes the number of symmetrically
distributed grid points along the displaced mode. Similarly, computing $\bm{X}$ 
necessitates $(N(N-1)/2)n_{\text{gp}}^2$ calculations, with 
$n_{\text{gp}}>2$. Although this procedure is computationally demanding, 
it remains feasible on contemporary high-performance computing resources, 
particularly when molecular symmetry is utilized and the embarrassingly parallel nature
of the electronic structure calculations is exploited.

A fundamental technical obstacle involves the arbitrary global phase 
associated with the wavefunction in electronic structure software.
For properties that relate to expectation values, this phase cancels; 
however, for properties involving the transition from one electronic state
to another, the unrelated, relative phases of the initial and final wavefunctions are important
to keep fixed.
As electronic structure calculations for different arrangements of the nuclei
are usually performed independently, determining a smooth hypersurface
of transition properties
requires selecting consistent phases (or the sign in case of real wavefunctions)
from the $2^{N n_{\text{gp}}}$ and $2^{(N(N-1)/2)n_{\text{gp}}^2}$ possible 
permutations. If a related property (\eg the magnetic transition dipole 
moment) varies slowly and does not change sign across the displacement 
range, phase determination becomes straightforward~\cite{kerber:2025}. 
However, if the electric transition dipole moment vanishes due to symmetry 
(as in the first-order Herzberg--Teller effect with symmetry-forbidden
Franck--Condon transition), the conventional approach 
involves first aligning phases within each displaced mode and subsequently 
correcting the relative phases of the modes compared to a chosen reference. 
This manual alignment is very tedious for the determination of second-order 
parameters; thus, a partially automated procedure becomes necessary.

Here, we report the procedure employed to obtain parameters for
large-dimensional systems, relying on the assumption that the property
hypersurface spanned by the data points is smooth. We employed
the Turbomole program package~\cite{turbomole:2020,furche:2016} and Time-Dependent
Density Functional Theory (TD-DFT), but the general procedure is transferable to
other electronic structure codes.

The components of the electronic part of the electric transition dipole moment are calculated
by contracting the TD-DFT transition vector with the occupied--virtual
block of the dipole moment matrix. The full matrix
representation of the dipole moment operator in the molecular orbital (MO) basis
$\bm{R}_{r}^{\text{MO}}\in
\mathbb{R}^{N_{\text{total}}^{\text{MO}}\times
N_{\text{total}}^{\text{MO}}}$, with
$N_{\text{total}}^{\text{MO}} = N_{\text{occ}}+N_{\text{virt}}$, is
related to its matrix representation in the atomic orbital (AO) basis by
\begin{equation}
  \bm{R}_{r}^{\text{MO}}
  =
  \left(\bm{C}^{\text{MO}\leftarrow\text{AO}}\right)^\intercal
  \bm{R}_{r}^{\text{AO}}
  \bm{C}^{\text{MO}\leftarrow\text{AO}},
  \label{eq:dipole-full-mo}
\end{equation}
where $\bm{C}^{\text{MO}\leftarrow\text{AO}}$ contains the
molecular orbital coefficients from an expansion as a linear combination of
atomic orbitals. The column vector
$\udl{c}_p\in\mathbb{R}^{N_{\text{AO}}}$ represents the AO expansion
coefficients of the $p$-th MO. In the following, $i$ labels occupied
molecular orbitals (MOs) and $a$ labels virtual (unoccupied) MOs. The real-valued
TD-DFT eigenvector of length $N_{\text{occ}}N_{\text{virt}}$, defined up to an arbitrary
phase, can be arranged as
\begin{equation}
  \bm{O}^{\text{MO}}
  \in
  \mathbb{R}^{N_{\text{occ}}\times N_{\text{virt}}},
\end{equation}
and,
\begin{equation}
  \mu_{r=x,y,z}
  =
  \trace
  \left[
  \left(\bm{O}^{\text{MO}}\right)^\intercal
  \bm{R}_{r,\text{occ},\text{virt}}^{\text{MO}}
  \right],
  \label{eq:tdm-calc-mo}
\end{equation}
where $\bm{R}_{r,\text{occ},\text{virt}}^{\text{MO}} \in
\mathbb{R}^{N_{\text{occ}}\times N_{\text{virt}}}$ is the
occupied--virtual block of the dipole moment operator of component $r$
in the MO basis. It is obtained from
\begin{equation}
  \bm{R}_{r,\text{occ},\text{virt}}^{\text{MO}}
  =
  \left(\bm{C}_{\text{occ}}^{\text{MO}\leftarrow\text{AO}}\right)^\intercal
  \bm{R}_r^{\text{AO}}
  \bm{C}_{\text{virt}}^{\text{MO}\leftarrow\text{AO}},
  \label{eq:dipole-ov-block}
\end{equation}
where $\bm{R}_{r}^{\text{AO}}\in
\mathbb{R}^{N_{\text{AO}}\times N_{\text{AO}}}$ represents the dipole
moment operator of component $r$ in the atomic orbital (AO) basis, and
$\bm{C}_{\text{occ}}^{\text{MO}\leftarrow\text{AO}}$ and
$\bm{C}_{\text{virt}}^{\text{MO}\leftarrow\text{AO}}$ are the occupied
and virtual blocks of the MO coefficient matrix,
$\bm{C}_{\text{occ}}^{\text{MO}\leftarrow\text{AO}} \in
\mathbb{R}^{N_{\text{AO}}\times N_{\text{occ}}}$ and
$\bm{C}_{\text{virt}}^{\text{MO}\leftarrow\text{AO}} \in
\mathbb{R}^{N_{\text{AO}}\times N_{\text{virt}}}$.
The molecular orbital coefficient matrix 
$\bm{C}^{\text{MO}\leftarrow\text{AO}} \in \mathbb{R}^{N_{\text{AO}}\times N_{\text{AO}}}$ 
comprises the columns of generalized eigenvectors that solve the Kohn--Sham eigenvalue 
problem~\cite{furche:2016,helgaker:2013}.
Because the AO basis is generally nonorthogonal,
$\bm{C}^{\text{MO}\leftarrow\text{AO}}$ is not an orthogonal matrix in
the Euclidean sense. Instead,
\begin{equation}
  \left(\bm{C}^{\text{MO}\leftarrow\text{AO}}\right)^\intercal
  \bm{S}^{\text{AO}}
  \bm{C}^{\text{MO}\leftarrow\text{AO}}
  =
  \bm{I},
\end{equation}
where $\bm{S}^{\text{AO}} \in \mathbb{R}^{N_{\text{AO}} \times N_{\text{AO}}}$
denotes the atomic orbital overlap matrix with elements $S_{\mu\nu}^{\text{AO}}
= \braket{\chi_\mu}{\chi_\nu}$.

We note that each column vector $\udl{c}_p$ in $\bm{C}^{\text{MO}\leftarrow\text{AO}}$
is defined up to an arbitrary overall phase. We remove phase ambiguity
by realizing that 
Eq.~\eqref{eq:tdm-calc-mo} can equivalently be expressed in the AO
basis by introducing the AO transition-density matrix
\begin{equation}
  \bm{O}^{\text{AO}}
  =
  \bm{C}_{\text{occ}}^{\text{MO}\leftarrow\text{AO}}
  \bm{O}^{\text{MO}}
  \left(\bm{C}_{\text{virt}}^{\text{MO}\leftarrow\text{AO}}\right)^\intercal .
  \label{eq:o-ao-definition}
\end{equation}
This real matrix $\bm{O}^{\text{AO}}$ has dimension $N_{\text{AO}}\times N_{\text{AO}}$.
The electric transition dipole moment can therefore be written as
\begin{equation}
  \mu_{r=x,y,z}
  =
  \trace
  \left[
  \left(\bm{O}^{\text{AO}}\right)^\intercal
  \bm{R}_r^{\text{AO}}
  \right].
  \label{eq:tdm-calc-ao}
\end{equation}
In this form, the phase information of the TD-DFT eigenvector is
contained in $\bm{O}^{\text{AO}}$.

Because the displacements of the molecular structure of the nuclei along modes
are typically chosen sufficiently small for accurate numerical differentiation, we may assume
that $\bm{O}^{\text{AO}}$ and $\bm{R}_r^{\text{AO}}$ are only perturbed
slightly from one displaced structure to the next. This assumption also underlies
the rationale for the Taylor expansion of the transition
dipole moment, whose derivatives with respect to the displacement are calculated here.
To compute the derivatives we require the relative phase of the
transition-density matrices at two neighboring displaced structures $n$
and $n+1$ to be consistent, \ie
\begin{equation}
  s_{\mu\nu}^{(n+1,n)}
  =
  \mathrm{sgn}
  \left(
  O_{\mu\nu}^{(n+1)\text{AO}}/
  O_{\mu\nu}^{(n)\text{AO}}
  \right)
  =
  1
  \qquad
  \forall \mu,\nu ,
  \label{eq:phase-condition}
\end{equation}
up to the numerical noise of individual matrix elements. In practice, a
threshold is applied to the elements of $\bm{O}^{(n)\text{AO}}$ to avoid
divisions by values close to zero. Each retained $\mu,\nu$ pair provides a binary
phase vote $s_{\mu\nu}^{(n+1,n)}$. From the numbers of positive votes $N_+$
and negative votes $N_-$, we define a phase-consistency measure
$C=\frac{|N_+-N_-|}{N_++N_-}$.
Thus, $C=1$ indicates full agreement on the relative phase across all evaluated components, whereas $C=0$ corresponds to a split distribution of votes and hence
to a completely ambiguous phase assignment.
To determine whether an overall phase flip occurred between $\bm{O}^{(n)\text{AO}}$ and
$\bm{O}^{(n+1)\text{AO}}$, the sign is assigned by majority vote ($N_+ > N_-$ or $N_- > N_+$).
This measure does not assess the
absolute sign, which is arbitrary, but only the consistency of the relative
phase inferred from redundant information.

The least reliable situation occurs when the reference structure is the
equilibrium structure and belongs to a point group in which the considered
transition is symmetry-forbidden. In the exact high-symmetry limit, all
components used for the phase comparison vanish (\ie the transition moment is zero),
so that no well-defined sign can be assigned to them.
The phase-alignment procedure is then singular in exact arithmetic:
the relative phase between the equilibrium structure and neighboring distorted structures cannot be inferred
from a vector whose norm is zero. In practice, this singular case can be avoided
either by computing the equilibrium structure without imposing point-group
symmetry, i.e. in $C_1$, thereby introducing small numerical symmetry breaking,
or by using a slightly distorted structure as the reference structure. In our
benchmarks, treating benzene and coronene in $C_1$ rather than in
$D_{6\mathrm{h}}$ symmetry was sufficient.

To maximize the overlap between nearby displaced structures $n$ and
$n+1$, a basis-set projection can be applied. We define the mixed AO
overlap matrix as
\begin{equation}
  S_{\mu\nu}^{m,n}
  =
  \braket{
  \chi_\mu^{(m)}}{
  \chi_\nu^{(n)}} ,
\end{equation}
where $\chi_\mu^{(m)}$ and $\chi_\nu^{(n)}$ are scalar-valued AO basis
functions in real space associated with displaced structures $m$ and $n$,
respectively.
In general, $\bm{S}^{m,n} \in \mathbb{R}^{N_{\text{AO}}^{(m)} \times N_{\text{AO}}^{(n)}}$, but here we can take $N_{\text{AO}}^{(m)}$ as identical to $N_{\text{AO}}^{(n)}$ ($= N_{\text{AO}}$).
The MO coefficient matrix from displaced structure $n+1$, projected into
the AO basis of displaced structure $n$, is then obtained as
\begin{equation}
  \widetilde{\bm{C}}^{(n+1)\rightarrow n}
  =
  \left(\bm{S}^{n,n}\right)^{-1}
  \bm{S}^{n,n+1}
  \bm{C}^{(n+1)} .
  \label{eq:basis-projection-ansatz}
\end{equation}

The real projected coefficient matrix $\widetilde{\bm{C}}^{(n+1)\rightarrow n}$ has dimension
$N_{\text{AO}}^{(n)} \times N_{\text{MO}}^{(n+1)}$.
The occupied and virtual blocks of
$\widetilde{\bm{C}}^{(n+1)\rightarrow n}$, denoted by
$\widetilde{\bm{C}}_{\text{occ}}^{(n+1)\rightarrow n}$ and
$\widetilde{\bm{C}}_{\text{virt}}^{(n+1)\rightarrow n}$, are inserted in
Eq.~\eqref{eq:o-ao-definition} to construct the projected AO transition
density
\begin{equation}
  \widetilde{\bm{O}}^{(n+1)\rightarrow n,\text{AO}}
  =
  \widetilde{\bm{C}}_{\text{occ}}^{(n+1)\rightarrow n}
  \bm{O}^{(n+1)\text{MO}}
  \left(
  \widetilde{\bm{C}}_{\text{virt}}^{(n+1)\rightarrow n}
  \right)^\intercal,
  \label{eq:basis-projection}
\end{equation}
which may be utilized in Eq.~\eqref{eq:phase-condition} instead of $\bm{O}^{(n+1)\text{AO}}$
of Eq.~\eqref{eq:o-ao-definition}.
This real matrix $\widetilde{\bm{O}}^{(n+1)\rightarrow n,\text{AO}}$ is expressed
in the AO basis of displaced structure $n$ and has dimension $N_{\text{AO}}^{(n)} \times N_{\text{AO}}^{(n)}$.

We implemented the above procedure in the \texttt{nonorth} code~\cite{klues:2016,hansmann:2021}.
The routine processes two separate TD-DFT calculations and computes their
relative phases of the transition-density matrices.
We utilize these results to first compute all relative phases within a 
displacement along a single vibrational normal mode
with reference to the equilibrium structure without imposing point-group symmetry.
Subsequently, we determine the relative phases of the two-dimensional
hypersurfaces spanned by a combination of two simultaneously displaced modes,
where the displacement of one of the two modes always provides a reference phase.
An inconsistent phase assignment with this method
would produce a discontinuity in the property hypersurface, which would be readily
apparent upon inspection.
In practice, the procedure proved robust for all test cases,
indicating that the perturbations induced by structural distortions
are indeed sufficiently small. Consequently, more sophisticated methods
based on basis projection are currently not required, but are available.

In this first application of our procedure, we obtained visual confirmation
of phase-consistency of the phase-corrected hypersurfaces.
An alternative approach is to use fit residuals as a diagnostic:
large or systematic deviations between the fitted model
and the computed hypersurface data can indicate remaining phase inconsistencies
or other problematic data points.
\subsection{Symmetry considerations for the computation of electric transition dipole derivatives in coronene and benzene}
For the computation of the first and second derivatives
of the electronic electric transition dipole moment with respect
to displacements along the normal modes, the symmetry of the compounds was taken into account.
Both benzene and coronene belong to the D$_{6\mathrm{h}}$ point group in their 
equilibrium structures of the two electronic states considered in this work.
The electronic ground state transforms
according to the \state{}{A}{1}{g} irreducible representation.
The lowest excited singlet state transforms according 
to the \state{}{B}{2}{u} irreducible representation, assuming the standard 
orientation where the molecule lies in the $xy$-plane and the $z$-axis 
coincides with the principal $C_6$ axis (out-of-plane). Moreover, the 
molecule is oriented within the $xy$-plane such that the $x$-axis
intersects with more atoms, whereas the $y$-axis bisects the C--C bonds
orthogonally. Within the 
D$_{6\mathrm{h}}$ symmetry framework, the electric dipole moment operator 
transforms as \sym{e}{1}{u} for the in-plane components ($x,y$) and 
\sym{a}{2}{u} for the out-of-plane component ($z$). Consequently, the 
linear term of the first-order Herzberg--Teller (HT) effect
is symmetry-allowed for vibrational modes 
of \sym{e}{2}{g} symmetry (yielding non-zero $x/y$ electric transition dipole 
components) and \sym{b}{1}{g} symmetry (yielding a non-zero $z$ 
component).

The quadratic terms or the bilinear terms of the second-order HT effect are
allowed for any pair of normal modes whose direct 
product contains the \sym{e}{2}{g} or \sym{b}{1}{g} irreducible 
representations. Combinations which yield irreducible representations containing
e$_{2\mathrm{g}}$ symmetry are \cite{wilson:1955}:
$\mathrm{a}_{1\mathrm{g}}\otimes\mathrm{e}_{2\mathrm{g}}$, 
$\mathrm{a}_{2\mathrm{g}}\otimes\mathrm{e}_{2\mathrm{g}}$, 
$\mathrm{b}_{1\mathrm{g}}\otimes\mathrm{e}_{1\mathrm{g}}$, 
$\mathrm{b}_{2\mathrm{g}}\otimes\mathrm{e}_{1\mathrm{g}}$, 
$\mathrm{e}_{1\mathrm{g}}\otimes\mathrm{e}_{1\mathrm{g}}$, 
$\mathrm{e}_{2\mathrm{g}}\otimes\mathrm{e}_{2\mathrm{g}}$, 
$\mathrm{a}_{1\mathrm{u}}\otimes\mathrm{e}_{2\mathrm{u}}$, 
$\mathrm{a}_{2\mathrm{u}}\otimes\mathrm{e}_{2\mathrm{u}}$, 
$\mathrm{b}_{1\mathrm{u}}\otimes\mathrm{e}_{1\mathrm{u}}$,
$\mathrm{b}_{2\mathrm{u}}\otimes\mathrm{e}_{1\mathrm{u}}$, 
$\mathrm{e}_{1\mathrm{u}}\otimes\mathrm{e}_{1\mathrm{u}}$, and
$\mathrm{e}_{2\mathrm{u}}\otimes\mathrm{e}_{2\mathrm{u}}$.
Combinations which yield irreducible representations containing
b$_{1\mathrm{g}}$ symmetry include \cite{wilson:1955}
$\mathrm{a}_{1\mathrm{g}}\otimes\mathrm{b}_{1\mathrm{g}}$, 
$\mathrm{a}_{2\mathrm{g}}\otimes\mathrm{b}_{2\mathrm{g}}$, 
$\mathrm{a}_{1\mathrm{u}}\otimes\mathrm{b}_{1\mathrm{u}}$, 
$\mathrm{a}_{2\mathrm{u}}\otimes\mathrm{b}_{2\mathrm{u}}$, 
$\mathrm{e}_{1\mathrm{u}}\otimes\mathrm{e}_{2\mathrm{u}}$, and 
$\mathrm{e}_{1\mathrm{g}}\otimes\mathrm{e}_{2\mathrm{g}}$.
Consequently, non-vanishing second-order derivatives of the electric transition
dipole moment occur exclusively for displacements along these 
specific mode combinations.
Due to the restricted size of benzene, no vibrational normal modes of
$\mathrm{b}_{1\mathrm{g}}$ symmetry exist.
\section{Additional application results}
\label{sec:examples}
\subsection{Illustrative examples}
The present implementation is benchmarked by comparison with 
results reported by Jnane \etal~\cite{jnane:2021},
utilizing parameters of a two-mode reduced-dimensional model system of phenanthrene~\cite{craig:1969,small:1971},
and a three-mode reduced-dimensional system of benzene using parameters of~\cite{fischer:1981}
and~\cite{berger:1998}.

In Ref.~\cite{jnane:2021} a 2-mode model of the 
S$_1$ ($^1\mathrm{A}_1$) $\leftarrow$ S$_0$ ($^1\mathrm{A}_1$) transition of phenanthrene,
which is a weakly allowed FC transition in the electric dipole approximation,
and a truncated 3-mode model of the  
S$_1$ ($^1$B$_{2\mathrm{u}}$) $\leftarrow$ S$_0$ ($^1$A$_{1\mathrm{g}}$)
transition of benzene, which is FC-forbidden in the electric dipole approximation,
were chosen to illustrate the auxiliary function method~\cite{jnane:2021}. 

In this section, the zero-temperature implementation of the auxiliary function method
and our finite-temperature extension of the auxiliary function method for the 
first-order HT effect are validated against the analytic 
implementation in \texttt{hotFCHT}~\cite{huh:2011a}.
Moreover, we test our implementation on the three-mode parametrization of the
benzene system.

\begin{figure*}
  \begin{minipage}{.45\linewidth}
    \includegraphics[width=\linewidth]{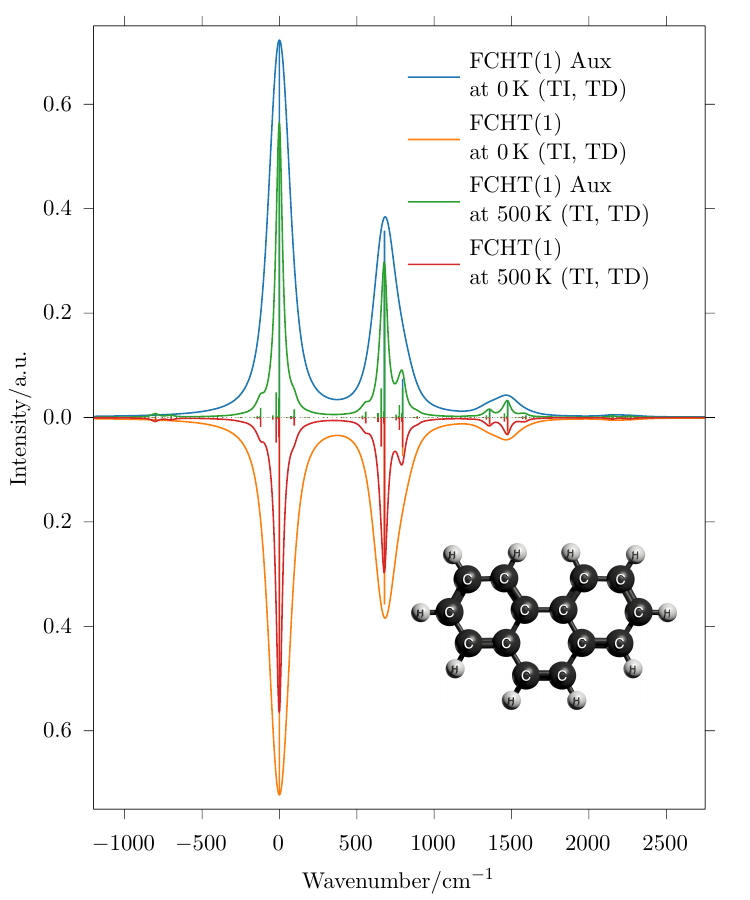}
  \end{minipage}
  \begin{minipage}{.45\linewidth}
    \includegraphics[width=\linewidth]{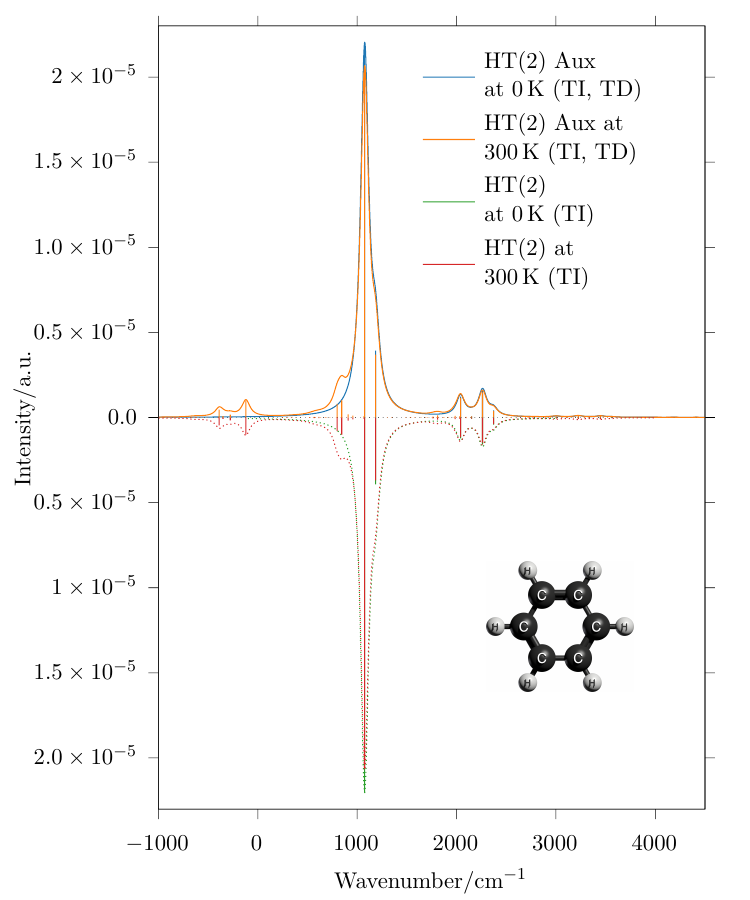}
  \end{minipage}
  \caption{Left: computed absorption profiles of the
    $\mathrm{S}_1\leftarrow\mathrm{S}_0$ transition of phenanthrene
    utilizing a reduced two-mode parametrization from Ref.~\cite{small:1971}
    within the time-independent (TI) and time-dependent (TD) frameworks 
    at (\SI{0}{\kelvin},\SI{500}{\kelvin}) using parameters from Ref.~\cite{small:1971}.
  Franck--Condon (FC) and linear Herzberg--Teller (HT) effects are considered.
  A Lorentzian line shape with a full width at half maximum (FWHM) of (\SI{100}{\per\centi\meter}, \SI{50}{\per\centi\meter})
  was employed for the TD framework.
  Right: Computed absorption profiles of the ($\mathrm{S}_1\leftarrow\mathrm{S}_0$)
  transition of benzene utilizing a reduced three-mode parametrization with parameters
  from Ref.~\cite{fischer:1981} within the TI and TD frameworks at \SI{0}{\kelvin} and \SI{300}{\kelvin}.
  Only quadratic HT effects are considered.
  A Lorentzian line shape with a FWHM of \SI{100}{\per\centi\meter}
  was employed for the TD framework. The dotted lines are the result of a manual
  convolution using the analytical TI results.
  In both panels: Comparison between auxiliary function method (Aux) and the analytical method.
  As scale on the x-axis the wavenumber in excess of the $0'-0$ transition is chosen.
  A Fast Fourier Transform (FFT) grid size of $2^{13}$, a time increment of $\Delta t\sim\SI{2}{\femto\second}$, and a
  time interval of $(-\SI{16.676169}{\pico\second},\SI{16.676169}{\pico\second})$
  were employed for the TD framework. A wavenumber graining of \SI{1}{\per\centi\meter}
  was employed for the TI framework.
}
  \label{fig:phenanthrene-benze1g}
\end{figure*}

Figure~\ref{fig:phenanthrene-benze1g} displays the computed 
zero-temperature transition profiles, calculated via the 
auxiliary function TI and TD methods, alongside their respective finite-temperature profiles
for both the two-mode model for the phenanthrene system and three-mode model for the benzene system
employed in Ref.~\cite{jnane:2021}. Comparison with the analytic implementation reveals that 
the auxiliary function method reproduces the transition profiles with 
negligible deviation.

It should be noted that the intensities calculated
in the original phenanthrene publication~\cite{small:1971}
deviate from our results. Having thoroughly validated our linear FCHT integral calculation
approach~\cite{berger:1998,coriani:2010,huh:2012}, we attribute these deviations to finite
numerical precision in the reported parameter sets and to truncation errors
in the calculation of multidimensional FCIs as performed in Ref.~\cite{small:1971}.
Our implementation of the auxiliary function method is consistent with the 
results of Ref.~\cite{jnane:2021} at \SI{0}{\kelvin}.
Minor discrepancies in the spectra are 
attributable to the use of the StrawberryFields package~\cite{killoran:2019} 
in the reference work, which expresses the Gaussian operator decomposition 
as a product of gates that are each represented in a finite basis,
thereby introducing basis set truncation errors. 
In contrast, the present implementation utilizes
recurrence relations that are explicitly described in Sec.~\ref{ssec:rr}, which
do not suffer from these truncation errors.
\subsection{Full-mode benzene model}

The S$_1$ ($^1$B$_{2\mathrm{u}}$) $-$ S$_0$ ($^1$A$_{1\mathrm{g}}$) 
transition of benzene has been studied extensively, both experimentally and 
theoretically, see \eg Refs.~\cite{duch:1983,turkevich:1942,knight:1975a,knight:1975b,
stephenson:1984a,berger:1998,coriani:2010,fally:2009}. In absorption spectroscopy experiments
of jet-cooled benzene that probe the ($^1\mathrm{B}_{2\mathrm{u}} \leftarrow {}^1\mathrm{A}_{1\mathrm{g}}$)
electronic transition, specific vibronic transitions could not be explained by linear HT coupling 
alone~\cite{stephenson:1984a}, rendering this system suitable for benchmarking 
our method, which includes HT effects up to second order.
For a direct comparison with our computed transition absorption profiles, we
benchmark our results against more recent experimental UV absorption
cross-section measurements of benzene~\cite{fally:2009}.

In Tab.~\ref{tab:benzene-energies-comparison} we list the computed harmonic vibrational
wavenumbers from our quantum-chemical calculations and compare them, where possible,
to experimentally obtained values for the vibrational fundamental transition wavenumbers.
The results are consistent with 
those obtained using similar density functional theory (DFT) 
methods at the B3LYP/def2-TZVP level of theory~\cite{coriani:2010}.
We therefore expect the computed absorption spectral profiles to be of comparable quality.
\begin{table}\centering
  \caption{Harmonic vibrational wavenumbers $\tilde{\omega}$ of benzene for the ground
    ($\mathrm{S}_0$, A$_{1\mathrm{g}}$ symmetry) and excited state ($\mathrm{S}_1$, B$_{2\mathrm{u}}$ symmetry)
  arranged in descending order within each irreducible representation $\Gamma$, compared to experiment.
  Unless otherwise noted, experimental values of corresponding
  fundamental transition wavenumbers $\tilde{\nu}$ are from Ref.~\cite{atkinson:1978}.
The Wilson numbering scheme~\cite{wilson:1934} is given in parentheses. }
  \label{tab:benzene-energies-comparison}
  \begin{threeparttable}
  \begin{tabular}{cc
      S[table-format=4.2]
      c
      S[table-format=4.2]
      c
    }\toprule
    &&\multicolumn{2}{c}{S$_0$}&\multicolumn{2}{c}{S$_1$}\\
  Mode&$\Gamma$&\multicolumn{2}{c}{{$\tilde{\omega}$/cm$^{-1}$} }&
  \multicolumn{2}{c}{{$\tilde{\omega}'$/cm$^{-1}$} }\\\midrule
  $\nu_{1}$ ($\nu_2$)&a$_{1\mathrm{g}}$&3195.66   &(3073)&3221.13&(3130.1)\\
    $\nu_{2}$ ($\nu_1$)&a$_{1\mathrm{g}}$&1018.09 &(993.1)& 960.95&(923.0)\\
    $\nu_{3}$ ($\nu_3$)&a$_{2\mathrm{g}}$&1387.08 &(1350)&1369.38&\\
    $\nu_{4}$ ($\nu_5$)&b$_{2\mathrm{g}}$& 993.14 &(990)& 761.76&(744.6)\\
    $\nu_{5}$ ($\nu_4$)&b$_{2\mathrm{g}}$& 718.61 &(707)& 326.02&(365)\\
    $\nu_{6}$ ($\nu_{10}$)&e$_{1\mathrm{g}}$ &861.36 & (846)& 575.50&(580.7)\\
    $\nu_{7}$ ($\nu_{7}$)&e$_{2\mathrm{g}}$&3170.31  &(3056)&3194.57&(3077.2)\\
    $\nu_{8}$ ($\nu_{8}$)&e$_{2\mathrm{g}}$&1637.06  &(1599)&1569.29&(1516\tnote{a}~)\\
    $\nu_{9}$ ($\nu_{9}$)&e$_{2\mathrm{g}}$&1199.74  &(1178)&1179.40&(1147.7)\\
    $\nu_{10}$ ($\nu_6$)&e$_{2\mathrm{g}}$& 624.60 &(608.3)& 533.49&(521.4)\\
    $\nu_{11}$ ($\nu_{11}$)&a$_{2\mathrm{u}}$& 687.20 &(674.0)& 547.45&(515--518)\\
   $\nu_{12}$ ($\nu_{13}$)&b$_{1\mathrm{u}}$&3160.74 &(3057)&3190.15&\\
   $\nu_{13}$ ($\nu_{12}$)&b$_{1\mathrm{u}}$&1030.09 &(1010)&1010.13&\\
   $\nu_{14}$ ($\nu_{14}$)&b$_{2\mathrm{u}}$&1336.87 &(1309)&1468.70&(1570)\\
   $\nu_{15}$ ($\nu_{15}$)&b$_{2\mathrm{u}}$&1176.86 &(1146)&1176.58&(1148)\\
   $\nu_{16}$ ($\nu_{20}$)&e$_{1\mathrm{u}}$&3185.75 &(3064)&3209.87&\\
   $\nu_{17}$ ($\nu_{19}$)&e$_{1\mathrm{u}}$&1517.76 &(1482)&1451.29&($\approx 1330$\tnote{b}~)\\
   $\nu_{18}$ ($\nu_{18}$)&e$_{1\mathrm{u}}$&1061.89 &(1037)& 968.12&(919)\\
   $\nu_{19}$ ($\nu_{17}$)&e$_{2\mathrm{u}}$& 962.25 &(967)& 725.13&(718.2)\\
   $\nu_{20}$ ($\nu_{16}$)&e$_{2\mathrm{u}}$& 410.93 &(398.8)& 251.63&(237.5)\\
    \bottomrule
  \end{tabular}
  \begin{tablenotes}
    \item[a] Ref.~\cite{stephenson:1984a}.
    \item[b] Ref.~\cite{aron:1980}.
  \end{tablenotes}
\end{threeparttable}
\end{table}
\begin{figure*}\centering
  \begin{minipage}{.45\linewidth}
    \includegraphics[width=\linewidth]{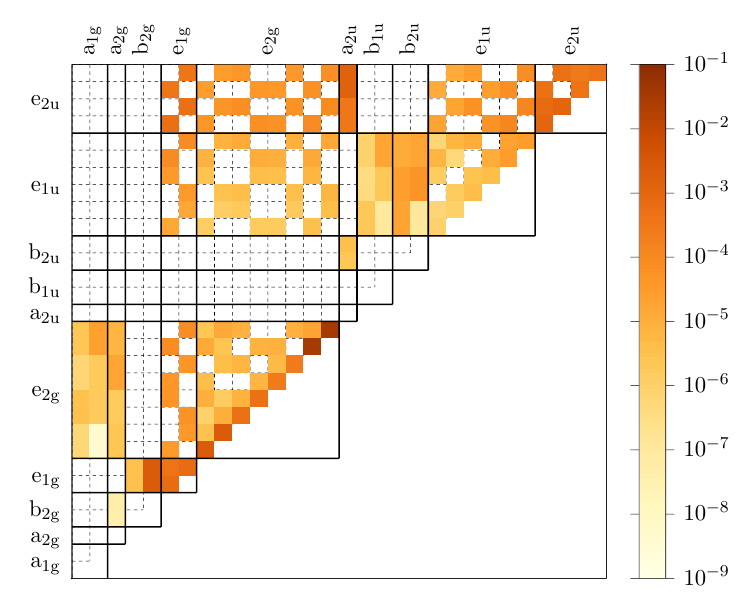}
  \end{minipage}
  \begin{minipage}{.45\linewidth}
    \includegraphics[width=\linewidth]{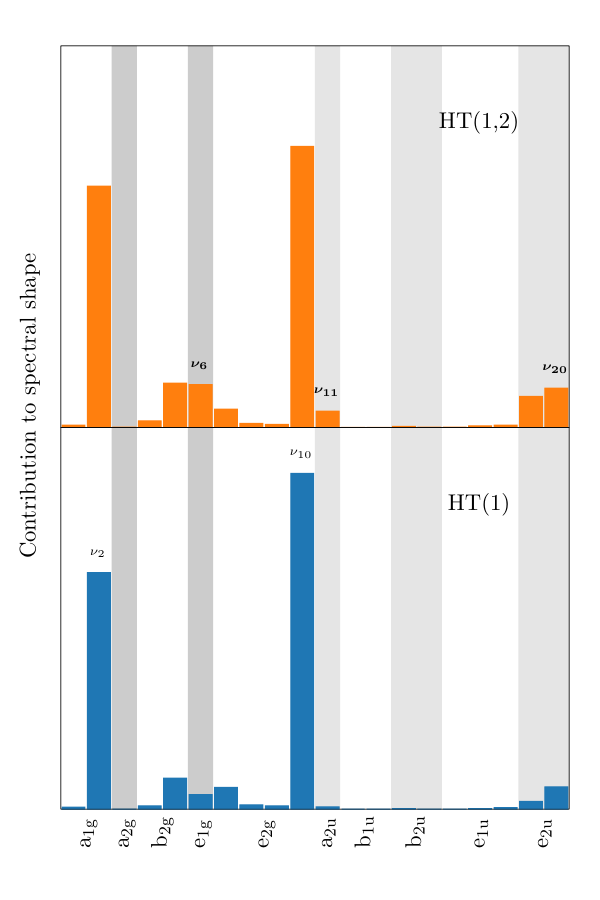}
  \end{minipage}
  \caption{
    Left: Ground state mode-resolved contributions to the total intensity derived 
  from Eq.~\eqref{eq:norm-constant} with $\mu_r^{(0)}=0$ due to the Franck--Condon
  symmetry-forbidden nature of the $(\mathrm{S}_1\leftarrow\mathrm{S}_0)$ transition
  in benzene. Linear terms $(\mathrm{e}_{2\mathrm{g}})$ are added to the diagonal elements. The vibrational 
  modes are arranged in descending order of their harmonic vibrational wavenumbers within each 
  irreducible representation. Both components of degenerate modes are shown.
  Solid lines mark the boundaries between different
  irreducible representations; dotted lines indicate individual vibrational 
  modes. Right: Mode contributions calculated by evaluating the auxiliary 
  function linear combination of Eq.~\eqref{eq:sum-rule-2} with $n=0$ for 
  benzene at \SI{0}{\kelvin}.
  Contributions from degenerate modes with identical harmonic vibrational wavenumbers are combined.
  Treatment of the linear Herzberg--Teller (HT) 
  effect is denoted HT(1); inclusion of both linear and quadratic effects 
  is denoted HT(1,2).
  Select mode contributions with considerable increase due to second-order HT effects are denoted in bold.
}
  \label{fig:benzene-mode-contributions}
\end{figure*}
\begin{figure*}
  \begin{minipage}{.45\linewidth}
    \includegraphics[width=\linewidth]{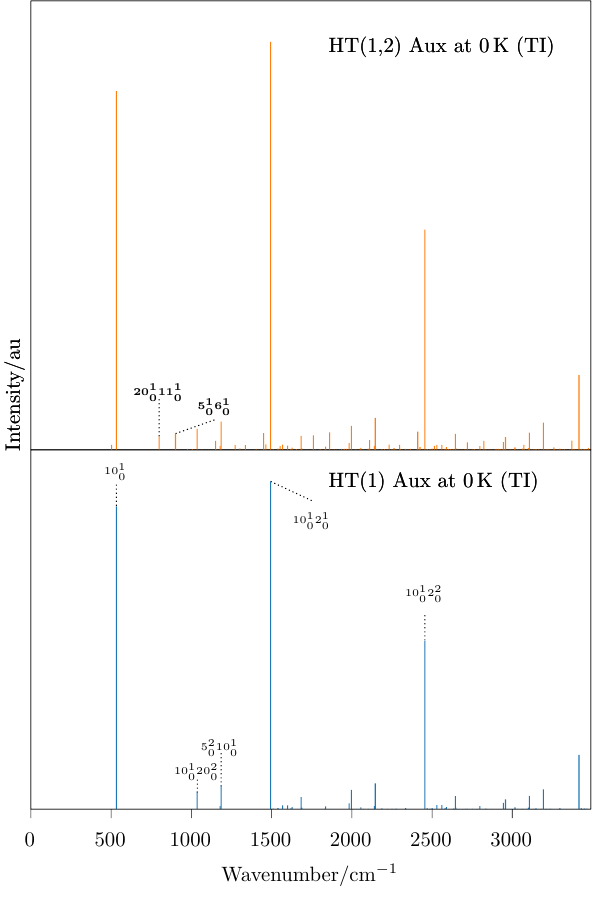}
  \end{minipage}
  \begin{minipage}{.45\linewidth}
    \includegraphics[width=\linewidth]{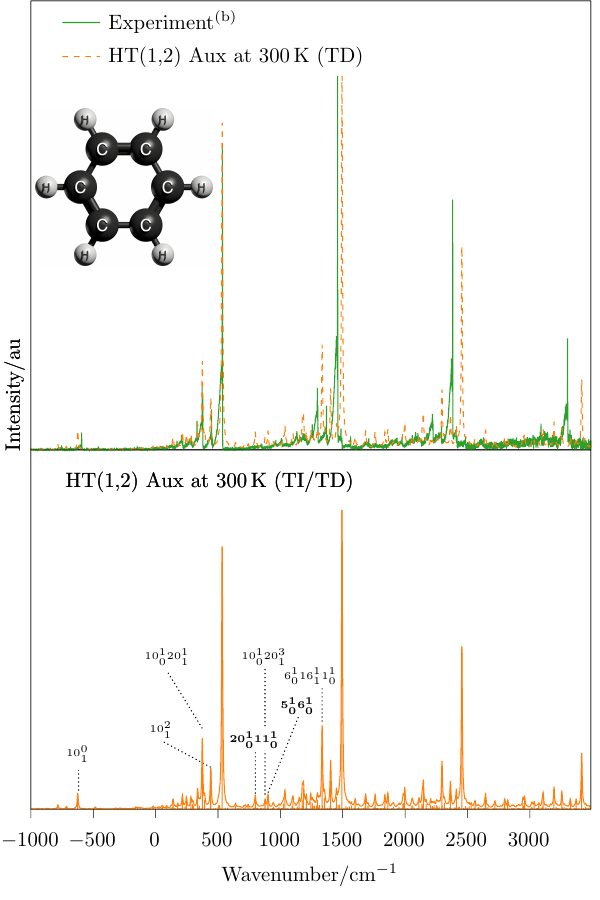}
  \end{minipage}
\caption{
  Computed absorption profiles of the $\mathrm{S}_1\leftarrow \mathrm{S}_0$
transition of benzene at various orders of Herzberg--Teller (HT) coupling
employing the auxiliary function method within the time-independent (TI)
framework at \SI{0}{\kelvin} (left) as well as TI and time-dependent (TD) frameworks at
\SI{300}{\kelvin} (right), compared to the experimental UV Fourier transform
absorption cross sections $(\mathrm{b})$ of Fally \etal from Ref.~\cite{fally:2009} at \SI{293}{\kelvin}.
Additionally, the TD \SI{300}{\kelvin} absorption profile is superimposed onto the experimental spectrum in the top right.
  The experimental spectrum was normalized with respect to the most intense peak.
A Lorentzian line shape with a FWHM of $\SI{10}{\per\centi\meter}$, a
Fast Fourier Transform (FFT) grid size of $2^{14}$, a time increment of
$\Delta t\sim\SI{2}{\femto\second}$, and a time interval of
$(-\SI{16.677187}{\pico\second},\SI{16.677187}{\pico\second})$ were employed for the TD framework.
Peak assignments correspond to the notation of Tab.~\ref{tab:benzene-energies-comparison}.
Treatment of the linear Herzberg--Teller (HT) effect is denoted HT(1);
inclusion of both linear and quadratic effects HT(1,2).
Select transitions due to second-order HT effects are denoted in bold letters.
For a detailed peak analysis considering HT(1) effects, see Ref.~\cite{berger:1998}.
}
\label{fig:benzene-abs-spec}
\end{figure*}

Analogously to the coronene results we display
the mode-resolved contributions to the total summed intensity $\mathcal{N}$
in terms of the ground-state modes and the importance of each final-state mode
for the distribution of intensity over the vibronic transitions; they are shown
in Fig.~\ref{fig:benzene-mode-contributions}. 
The dominant contribution from ground-state modes arises from mode $\nu_{10}$,
which was an expected finding from previous
analysis~\cite{berger:1998}. In our calculations, the largest contribution that is attributable
solely to the second-order HT effect is from the mode combination $\nu_4\otimes\nu_6$.
The total intensity at \SI{0}{\kelvin} considering first-order HT
effects is computed to be $\SI{8.17E-02}{\mathrm{D}^2}$, and considering up to second-order HT effects 
is computed to be $\SI{9.95E-02}{\mathrm{D}^2}$. In our calculations, quadratic and bilinear HT effects
therefore account for about 18\% of the total intensity in an HT expansion of up to second-order.

Comparing the contributions
of excitations in specific modes
between the linear and quadratic HT models, 
we see a noticeable increase in importance for the \sym{b}{2}{g}, \sym{e}{1}{g}
and \sym{e}{2}{u} modes. This finding is consistent with the 
rationale that inclusion of quadratic terms allows direct contributions 
from overtone and combination bands satisfying the selection rules (\eg 
$\mathrm{e}_{2\mathrm{u}} \otimes \mathrm{e}_{2\mathrm{u}}$ containing 
$\mathrm{e}_{2\mathrm{g}}$).
The increase in contributions is indicative of quadratic HT effects in the 
absorption spectra. However, we note that the mode contributions in the
excited state do not differ substantially between the HT(1) and HT(1,2) treatments;
we therefore expect the differences
in computed absorption spectra to be small.

In Fig.~\ref{fig:benzene-abs-spec} we show the absorption spectra of the
full-mode benzene system at various orders of HT coupling within the
time-independent and time-dependent frameworks, compared to an experimental absorption spectrum.
Overall, agreement with experiment~\cite{fally:2009} is satisfactory given the inherent 
limitations of the DFT functionals~\cite{coriani:2010}.
In particular, signals involving the $\nu_{10}$ mode and $\nu_2$ are well-represented.
The additional intensity recovered by including quadratic HT effects is minor,
which was anticipated from the prescreening analysis. Furthermore, the contributions 
from quadratic HT coupling appear in spectral regions already densely 
populated by linear HT intensity.

We restrict
our investigation of the peaks to the energy region below \SI{1000}{\per\centi\meter}, 
because the higher energies are mostly the result of progressions
with the \sym{a}{1}{g} mode $\nu_2$. 
The peaks assigned to $20_0^111_0^1$ 
in Ref.~\cite{stephenson:1984a} ($11_0^116_0^1$ in their notation)
resulting from an 
$\mathrm{e}_{2\mathrm{u}} \otimes \mathrm{a}_{2\mathrm{u}}$ coupling 
as well as $5_0^16_0^1$ ($5_0^110_0^1$ in the numbering scheme of Ref.~\cite{stephenson:1984a}) resulting from a 
$\mathrm{b}_{2\mathrm{g}} \otimes \mathrm{e}_{1\mathrm{g}}$ coupling
are also found in our data.
Similarly, the $5_0^16_0^1$ peak is present in the reduced three-mode model
employed in Ref.~\cite{jnane:2021}, albeit at a higher transition wavenumber. The reason
is that Ref.~\cite{jnane:2021} employed the wavenumbers of Ref.~\cite{berger:1998},
which are larger for the \sym{b}{2}{g} mode than in our present work (see Tab.~\ref{tab:benzene-energies-comparison}). 
Moreover, because Ref.~\cite{jnane:2021} normalized its computed spectral shapes, the importance
of this peak is not put into the context of the full spectral shape.
Our full-dimensional calculations show that, similar to the experiment~\cite{fally:2009},
the $5_0^16_0^1$ peak is comparatively small. 
As the linear terms dominate, quadratic 
terms are not strictly required to assign the major peaks in the experimental 
spectrum, though they refine the intensity distribution. For a detailed peak 
analysis considering linear HT effects, see Ref.~\cite{berger:1998}.

\end{document}